\documentstyle[preprint,tighten,eqsecnum,floats,aps,epsfig]{revtex}

\begin{document}

\draft

\title{ Electroweak quasielastic response functions in nuclear matter
}
\author{
M. B. Barbaro$^a$, A. De Pace$^a$, T.W. Donnelly$^b$ and A. Molinari$^a$
}
\address{
\em $^a$ Dipartimento di Fisica Teorica dell'Universit\`a
 di Torino and \\
 Istituto Nazionale di Fisica Nucleare, Sezione di Torino, \\
 via P.Giuria 1, I-10125 Torino, Italy \\
 $^b$ Center for Theoretical Physics, \\
 Laboratory for Nuclear Science and Department of Physics, \\
 Massachusetts Institute of Technology, Cambridge, MA 02139, USA
}

\maketitle

\begin{abstract}
Quasielastic electromagnetic and parity-violating electron scattering
response functions of relativistic nuclear matter are reviewed. 
The roles played by the Hartree-Fock field and by nuclear correlations
in the Random Phase Approximation (treated within the continued
fraction scheme) are illustrated. The parity-violating responses of
nuclei to polarized electrons are also revisited, stressing in
particular the crucial role played by the pion in the nuclear dynamics.
Finally, some issues surrounding scaling and sum rules are addressed.
\end{abstract}
\bigskip\bigskip

Content

\begin{itemize}
  \item[-] \ref{sec:intro}. Introduction

  \item[-] \ref{sec:QEP}.
    Quasielastic response functions for inclusive electron scattering
    {\em 
    \begin{itemize}
      \item[] A. Response functions
      \item[] B. Non-relativistic vs relativistic kinematics
      \item[] C. Free response
      \item[] D. Hartree-Fock response
      \item[] E. Random phase approximation response
      \item[] F. Effective particle-hole interaction
      \item[] G. Testing the model
    \end{itemize}
    }

  \item[-] \ref{sec:pv}. Parity-violating electron scattering  
    {\em 
    \begin{itemize}
      \item[] A. The asymmetry, the currents and the RFG responses
      \item[] B. The role of the pion and other mesons
      \item[] C. The axial response and the asymmetry
    \end{itemize}
    }

  \item[-] \ref{sec:sumrules}. Scaling and sum rules

  \item[-] \ref{sec:outlook}. Outlook and perspectives

\end{itemize}

\eject

\section{ Introduction }
\label{sec:intro}

Traditionally, studies of excitations of the nucleus via the
interactions between leptons and the nucleus have been centered mostly
on the longitudinal ($R_{\text{L}}$) and transverse ($R_{\text{T}}$) 
nuclear response functions explored in unpolarized, inelastic, 
{\em inclusive} $(e,e')$ electron scattering. The present paper will 
also be largely restricted to the investigation of these quantities, 
although the set of experimentally accessible responses is much
larger, including also semi-inclusive responses and responses that
arise when initial or final hadronic degrees of freedom are active. 
In the former, particles are detected in coincidence with the
scattered electron $((e,e'p),$ $(e,e'n),$ $(e,e'pn),$ etc.), whereas
the latter includes reactions such as ${\vec A}(e,e'),$ 
$A(e,e'{\vec p}\ ),$ etc. Of course, in all cases the incident electron
may also be polarized. A concerted effort is presently being placed on
experimental studies of this extended set of responses where special 
sensitivities to normally hidden aspects of nuclear structure are
expected to exist. Such studies are very challenging and thus only 
relatively recently have they been made feasible by advances in 
accelerators, in developments of polarized electron beams and polarized nuclear
targets, and in the construction of the required hadron polarimetry. 
Issues still surround the inclusive unpolarized responses themselves, however,
and since a successful level of understanding of the underlying
nuclear dynamics would be incomplete without a coherent picture of the
entire set of responses, unpolarized inclusive {\em and}
semi-inclusive/polarized, the former still deserve continued
study and so provide the focus for the present article.

Beyond the electromagnetic (EM: parity-conserving, vector) responses 
$R_{\text{L}}$ and $R_{\text{T}}$ our study will also include their 
parity-violating analogs $R_{\text{L}}^{\text{AV}}$ and 
$R_{\text{T}}^{\text{AV}}$ as well as the nuclear parity-violating 
axial response $R_{\text T'}^{\text{VA}}$. Here AV indicates that
the axial leptonic and vector hadronic currents enter; VA indicates 
the converse. This larger set of inclusive responses may be explored
via the inclusive scattering of longitudinally polarized electrons
from unpolarized nuclei, since the electron helicity asymmetry is 
parity-violating. The three new responses all arise from interferences
between the weak neutral current (WNC) and EM
current. In the cases of  $R_{\text{L}}^{\text{AV}}$ and 
$R_{\text{T}}^{\text{AV}}$ it is the vector part of the WNC that
enters and this is believed to be closely related to the EM current
(in the absence of strangeness content in the nucleus these 
two responses are tied to $R_{\text{L}}$ and $R_{\text{T}}$); in the
case of  $R_{\text T'}^{\text{VA}}$ it is the {\em axial} part of 
the WNC that enters, namely an interesting new inclusive nuclear
response function.

The vector responses (the four labeled either L or T) are, of
course, interesting in their own right, since disentangling them 
through measurements of both parity-conserving and -violating 
inclusive electron scattering would permit the isolation of the 
isoscalar and isovector contributions they contain, as will 
be discussed in detail later. Accomplishing this separation would 
represent a significant step forward in our understanding of nuclear 
structure: indeed, for a long time researchers have sought possible
ways of ``measuring'' how nuclear correlations work in the isoscalar 
and isovector channels.

A further point worth noting is that analogous responses also play a
role in the scattering of hadrons from nuclei. However, in
order to interpret the experimental data properly, in addition to the 
response functions there one also needs a reliable description of the
reaction mechanism. In fact, unlike either real or virtual 
photons, hadrons are mostly absorbed or scattered at the surface of the
nucleus. Moreover, the hadrons disrupt the nucleus to a much larger 
extent than do photons, and therefore the interpretation of reactions 
induced by photons is generally felt to be under better control than 
those induced by hadrons. Moderating this statement to some degree is 
the fact that electron scattering is still somewhat flawed by a not
yet fully satisfactory understanding of dispersive effects and of the 
distortion of the electron waves moving in the EM potential of the 
nucleus. Accurate knowledge of the nuclear response functions
gained with electron scattering provides a way to test the reaction 
mechanisms of the models employed in hadron scattering. However, the 
electroweak studies do not provide all of the information we seek and 
in this regard it is worth noting that in some cases reactions induced
by hadrons give access to nuclear responses that are not easily
extracted in electron scattering, the best example in this connection 
being offered by the long sought after spin-longitudinal isovector response.

Turning now to the problem of modeling the nuclear response
functions, in the present work we confine our attention to the
quasifree region, which is well
suited for a microscopic treatment in terms of nucleons and mesons, 
specifically in terms of the standard field theoretical techniques
that we employ. Although the $\Delta$ peak may also be treated in the 
same framework, it will not be dealt with here to curb the length of
this article. Since we are concerned with kinematical regions where 
relativistic effects are relevant, a good starting point for the 
development of our approach is given by the 
{\em Relativistic Fermi Gas} (RFG), a covariant model in the sense
that its ingredients are the fully relativistic nucleon propagators
and EM/WNC vertices. Of course the RFG misses surface and finite-size
effects. First of all, these are of secondary importance in obtaining 
a general understanding of the scattering of electrons in the
quasielastic and $\Delta$ peak domains. Secondly, they can
be accounted for within the semiclassical approach, which exploits the 
advantages offered by the translational invariance of the
RFG and yet is able to incorporate some of the physics of a finite system.

As discussed in detail later, the perturbative approach we follow
requires the setting up of a nuclear mean field (Hartree-Fock, HF) and
the treatment of the residual interaction effects in the fully 
antisymmetrized {\em Random Phase Approximation} (RPA). In addition 
one needs as preliminary input the nucleon-nucleon force: since we 
would like to view the nucleus as an interacting system of baryons and
mesons, a natural choice in this connection is given by a
meson-exchange interaction such as the Bonn potential, which can be 
cast in the framework of an effective field theory. Another
preliminary problem relates to the short-range nuclear correlations
induced by the violent repulsion present in the nucleon-nucleon force
at small distances. A technique for their treatment is indeed
available, namely the summation of the Brueckner ladder diagrams;
however, it is not yet possible to employ this technique 
covariantly especially at 
high density, where on the one hand ladder diagrams are increasingly 
important and on the other the role of relativity cannot be ignored. 
In lieu of this, in a few of the results discussed in the following 
section we shall indicate what insight can be gained by employing a 
parameterization of a non-relativistic $G$-matrix based on the Bonn potential.

\section{ Quasielastic response functions for inclusive electron scattering }
\label{sec:QEP}

As mentioned in the Introduction, in past years quasielastic electron 
scattering from nuclei has been the subject of intense experimental 
\cite{Jou96,Ang96,Wil97} and theoretical (see, e.~g., 
Refs.~\cite{Del85}--\cite{Gil97}) investigations. The first aim of the
theoretical studies is to test the available nuclear models; once the 
nuclear physics issues are well understood, one might then hope to gain
insight into other aspects of the problem, for instance into the form 
factors of the nucleon, which can be extracted from the data with an 
accuracy that is strictly connected to our ability to handle the 
nuclear physics.

In principle, the quasifree regime is thought to be the obvious place
to focus on, as one hopes that there the
physical quantities of interest may be computed in a reliable way, 
while also in this case in practice one has to cope with significant 
problems. Many diverse techniques have been employed in the
literature. Each of them has its own relative merits and deficiencies
and clearly it would be highly desirable to be able to reach some
degree of convergence in their outcomes.

In the following \cite{DeP98}, we shall be concerned with Green's
function techniques as introduced, e.~g., in Ref.~\cite{Fet71}. 
This method can be, and has been, applied both to finite nuclei and 
nuclear matter. Here, we shall focus on nuclear matter, having in mind
applications to electron scattering (that is, without the
complications introduced by the reaction mechanism of hadronic probes)
in a range from a few hundreds to about 1 GeV/c of transferred
momentum where the quasielastic peak is far from low-energy resonances
and not too much affected by finite-size effects. The use of nuclear 
matter reduces the computational load, thus allowing a more
straightforward implementation of more sophisticated theoretical
schemes than would otherwise be feasible, and this makes it easier to 
develop and test approximation methods that might subsequently also 
be utilized for calculations in finite nuclei.

Let us now briefly summarize the theoretical framework that we shall
discuss in detail in the following subsections.

A first issue one has to confront in setting up the formalism concerns
the treatment of relativistic effects. Kinematical effects, while 
obviously rather important, can be included in a straightforward way.
The treatment of dynamical effects is more delicate. Two main paths
have been followed in the literature,  either using field theoretical 
methods (as done, e.~g., in the Walecka model and its derivations 
\cite{Wal95}) or using potential techniques (i.~e., employing 
phenomenological potentials truncated at some order in the 
non-relativistic expansion). Here we shall take the second path, but 
to limit the amount of material to be covered, we shall discuss only 
non-relativistic potentials. 

The extensions necessary to include higher-order relativistic terms 
are discussed in Ref.~\cite{Bar96a}, where the nuclear response
functions have been calculated using techniques similar to the ones 
explained below, using as an input the relativistic Bonn potential 
\cite{Mac87} expanded in powers of $P/m_N$ and $q/m_N$ up to second
order --- $P$ and $q$ being the average of the incoming and outgoing 
nucleon momenta and the exchanged momentum, respectively. As shown in 
\cite{Bar96a}, the effect of these dynamical relativistic corrections 
is significant; indeed, the validity of that expansion at high
momenta and the inclusion in that framework of short-range nucleon-nucleon
correlations has yet to be explored (see, however, 
Refs.~\cite{Bar96a,Bar96b,Amo96}).

Next, one should choose the phenomenological input potential and, 
in connection with this choice, attempt to cope with the problem of 
dealing with short-range correlations. All of the formulae given 
in the following sections are based on a generic one-boson-exchange 
potential. They can thus be used both with a bare phenomenological 
interaction --- such as one of the Bonn potential variants --- or
with a one-boson-exchange parameterization of the $G$-matrix generated
from some potential. The use of an effective interaction derived from a 
$G$-matrix is a common way of including short-range
correlations. However, apart from the relativistic issue, one should be
aware of possible problems due to the use of a local potential to fit 
non-local matrix elements. At least in a few cases discussed in the 
literature this does not appear to be a reason for concern 
\cite{Nak84,Nak87}. On the other hand, possible effects arising only
in the quasielastic regime remain completely unexplored. Indeed, 
$G$-matrices employed in quasielastic calculations are usually
generated using bound-state boundary conditions, which make them 
real and practically energy-independent, while in general they are 
both complex and energy-dependent.

Once we have fixed the effective interaction, we can proceed to
consider a hierarchy of approximation schemes.

The lowest-order approximation is, of course, given by the free Fermi 
gas. Then, one may include mean-field correlations at the HF level 
(or Brueckner-Hartree-Fock (BHF) if short-range correlations are
accounted for). In nuclear matter a HF calculation can be done exactly
without too much effort. Later we show how a quite accurate analytic 
approximation can be derived, and how this is needed to combine the 
HF and RPA schemes. The latter is the last resummation technique we
shall discuss. It should be noticed that even in nuclear matter the 
calculation of the {\em antisymmetrized} RPA response functions is not
trivial. Indeed, most calculations, labeled ``RPA'' in the literature,
are actually performed in the so-called ``ring approximation'', where
only the direct contributions are kept. For this case, in nuclear
matter one gets a simple algebraic equation for the response. Here, we
use the continued fraction (CF) technique to provide a semi-analytical
estimate of the full RPA response (see Refs.~\cite{Shi89} and 
\cite{Bub91} for alternative methods). Calculations with this method 
have been performed both in finite nuclei \cite{Del85,Del87} and in 
nuclear matter \cite{Alb93,Bar94,Bar96a,Bar96b}, always truncating the
CF expansion at first order because of the difficulty of the
numerical calculations involved. We have pushed the analytical calculation far 
enough to yield not only a fast and accurate estimate of the
first-order CF contribution, but also of the second-order one. Since 
the rate of convergence of the CF expansion cannot be assessed on the 
basis of general theorems, this is the only way of getting a
quantitative grip on the quality of the approximation. As mentioned 
before, HF (and kinematical relativistic) effects can then be
incorporated in the RPA calculation, yielding as the final 
approximation scheme a HF-RPA (or BHF-RPA) response function.

Of course, several many-body contributions have been left out in our
analysis. However the classes of many-body diagrams discussed here 
already allow one to capture the main features of
the quasielastic response and, since semi-analytical
methods have been developed for their computation, our formalism 
constitutes a valid starting point for the study of other 
many-body effects.

\subsection{ Response functions }
\label{subsec:Resp}

We consider an infinite system of interacting nucleons 
at some density fixed by the Fermi momentum $k_F$. For the kinetic 
energies of the nucleons we can choose either relativistic or 
non-relativistic expressions, whereas we assume that the interactions 
take place through a non-relativistic potential. For the latter the 
following expression in momentum space is assumed
\begin{eqnarray}
  V(\bbox{k}) &=& V_0(k) +
    V_{\tau}(k) \bbox{\tau}_1\cdot\bbox{\tau}_2 +
    V_{\sigma}(k) \bbox{\sigma}_1\cdot\bbox{\sigma}_2 +
    V_{\sigma\tau}(k) \bbox{\sigma}_1\cdot\bbox{\sigma}_2 \ 
      \bbox{\tau}_1\cdot\bbox{\tau}_2 \nonumber \\
  && \quad +
    V_t(k) S_{12}(\hat{\bbox{k}}) +
    V_{t\tau}(k) S_{12}(\hat{\bbox{k}}) \bbox{\tau}_1\cdot\bbox{\tau}_2 ,
\label{eq:pot}
\end{eqnarray}
where $S_{12}$ is the standard tensor operator and $V_{\alpha}(k)$ 
represents the momentum space potential in channel $\alpha$. Here 
$V_{\alpha}(k)$ has the general form of a static one-boson-exchange 
potential so that in each spin-isospin channel, namely 
$(0, \tau, \sigma, \sigma\tau, t, t\tau)$, it is represented as a 
sum of contributions from different mesons, 
$V_{\alpha}\equiv\sum_i V_{\alpha}^{(i)}$. In the central channels 
($0$, $\tau$, $\sigma$, $\sigma\tau$) the contribution from any meson 
can be expressed as the combination of a short-range (``$\delta$'') 
piece and a longer range (``momentum-dependent'') 
piece\footnote{ The nomenclature stems from the fact that, in the 
absence of form factors, $V_{\delta}$ is a constant and is represented
by a Dirac $\delta$-function in coordinate space, whereas 
$V_{\text{MD}}$ is the momentum-dependent piece.}:
\begin{mathletters}
\label{eq:mes-exch}
\begin{eqnarray}
  V_{\delta}^{(i)}(k) &=&  g_{\delta}^{(i)} 
    \left(\frac{\Lambda_i^2-m_i^2}{\Lambda_i^2+k^2}\right)^\ell \\
  V_{\text{MD}}^{(i)}(k) &=&  g_{\text{MD}}^{(i)} 
    \frac{m_i^2}{m_i^2+k^2}
    \left(\frac{\Lambda_i^2-m_i^2}{\Lambda_i^2+k^2}\right)^\ell , 
    \quad  \ell=0,1,2 ,
\end{eqnarray}
whereas in the tensor channels ($t$, $t\tau$) one has
\begin{equation}
  V_{\text{TN}}^{(i)}(k) =  g_{\text{TN}}^{(i)} 
    \frac{k^2}{m_i^2+k^2}
    \left(\frac{\Lambda_i^2-m_i^2}{\Lambda_i^2+k^2}\right)^\ell , 
    \quad \ell=0,1,2 .
\end{equation}
\end{mathletters}
In Eqs.~(\ref{eq:mes-exch}), $g_{\delta}^{(i)}$, $g_{\text{MD}}^{(i)}$
and $g_{\text{TN}}^{(i)}$ are the (dimensional) coupling constants 
of the $i$-th meson, $m_i$ is its mass and $\Lambda_i$ the cut-off; 
more generally, potentials without form factors or with monopole or 
dipole form factors are allowed.

\begin{figure}[t]
\begin{center}
\mbox{\epsfig{file=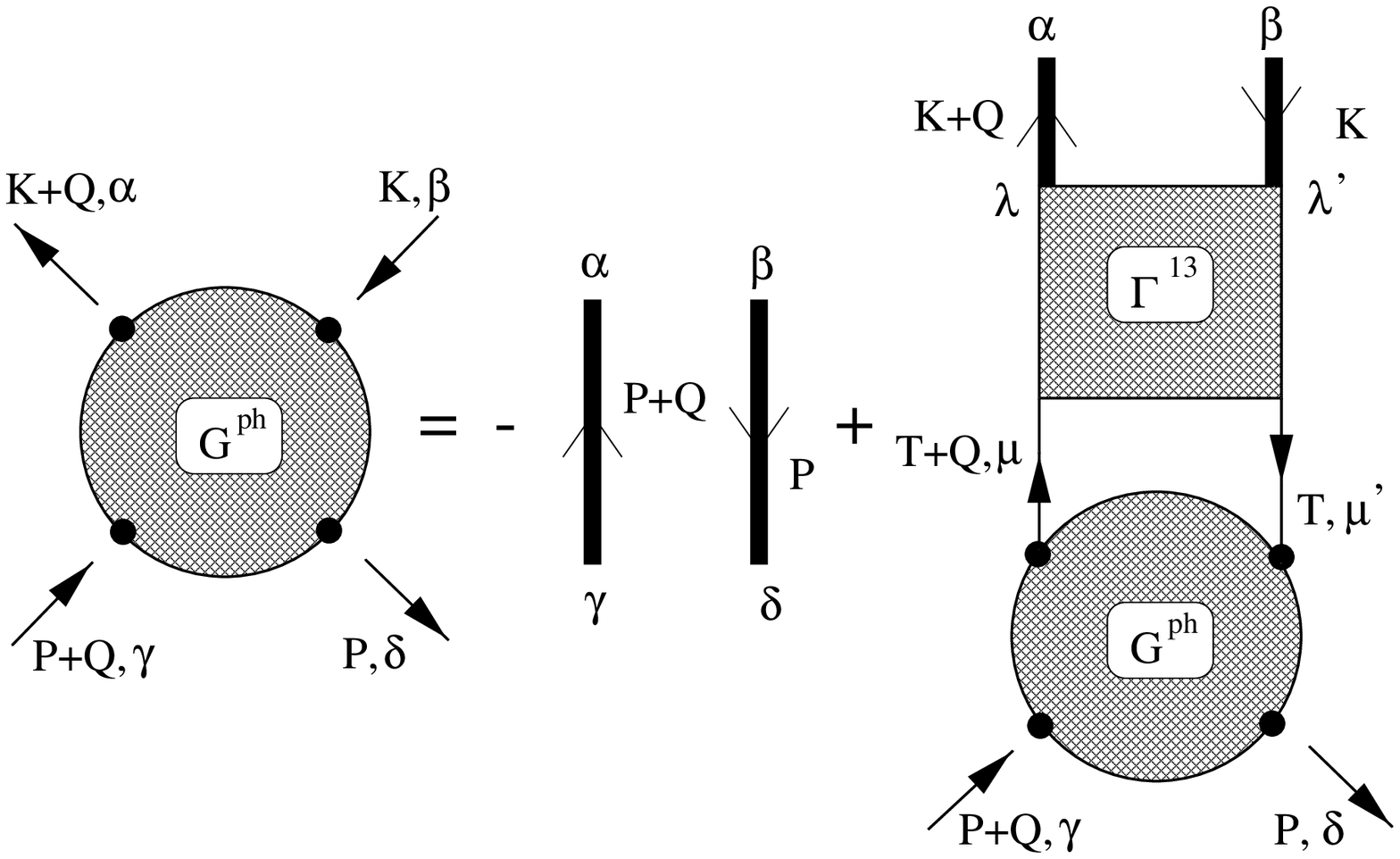,width=.7\textwidth}}
\vskip 2mm
\caption{ Diagrammatic representation of the Galitskii-Migdal integral equation
for the ph Green's function, $G^{\text{ph}}$; $\Gamma^{13}$ is the irreducible
vertex function in the ph channel; the heavy lines represent the exact one-body
Green's functions.
 }
\label{fig:Gph}
\end{center}
\end{figure}

Our starting point \cite{Gal58,Abr63,Alb91} is the Galitskii-Migdal 
integral equation for the particle-hole (ph) four-point Green's 
function\footnote{ Capital letters refer to four-vectors and lower-case 
letters to three-vectors; the Greek letters $\alpha,\beta,...$ refer 
to a set of spin-isospin quantum numbers.},
\begin{eqnarray}
  && G^{\text{ph}}_{\alpha\beta,\gamma\delta}(K+Q,K;P+Q,P) = 
   -G_{\alpha\gamma}(P+Q)\,G_{\delta\beta}(P)\,(2\pi)^4\delta(K-P) \nonumber \\
 && + i G_{\alpha\lambda}(K+Q)\,G_{\lambda'\beta}(K)\int\frac{d^{4}T}{(2\pi)^4}
    \Gamma^{13}_{\lambda\lambda',\mu\mu'}(K+Q,K;T+Q,T)\,
    G^{\text{ph}}_{\mu\mu',\gamma\delta}(T+Q,T;P+Q,P) , \nonumber \\
\label{eq:Gph}
\end{eqnarray}
diagrammatically illustrated in Fig.~\ref{fig:Gph}.
In Eq.~(\ref{eq:Gph}), $G$ represents the exact one-body Green's
function, whereas $\Gamma^{13}$ is the irreducible vertex function 
in the ph channel.

\begin{figure}[t]
\begin{center}
\mbox{\epsfig{file=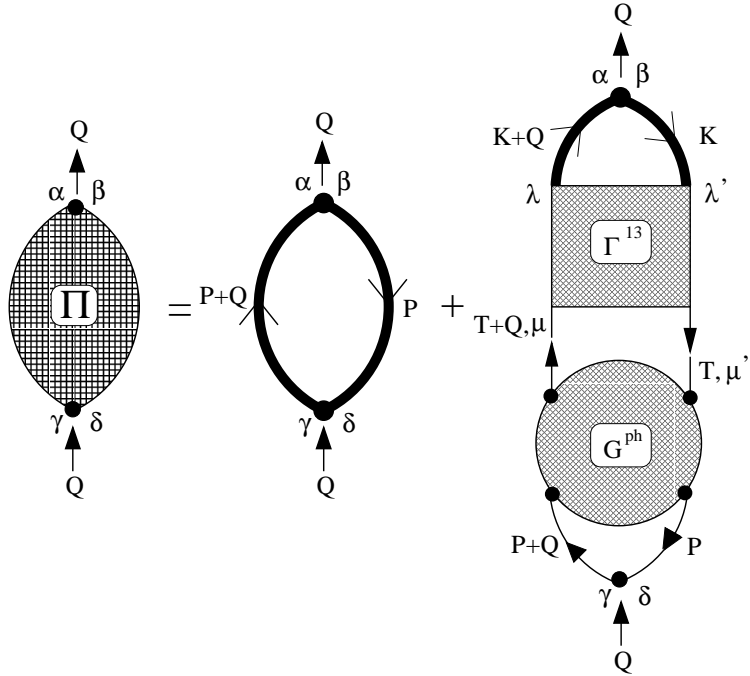,width=.6\textwidth}}
\vskip 2mm
\caption{ Diagrammatic representation of the polarization propagator 
$\Pi$ derived from the ph Green's function $G^{\text{ph}}$.
 }
\label{fig:Pi}
\end{center}
\end{figure}

Given $G^{\text{ph}}$ one can then define the {\em polarization propagator}
\begin{eqnarray}
  \Pi_{\alpha\beta,\gamma\delta}(Q) & \equiv & 
    \Pi_{\alpha\beta,\gamma\delta}(q,\omega) \nonumber \\
  &=& i\int\frac{d^{4}P}{(2\pi)^4}\frac{d^{4}K}{(2\pi)^4}
    G^{\text{ph}}_{\alpha\beta,\gamma\delta}(K+Q,K;P+Q,P) ,
\label{eq:Pi}
\end{eqnarray}
whose diagrammatic representation is displayed in Fig.~\ref{fig:Pi}.
Note that for $\Pi(q,\omega)$ one cannot in general write down an 
integral (or algebraic) equation.

In the case of electron scattering, one can define charge --- or {\em
longitudinal} --- and magnetic --- or {\em transverse} --- polarization
propagators. These, in the non-relativistic regime, read
\begin{mathletters}
\label{eq:PiLT}
\begin{eqnarray}
  &&\Pi^{I}_{\text{L}}(q,\omega) = \text{tr}[\hat{O}^{I}_{\text{L}}
    \hat{\Pi}(q,\omega) \hat{O}^{I}_{\text{L}}] \\
  &&\Pi^{I}_{\text{T}}(q,\omega) = \sum_{ij}\Lambda_{ji}\Pi^{I}_{ij}(q,\omega)
    , \qquad \Pi^{I}_{ij}(q,\omega) = \text{tr}[\hat{O}^{I}_{\text{T};i}
    \hat{\Pi}(q,\omega) \hat{O}^{I}_{\text{T};j}] \\
  &&\phantom{\Pi^{I}_{\text{T}}(q,\omega) = 
    \sum_{ij}\Lambda_{ji}\Pi^{I}_{ij}(q,\omega) , } \qquad \,
    \Lambda_{ij} = (\delta_{ij}-\bbox{q}_i\bbox{q}_j/q^2)/2 , 
    \nonumber
\end{eqnarray}
\end{mathletters}
where, for brevity, the dependence upon the spin-isospin indices has 
been represented in matrix form, introducing hats to indicate matrices.
In Eqs.~(\ref{eq:PiLT}), $I$ labels the isospin channel and the
longitudinal and transverse vertex operators are given by:
\begin{equation}
  \left\{ 
    \begin{array}{l}
      \hat{O}^{I=0}_{\text{L}} = 1/2 \\
      \hat{O}^{I=1}_{\text{L}} = \tau_3/2 
    \end{array}
    \right. \qquad\qquad
  \left\{
    \begin{array}{l}
      \hat{O}^{I=0}_{\text{T};i} = \sigma_i/2 \\
      \hat{O}^{I=1}_{\text{T};i} = \sigma_i\tau_3/2 .
    \end{array}
    \right.
\label{eq:OLT}
\end{equation}

The inelastic inclusive scattering cross section where the
momentum $q$ and energy $\omega$ are transferred to the 
nucleus is a linear combination of the imaginary parts of
$\Pi_{\text{L,T}}(q,\omega)$. It is then customary to define
longitudinal and transverse response functions according to
\begin{equation}
  R_{\text{L,T}}(q,\omega) = R^{I=0}_{\text{L,T}}(q,\omega) + 
    R^{I=1}_{\text{L,T}}(q,\omega) ,
\end{equation}
which are related to $\Pi_{\text{L,T}}$ by
\begin{eqnarray}
  R^{I}_{\text{L,T}}(q,\omega) &=& 
    -\frac{V}{\pi} {f^{(I)}_{\text{L,T}}}^2(q,\omega)
    \text{Im}\Pi^{I}_{\text{L,T}}(q,\omega) \nonumber \\
  &=& -\frac{3\pi A}{2k_F^3} {f^{(I)}_{\text{L,T}}}^2(q,\omega) 
    \text{Im}\Pi^{I}_{\text{L,T}}(q,\omega) ,
\label{eq:RILT}
\end{eqnarray}
where $V$ is the volume, $A$ the mass number and the 
${f^{(I)}_{\text{L,T}}}^2$ embody the squared EM form 
factors of the nucleon. The latter are briefly discussed in 
Appendix~\ref{app:A}.

\subsection{ Non-relativistic vs relativistic kinematics}
\label{subsec:rel}

The response functions introduced above have been defined as functions
of the momentum transfer $q$ and energy transfer $\omega$. Actually, 
it is possible --- and convenient --- to define a scaling variable 
$\psi$ that is a function of $q$ and $\omega$ and may be used in place of
$\omega$. This variable is such that the responses of a free Fermi gas
in the non-Pauli-blocked region ($q>2 k_F$) can be expressed in terms 
of the variable $\psi$ only (apart from $q$-dependent multiplicative 
factors). We shall see that even in the Pauli-blocked region and for 
an interacting system it is convenient to use the pair of variables 
($q$,$\psi$) instead of ($q$,$\omega$).

Besides the obvious advantages related to the use of a scaling
variable (see Section \ref{sec:sumrules}), there is another reason 
for expressing the responses in terms of $\psi$: when the 
latter is used the responses viewed as functions 
of $\psi$ turn out to adjust to the form assumed for the nucleon 
kinetic energy. To be more specific, starting from either a 
non-relativistic or relativistic Fermi gas, one is always led to 
essentially the same dependence of the responses upon the 
corresponding $\psi$ variable.

We shall see in the following subsections that the energy denominators
of the free nucleon propagators appearing in the Feynman diagrams 
for the response functions are always given by 
$\omega-\epsilon^{(0)}_{\bbox{k}+\bbox{q}}+\epsilon^{(0)}_{\bbox{k}}$,
where $\epsilon^{(0)}_{\bbox{k}}$ is the kinetic energy of a nucleon 
of momentum $k$ and $k<k_F$. In the non-relativistic case
\begin{eqnarray}
  \omega-\epsilon^{(0)\text{nr}}_{\bbox{k}+\bbox{q}}
    +\epsilon^{(0)\text{nr}}_{\bbox{k}} &=& \omega 
    - \frac{(\bbox{k}+\bbox{q})^2}{2m_N} + \frac{k^2}{2m_N} \nonumber \\
  &=& \frac{qk_F}{m_N}\left(\psi_{\text{nr}}-
    \hat{\bbox{q}}\cdot\frac{\bbox{k}}{k_F}\right) ,
\label{eq:endenNR}
\end{eqnarray}
where
\begin{equation}
  \psi_{\text{nr}} = \frac{1}{k_F}\left(\frac{\omega m_N}{q}-\frac{q}{2}\right)
\label{eq:psiNR}
\end{equation}
is the standard scaling variable of the non-relativistic Fermi gas 
and $m_N$ the nucleon mass.

In the relativistic case, one would have 
\begin{equation}
  \omega-\epsilon^{(0)\text{r}}_{\bbox{k}+\bbox{q}}
    +\epsilon^{(0)\text{r}}_{\bbox{k}} = \omega -
    \sqrt{(\bbox{k}+\bbox{q})^2+m_N^2}+\sqrt{k^2+m_N^2} ;
\label{eq:endenR}
\end{equation}
however, in Ref.~\cite{Alb90} it was shown that at the pole 
(where the above vanishes) a very good approximation for 
Eq.~(\ref{eq:endenR}) obtains by using Eq.~(\ref{eq:endenNR})
with, instead of $\psi_{\text{nr}}$,
\begin{equation}
  \psi_{\text{r}} = \frac{1}{k_F}\left[
    \frac{\omega m_N(1+\omega/2m_N)}{q}-\frac{q}{2}\right] 
\label{eq:psiR}
\end{equation}
and then by multiplying the free response by $1+\omega/m_N$, 
which is proportional to the Jacobian of the transformation
from the variable $\omega$ to the variable $\psi$. Thus the use of 
the scaling variable in Eq.~(\ref{eq:psiR}) entails the substitution
\begin{equation}
  \omega-\epsilon^{(0)\text{r}}_{\bbox{k}+\bbox{q}}
    +\epsilon^{(0)\text{r}}_{\bbox{k}} \to 
    \omega \left( 1 + \frac{\omega}{2m_N} \right)
    -\epsilon^{(0)\text{nr}}_{\bbox{k}+\bbox{q}}
    +\epsilon^{(0)\text{nr}}_{\bbox{k}} .
\label{eq:endenRappr}
\end{equation}
In turn, this implies that the pole (which provides the contribution 
to the imaginary part of the propagator) is located at 
$\omega=\sqrt{m_N^2+q^2+2\bbox{q}\cdot\bbox{k}}-m_N$, namely at 
the place predicted  by the exact expression in Eq.~(\ref{eq:endenR}) 
when $k^2$ is neglected with respect to $m_N^2$. As stated above, 
since $k$ is always below $k_F$, this is a good approximation and, 
indeed, the free RFG response calculated using the scaling variable in 
Eq.~(\ref{eq:psiR}) reproduces that of the exact 
calculation accurately, the discrepancy being typically below 1\%.

\begin{figure}[t]
\begin{center}
\vskip 2mm
\mbox{\epsfig{file=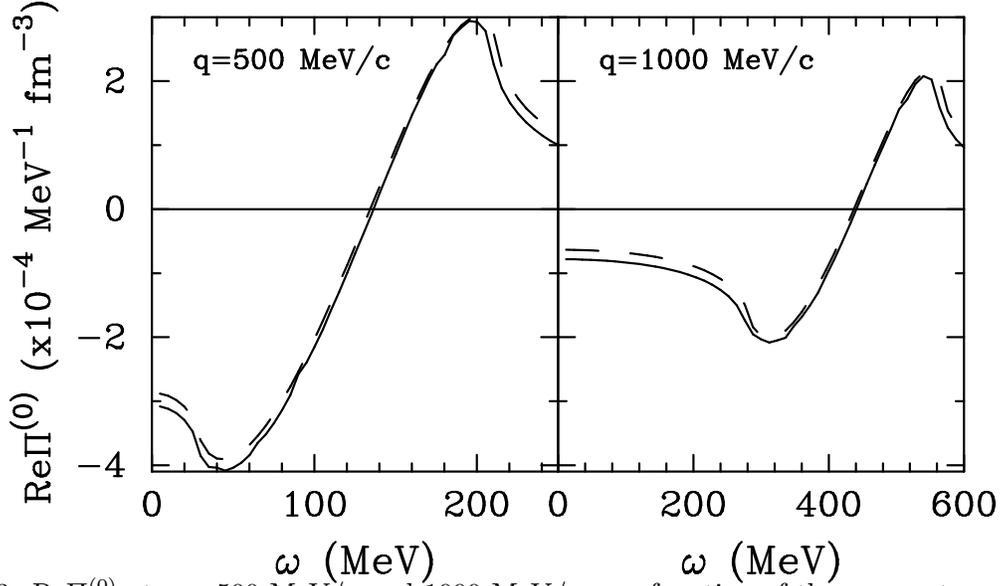,width=.8\textwidth}}
\caption{ $\text{Re}\Pi^{(0)}$ at $q=500$ MeV/c and 1000 MeV/c as a 
function of the energy transfer, using the exact relativistic kinetic 
energies (solid) and the approximation discussed in the text (dashed);
here $k_F=195$ MeV/c.
 }
\label{fig:RePi0}
\end{center}
\end{figure}

However, in the calculation of higher-order (RPA) contributions, the 
real part of the energy denominators also comes into play and the 
validity of the approximation far from the pole should also be checked. 
With some algebra --- and assuming $k^2/m_N^2\ll 1$ --- one can write 
\begin{eqnarray}
  \omega-\epsilon^{(0)\text{r}}_{\bbox{k}+\bbox{q}}
    +\epsilon^{(0)\text{r}}_{\bbox{k}} &\cong& 
    \frac{q k_F}{m_N} 
    \frac{\psi_r-\hat{\bbox{q}}\cdot\bbox{k}/k_F}{\displaystyle 
      \frac{\strut 1}{\displaystyle 2}
      \left(1+\frac{\strut\omega}{\displaystyle m_N}
        +\sqrt{1+\frac{\strut q^2+2\bbox{q}\cdot\bbox{k}}{\displaystyle m_N^2}}
        \right)}
  \nonumber \\
  &\cong& \frac{q k_F}{m_N} 
    \frac{\psi_r-\hat{\bbox{q}}\cdot\bbox{k}/k_F}{1+\omega/m_N} ,
\label{eq:endenRr}
\end{eqnarray}
where, in the last passage, we have replaced the square root with 
its value at the pole. In Fig.~\ref{fig:RePi0}, we display the real
part of the free polarization propagator (defined in the following 
subsection) using the exact relativistic dispersion relation and 
the prescription of Eq.~(\ref{eq:endenRr}) at $q=500$ MeV/c and 1
GeV/c as a function of $\omega$. The agreement between the two ways 
of calculating $\text{Re}\Pi^{(0)}$ is quite good at both momenta.

Equation~(\ref{eq:endenRr}) provides an approximation for the free ph 
propagator. A prescription to obtain the (kinematically) relativistic 
polarization propagators at any order in the RPA expansion (see 
Section \ref{subsec:RPA-resp}) can easily be obtained by noting that 
$\Pi^{(n)}$ --- the $n$-th order contribution to the RPA chain --- 
contains $n+1$ ph propagators; one then has
\begin{mathletters}
\label{eq:Pinrelnonrel}
\begin{equation}
  \Pi^{(n)\text{r}}(q,\omega) = \left(1+\frac{\omega}{m_N}\right)^{n+1}
    \Pi^{(n)\text{nr}}(q,\omega(1+\omega/2m_N)) .
\label{eq:Pinrelomega}
\end{equation}
Actually, all of the response functions derived below are expressed 
in terms of a generic scaling variable $\psi$, as
$\Pi^{(n)}(q,\psi)$. One can then get the non-relativistic response 
by using the (exact) expression in Eq.~(\ref{eq:psiNR}) for $\psi$ 
and the relativistic response by using the (approximate) form in 
Eq.~(\ref{eq:psiR}) and multiplying each polarization propagator 
by the appropriate power of $1+\omega/m_N$, i.~e.
\begin{equation}
  \Pi^{(n)\text{r}}(q,\omega) = \left(1+\frac{\omega}{m_N}\right)^{n+1}
    \Pi^{(n)\text{nr}}(q,\psi_{\text{r}}) .
\label{eq:Pinrelpsi}
\end{equation}
\end{mathletters}
Note that in the calculations of Refs.~\cite{Bar96a,Bar96b} only 
an overall Jacobian factor, $1+\omega/m_N$, has been applied to 
the RPA response functions. In typical kinematical conditions the 
size of the error introduced by this further approximation is of 
the order of a few percent.

\subsection{ Free response }
\label{subsec:free-resp}

Although the free Fermi gas response function is a subject for
textbooks (see, e.~g., Ref.~\cite{Fet71}), it is useful to derive 
it here using a slightly different approach, since it illustrates 
at the simplest level the method we have adopted to overcome a 
technical difficulty one meets in nuclear matter calculations --- 
namely the presence of $\theta$-functions, which considerably 
complicates analytic integrations. As a side effect, the expression 
for $\Pi^{(0)}$ also comes out to be much more compact than in 
standard treatments.

From Eqs.~(\ref{eq:PiLT}) and (\ref{eq:OLT}), one immediately finds that 
\begin{equation}
  \Pi^{(0)}_{\text{L};I=0} = \Pi^{(0)}_{\text{L};I=1} =
  \Pi^{(0)}_{\text{T};I=0} = \Pi^{(0)}_{\text{T};I=1} \equiv
  \Pi^{(0)} ,
\end{equation}
where following Eqs.~(\ref{eq:Gph}) and (\ref{eq:Pi}) we have defined 
\begin{equation}
  \Pi^{(0)}(q,\omega) = \int\frac{d\bbox{k}}{(2\pi)^3} 
    G^{(0)}_{\text{ph}}(\bbox{k},\bbox{q};\omega) ,
\label{eq:Pi0G0}
\end{equation}
having set
\begin{equation}
  G^{(0)}_{\text{ph}}(\bbox{k},\bbox{q};\omega) = 
    -i \int\frac{dk_0}{2\pi}G^{(0)}(\bbox{k}+\bbox{q},k_0+\omega)
    G^{(0)}(k,k_0) ,
\label{eq:G0ph}
\end{equation}
$G^{(0)}(k,k_0)$ being the free one-body propagator
\begin{equation}
  G^{(0)}(k,k_0) = 
    \frac{\theta(k-k_F)}{k_0-\epsilon^{(0)}_{\bbox{k}}+i\eta} +
    \frac{\theta(k_F-k)}{k_0-\epsilon^{(0)}_{\bbox{k}}-i\eta} .
\end{equation}
The integration over $k_0$ in Eq.~(\ref{eq:G0ph}) is straightforward, yielding
\begin{equation}
  G^{(0)}_{\text{ph}}(\bbox{k},\bbox{q};\omega) = 
    \frac{\theta(k_F-k)\theta(|\bbox{k}+\bbox{q}|-k_F)}
    {\omega-\epsilon^{(0)}_{\bbox{k}+\bbox{q}}
    +\epsilon^{(0)}_{\bbox{k}}+i\eta}+
    \frac{\theta(k-k_F)\theta(k_F-|\bbox{k}+\bbox{q}|)}
    {-\omega+\epsilon^{(0)}_{\bbox{k}+\bbox{q}}
    -\epsilon^{(0)}_{\bbox{k}}+i\eta}
    ,
\label{eq:G0phold}
\end{equation}
which, inserted back into Eq.~(\ref{eq:Pi0G0}), would give the 
standard definition of $\Pi^{(0)}$. Instead, let us rewrite 
$G^{(0)}_{\text{ph}}$ as
\begin{eqnarray}
  G^{(0)}_{\text{ph}}(\bbox{k},\bbox{q};\omega) &=&
    \frac{\theta(k_F-k)\theta(|\bbox{k}+\bbox{q}|-k_F)}
    {\omega-\epsilon^{(0)}_{\bbox{k}+\bbox{q}}
    +\epsilon^{(0)}_{\bbox{k}}+i\eta}+
    \frac{\theta(k-k_F)\theta(k_F-|\bbox{k}+\bbox{q}|)}
    {-\omega+\epsilon^{(0)}_{\bbox{k}+\bbox{q}}
    -\epsilon^{(0)}_{\bbox{k}}+i\eta}
    \nonumber \\
  && + \frac{\theta(k_F-k)\theta(k_F-|\bbox{k}+\bbox{q}|)}
    {\omega-\epsilon^{(0)}_{\bbox{k}+\bbox{q}}+\epsilon^{(0)}_{\bbox{k}}
      +i\eta_{\omega}} +
    \frac{\theta(k_F-k)\theta(k_F-|\bbox{k}+\bbox{q}|)}
    {-\omega+\epsilon^{(0)}_{\bbox{k}+\bbox{q}}-\epsilon^{(0)}_{\bbox{k}}
      -i\eta_{\omega}} ,
\end{eqnarray}
having added and subtracted the quantity in the second line, where 
we have set $\eta_{\omega}=\text{sign}(\omega)\eta$. A few algebraic 
manipulations then yield
\begin{equation}
  G^{(0)}_{\text{ph}}(\bbox{k},\bbox{q};\omega) =
    \frac{\theta(k_F-k)-\theta(k_F-|\bbox{k}+\bbox{q}|)}
    {\omega-\epsilon^{(0)}_{\bbox{k}+\bbox{q}}+\epsilon^{(0)}_{\bbox{k}}
      +i\eta_{\omega}} .
\label{eq:G0phnew}
\end{equation}
Hence, from Eq.~(\ref{eq:Pi0G0}) one gets
\begin{eqnarray}
  \Pi^{(0)}(q,\omega) &=& \int\frac{d\bbox{k}}{(2\pi)^3}\theta(k_F-k)
    \left[\frac{1}
    {\omega-\epsilon^{(0)}_{\bbox{k}+\bbox{q}}+\epsilon^{(0)}_{\bbox{k}}
      +i\eta_{\omega}} +
    \frac{1}
      {-\omega-\epsilon^{(0)}_{\bbox{k}+\bbox{q}}+\epsilon^{(0)}_{\bbox{k}}
      -i\eta_{\omega}}\right] \nonumber \\
  &=& \frac{m_N}{q}\frac{k_F^2}{(2\pi)^2}
    \left[{\cal Q}^{(0)}(\psi)-{\cal Q}^{(0)}(\psi+\bar{q})\right] .
\label{eq:Pi0}
\end{eqnarray}
Note that only one $\theta$-function forcing $k$ below $k_F$ is left, 
Pauli blocking being enforced by cancellations between the energy 
denominators. In Eq.~(\ref{eq:Pi0}), we have introduced
$\bar{q}=q/k_F$ and the dimensionless function
\begin{equation}
  {\cal Q}^{(0)}(\psi) =
  \frac{1}{2}\int_{-1}^{1}dy\frac{1-y^2}{\psi-y+i\eta_{\omega}} ,
\end{equation}
which is easily evaluated, yielding 
\begin{mathletters}
\begin{eqnarray}
  \text{Re}{\cal Q}^{(0)}(\psi) &=& \psi+\frac{1}{2}(1-\psi^2)
    \text{ln}\left|\frac{1+\psi}{1-\psi}\right| = 
    \frac{2}{3}[Q_0(\psi)-Q_2(\psi)] \\
  \text{Im}{\cal Q}^{(0)}(\psi) &=& -\text{sign}(\omega)\theta(1-\psi^2)
    \frac{\pi}{2}(1-\psi^2) = -\text{sign}(\omega)\theta(1-\psi^2)
    \frac{\pi}{3}[P_0(\psi)-P_2(\psi)] ,
\end{eqnarray}
\end{mathletters}
where $P_n$ and $Q_n$ are Legendre polynomials and Legendre functions 
of the second kind, respectively.

\begin{figure}[t]
\begin{center}
\mbox{\epsfig{file=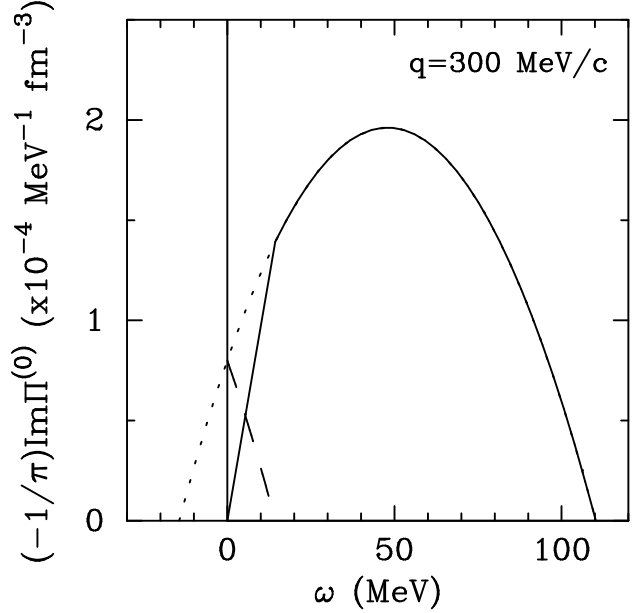}}
\vskip 2mm
\caption{ Free response function at $q=300$ MeV/c and $k_F=195$ MeV/c.
The parabola given by the dotted line plus the parabolic part of 
the solid line represents the response of a Fermi sphere without 
Pauli blocking; the dashed line represents the response at 
negative energies reflected in the vertical axis and, once 
subtracted from the dotted line, yields the solid straight line, 
that is, the Pauli-blocked part of the response.
 }
\label{fig:Pi0PB}
\end{center}
\end{figure}

The expression in Eq.~(\ref{eq:Pi0}) has a simple physical 
interpretation. If one switches off Pauli blocking, the response 
of a Fermi sphere, with four particles per momentum state up to 
$k_F$, is given by a parabola over the response region 
$q^2/2m_N-qk_F/m_N<\omega<q^2/2m_N+qk_F/m_N$, that is, the curve 
obtained joining the dotted line and the parabolic section of the solid
line in Fig.~\ref{fig:Pi0PB}. With respect to the Pauli blocking, two
kinds of spurious terms arise when $k$ and $|\bbox{k}+\bbox{q}|$ 
are {\em both} below the Fermi surface. If $|\bbox{k}+\bbox{q}|>k$, 
then a spurious contribution occurs in the Pauli-forbidden region 
$0<\omega<qk_F/m_N-q^2/2m_N$, whereas if $|\bbox{k}+\bbox{q}|<k$, 
then a contribution occurs with the {\em same} strength for 
$q^2/2m_N-qk_F/m_N<\omega<0$. Hence, in order to get the correct 
response function, one simply subtracts --- for a given $\omega>0$ 
in the Pauli-forbidden region --- the total of the spurious
contributions at $-\omega$, thus getting the familiar linear 
dependence on $\omega$. Graphically, as illustrated in 
Fig.~\ref{fig:Pi0PB}, this amounts to reflecting the response 
at negative transferred energies in the vertical axis and then subtracting it.

\subsection{ Hartree-Fock response }
\label{subsec:HF-resp}

The HF polarization propagator in nuclear matter is obtained by 
dressing the one-body propagators appearing in $\Pi^{(0)}$ with 
the first-order self-energy $\Sigma^{(1)}$, so that one can 
follow essentially the same derivation of the previous subsection. The 
spin-isospin matrix elements are the same as for the free response, yielding
\begin{equation}
  \Pi^{\text{HF}}_{\text{L};I=0} = \Pi^{\text{HF}}_{\text{L};I=1} =
  \Pi^{\text{HF}}_{\text{T};I=0} = \Pi^{\text{HF}}_{\text{T};I=1} \equiv
  \Pi^{\text{HF}} ,
\end{equation}
where 
\begin{equation}
  \Pi^{\text{HF}}(q,\omega) = \int\frac{d\bbox{k}}{(2\pi)^3} 
    G^{\text{HF}}_{\text{ph}}(\bbox{k},\bbox{q};\omega)
\label{eq:PiHFGHF}
\end{equation}
and
\begin{equation}
  G^{\text{HF}}_{\text{ph}}(\bbox{k},\bbox{q};\omega) = 
    -i \int_{-\infty}^\infty\frac{dk_0}{2\pi}
  G^{\text{HF}}(\bbox{k}+\bbox{q},k_0+\omega)
    G^{\text{HF}}(k,k_0) ,
\label{eq:GHFph}
\end{equation}
$G^{\text{HF}}(k,k_0)$ being the HF one-body propagator
\begin{equation}
  G^{\text{HF}}(k,k_0) = 
    \frac{\theta(k-k_F)}{k_0-\epsilon^{(1)}_{\bbox{k}}+i\eta} +
    \frac{\theta(k_F-k)}{k_0-\epsilon^{(1)}_{\bbox{k}}-i\eta} , 
\end{equation}
with
\begin{equation}
  \epsilon^{(1)}_{\bbox{k}} = \epsilon^{(0)}_{\bbox{k}} +
    \Sigma^{(1)}(k) .
\end{equation}

Since the first-order self-energy does not depend on the energy, the 
integration over $k_0$ can be performed along the lines of 
Eqs.~(\ref{eq:G0phold})--(\ref{eq:G0phnew}), yielding 
\begin{equation}
  G^{\text{HF}}_{\text{ph}}(\bbox{k},\bbox{q};\omega) =
    \frac{\theta(k_F-k)-\theta(k_F-|\bbox{k}+\bbox{q}|)}
    {\omega-\epsilon^{(1)}_{\bbox{k}+\bbox{q}}+\epsilon^{(1)}_{\bbox{k}}
      +i\eta_{\omega}} 
\label{eq:GHFphnew}
\end{equation}
and, finally, 
\begin{equation}
  \Pi^{\text{HF}}(q,\omega) = \int\frac{d\bbox{k}}{(2\pi)^3}\theta(k_F-k)
    \left[\frac{1}
    {\omega-\epsilon^{(1)}_{\bbox{k}+\bbox{q}}+\epsilon^{(1)}_{\bbox{k}}
      +i\eta_{\omega}} +
    \frac{1}
    {-\omega-\epsilon^{(1)}_{\bbox{k}+\bbox{q}}+\epsilon^{(1)}_{\bbox{k}}
      -i\eta_{\omega}}\right] .
\label{eq:PiHF}
\end{equation}
The HF response function is proportional to the imaginary part of 
$\Pi^{\text{HF}}$:
\begin{eqnarray}
  \text{Im}\Pi^{\text{HF}}(q,\omega) &=& -\text{sign}(\omega) \pi
    \int\frac{d\bbox{k}}{(2\pi)^3}\theta(k_F-k)
    \left[\delta\left(
      \omega-\epsilon^{(1)}_{\bbox{k}+\bbox{q}}+\epsilon^{(1)}_{\bbox{k}}
      \right) - 
    \delta\left(
      -\omega-\epsilon^{(1)}_{\bbox{k}+\bbox{q}}+\epsilon^{(1)}_{\bbox{k}}
      \right)\right] \nonumber \\
  &=& -\text{sign}(\omega) \pi \frac{m_N}{q}\frac{1}{(2\pi)^2} 
\label{eq:HFexact} \\
  &&\quad\times  \int_0^{k_F} dk\,k\frac{1}{m_N}\left[
    m_N^*(\sqrt{k^2+q^2+2 q y_0})
    -m_N^*(\sqrt{k^2+q^2+2 q \bar{y}_0}) \right] , \nonumber
\end{eqnarray}
having defined the effective mass as 
\begin{mathletters}
\begin{equation}
  m_N^{*\text{nr}}(k) = \frac{m_N}{\displaystyle 1 
    + \frac{\strut m_N}{\displaystyle k}
    \frac{\strut d\Sigma^{(1)}}{\displaystyle dk}} 
\end{equation}
or
\begin{equation}
  m_N^{*\text{r}}(k) = \frac{\sqrt{m_N^2+k^2}}{\displaystyle 1 
    + \frac{\strut \sqrt{m_N^2+k^2}}{\displaystyle k}
    \frac{\strut d\Sigma^{(1)}}{\displaystyle dk}} ,
\end{equation}
\end{mathletters}
for the non-relativistic or relativistic case, respectively,
whereas $y_0$ and $\bar{y}_0$ solve the equations
\begin{equation}
  \left\{
    \begin{array}{ccc}
       f_{\text{HF}}(\omega|k,y_0)         &=& 0 \\
       f_{\text{HF}}(-\omega|k,\bar{y}_0)  &=& 0 ,
    \end{array} \right.
\end{equation}
with
\begin{mathletters}
\begin{eqnarray}
  f_{\text{HF}}^{\text{nr}}(\omega|k,y) &=& \omega-\frac{q^2}{2m_N}
    -\frac{q y}{m_N}-\Sigma^{(1)}(\sqrt{k^2+q^2+2 q y})+\Sigma^{(1)}(k) \\
  f_{\text{HF}}^{\text{r}}(\omega|k,y) &=& \omega-\sqrt{m_N^2+k^2+q^2+2 q y}
    +\sqrt{m_N^2+k^2}
    -\frac{q y}{m_N} \nonumber \\
  &&\qquad -\Sigma^{(1)}(\sqrt{k^2+q^2+2 q y})+\Sigma^{(1)}(k) .
\end{eqnarray}
\end{mathletters}
Although the evaluation of the HF response is numerically quite 
straightforward, in Ref.~\cite{Bar96a} an analytic approximation 
for $\text{Im}\Pi^{\text{HF}}$ has been worked out, with the aim 
of using it to include the HF field in RPA calculations. Here, it 
will be shown that the analytic approximation is valid not only for 
the HF response, but more generally, although in the HF case one can 
directly assess the good accuracy of the procedure. 

In any Feynman diagram considered here and in the following, the
nucleon self-energy enters through the ph energy denominators,
\begin{equation}
  \omega-\epsilon^{(1)\text{nr}}_{\bbox{k}+\bbox{q}}+
    \epsilon^{(1)\text{nr}}_{\bbox{k}} =
    \omega - \frac{(\bbox{k}+\bbox{q})^2}{2m_N} + \frac{k^2}{2m_N} 
    -\Sigma^{(1)}(|\bbox{k}+\bbox{q}|)+\Sigma^{(1)}(k) ,
\label{eq:endenSE}
\end{equation}
where the non-relativistic expression for the nucleon kinetic energy 
has been used. In Eq.~(\ref{eq:endenSE}), one can always assume that 
$k<k_F$ and $|\bbox{k}+\bbox{q}|>k_F$. Although the latter inequality 
is not immediately apparent from, e.~g., Eq.~(\ref{eq:PiHF}), remember
that cancellations between the energy denominators are such as to 
enforce the Pauli principle; the same will also be true for the RPA 
diagrams\footnote{ It should also be noted that the infinite Fermi 
gas is more in touch with the physics for relatively large momenta 
($q\gtrsim 2k_F$), where the above conditions are satisfied by definition.}.

Clearly, if $\Sigma^{(1)}(k)$ were parabolic in the momentum, the 
inclusion of the self-energy would be achieved simply by substituting 
an effective mass for $m_N$. For realistic potentials, a
parabolic fit for the self-energy over the whole range of momenta 
is in general not a good approximation. It is a good approximation, 
on the other hand, to fit {\em separately} the particle and hole 
parts of the self-energy, the fit being restricted to the range of 
momenta actually involved in the integration. Since in 
Eq.~(\ref{eq:PiHF}) (but also in the RPA diagrams discussed later) 
$k$ is integrated from 0 to $k_F$ and, furthermore, 
$|\bbox{k}+\bbox{q}|>k_F$, one can set
\begin{eqnarray}
  \Sigma^{(1)} &\cong& \bar{A} + \bar{B}\frac{k^2}{2m_N} ,
    \quad 0<k<k_F , \nonumber \\
    \\
  \Sigma^{(1)} &\cong& A + B\frac{k^2}{2m_N} ,
    \quad \text{max}(q-k_F,k_F)<k<q+k_F . \nonumber
\end{eqnarray}
Inserting this ``biparabolic approximation'' back into 
Eq.~(\ref{eq:endenSE}), and setting $\varepsilon=\bar{A}-A$ 
and $m_N^{*\text{nr}}=m_N/(1+B)$, one gets 
\begin{eqnarray}
  \omega-\epsilon^{(1)\text{nr}}_{\bbox{k}+\bbox{q}}+
    \epsilon^{(1)\text{nr}}_{\bbox{k}} &\cong&
    \omega - (1+B)\frac{q^2}{2m_N}-(1+B)\frac{\bbox{q}\cdot\bbox{k}}{m_N} + 
    \bar{A}-A + (\bar{B}-B)\frac{k^2}{2m_N} \nonumber \\
  &=& \frac{qk_F}{m_N^{*\text{nr}}}\left\{\frac{1}{k_F}\left[
    (\omega+\varepsilon)\frac{m_N^{*\text{nr}}}{q}-\frac{q}{2}\right]-
    \hat{\bbox{q}}\cdot\frac{\bbox{k}}{k_F}+\frac{\bar{B}-B}{1+B}\frac{k_F}{2q}
    \left(\frac{k}{k_F}\right)^2 \right\} \nonumber \\
  &\cong& \frac{qk_F}{m_N^{*\text{nr}}}\left[
    \psi_{\text{nr}}^*-\hat{\bbox{q}}\cdot\frac{\bbox{k}}{k_F} \right] .
\label{eq:endenSEappr}
\end{eqnarray}
To go from the second to the last line in Eq.~(\ref{eq:endenSEappr}), 
we have neglected the term proportional to $k^2$, which is expected 
to be small, since $k<k_F$ and, typically, $q>k_F$. However, this 
approximation depends upon the interaction and one should check its 
validity, since it affects both the parameters $B$ and $\bar{B}$.
In Ref.~\cite{Bar96a} the term neglected has been shown to be small 
for the Bonn potential; the same turns out to be true also for the 
effective interaction employed in the next section.

Equation~(\ref{eq:endenSEappr}) is similar to the expression 
(\ref{eq:endenNR}) for the free energy denominator, but for the substitutions
\begin{eqnarray}
  m_N &\to& m_N^{*\text{nr}} = \frac{m_N}{1+B} \nonumber \\
  \psi_{\text{nr}} &\to& \psi_{\text{nr}}^* = 
    \frac{1}{k_F}\left[
    (\omega+\varepsilon)\frac{m_N^*}{q}-\frac{q}{2}\right] 
    \label{eq:mNstar} \\
  && \phantom{\psi_{\text{nr}}^*} = \frac{\psi_{\text{nr}}+\chi}{1+B} ,
    \qquad \chi = \frac{1}{k_F}
      \left(\frac{\varepsilon m_N}{q}-B\frac{q}{2}\right) ,
    \nonumber
\end{eqnarray}
(or $\omega\to\omega+\varepsilon$).

In Ref.~\cite{Bar96a} relativistic kinematics had been accounted for 
by applying to the above formulae the substitution 
$\omega\to\omega(1+\omega/2m_N)$ previously discussed. The correct 
approximation can be worked out by starting again from the ph 
propagator by defining
($\Delta\Sigma^{(1)}(\bbox{k},\bbox{q})\equiv
\Sigma^{(1)}(k)-\Sigma^{(1)}(|\bbox{k}+\bbox{q}|)$) and rewriting it as 
\begin{eqnarray}
  && \frac{1}{\omega-\epsilon^{(1)\text{r}}_{\bbox{k}+\bbox{q}}+
    \epsilon^{(1)\text{r}}_{\bbox{k}}} = \nonumber \\
  && \quad =
    \frac{\omega+\sqrt{k^2+m_N^2}+\Delta\Sigma^{(1)}(\bbox{k},\bbox{q})+
    \sqrt{(\bbox{k}+\bbox{q})^2+m_N^2}}{\omega^2+2\omega\sqrt{k^2+m_N^2}+
    2(\omega+\sqrt{k^2+m_N^2})\Delta\Sigma^{(1)}(\bbox{k},\bbox{q})+
    [\Delta\Sigma^{(1)}(\bbox{k},\bbox{q})]^2-q^2-2\bbox{q}\cdot\bbox{k}} 
    \nonumber \\
  && \quad \cong 
    \frac{m_N^{*\text{r}}}{q k_F}\frac{1+\omega/m_N+\Delta^{(1)}/m_N}
    {\psi^*_{\text{r}}-\hat{\bbox{q}}\cdot\bbox{k}/k_F} ,
\label{eq:endenSErel}
\end{eqnarray}
where
\begin{eqnarray}
  m_N^{*\text{r}} &=& \frac{m_N}{1+B(1+\omega/m_N+\Delta^{(1)}/m_N)} 
    \nonumber \\
  \psi^*_{\text{r}} &=& \frac{\psi_{\text{r}}+
  \chi[1+B(1+\omega/m_N+\Delta^{(1)}/2m_N)]}{1+
  B(1+\omega/m_N+\Delta^{(1)}/m_N)}
  \label{eq:mNstarrel} \\
  \Delta^{(1)} &=& \varepsilon-B\frac{q^2}{2m_N}\equiv\frac{q k_F}{m_N}\chi , 
    \nonumber
\end{eqnarray}
with $\chi$ already defined in Eq.~(\ref{eq:mNstar}). In deriving 
Eq.~(\ref{eq:endenSErel}), we have assumed that $k^2\ll m_N^2$, have 
evaluated the numerator at the pole thus discarding any angular 
dependence and, in the denominator, have retained only terms at 
most linear in $\hat{\bbox{q}}\cdot\bbox{k}/k_F$. As one can see, 
besides the transformation $\omega\to\omega(1+\omega/2m_N)$ there 
are other relativistic corrections, both to the effective scaling variable and 
to the Jacobian.

The quality of the approximations introduced above is good: indeed 
the HF response is reproduced with at most a few percent discrepancy 
(except on the borders of the response region, where the Fermi gas 
is anyway unrealistic). Thus, we see that in either the
non-relativistic or relativistic case, the prescription to include 
HF correlations in a response function is simply to replace $\psi$ 
with $\psi^*$ and $m_N$ with $m_N^*$ (and to multiply by a 
normalization factor when employing relativistic kinematics (see 
Eqs.~(\ref{eq:Pinrelnonrel})). For instance, from 
Eq.~(\ref{eq:Pi0}) one gets
\begin{equation}
  \Pi^{\text{HF}}(q,\omega) \cong J \frac{m_N^*}{q}\frac{k_F^2}{(2\pi)^2}
    \left[{\cal Q}^{(0)}(\psi^*)-{\cal Q}^{(0)}(\psi^*+\bar{q})\right] ,
\label{eq:HFappr}
\end{equation}
with
\begin{equation}
  J_{\text{nr}} = 1 
\end{equation}
and
\begin{equation}
  \label{eq:JHF} 
  J_{\text{r}}  = 1+\frac{\omega}{m_N}+\frac{\Delta^{(1)}}{m_N} . \nonumber
\end{equation}

In Appendix~\ref{app:B}, we give the explicit expressions for the 
first-order self-energy, based on the generic potential in 
Eqs.~(\ref{eq:pot})--(\ref{eq:mes-exch}).

\subsection{ Random phase approximation response }
\label{subsec:RPA-resp}

\begin{figure}[t]
\begin{center}
\vskip 2mm
\mbox{\epsfig{file=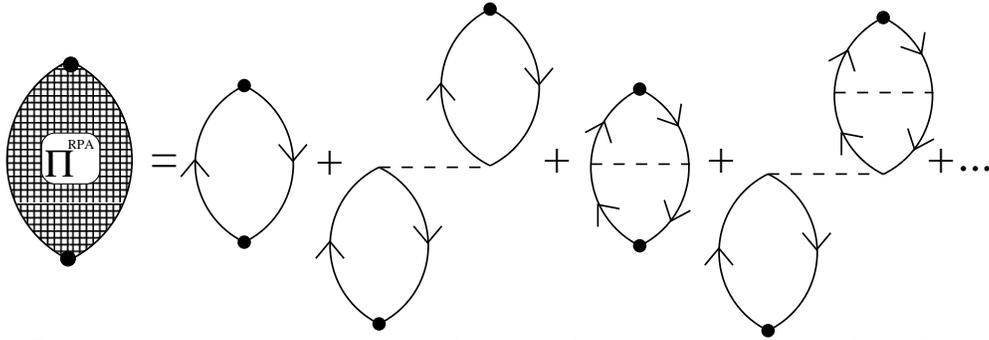,width=.8\textwidth}}
\caption{ Diagrammatic representation of the perturbative expansion for the
polarization propagator in the random phase approximation.
 }
\label{fig:PiRPA}
\end{center}
\end{figure}

If in Eq.~(\ref{eq:Gph}) one substitutes the irreducible vertex
function $\Gamma^{13}$ with the matrix elements of the bare 
potential, one gets the so-called {\em random phase approximation} 
to $G^{\text{ph}}$. In terms of the polarization propagator in 
Eq.~(\ref{eq:Pi}) one would get an infinite sum of diagrams such 
as those shown in Fig.~\ref{fig:PiRPA}.

We have already noted at the beginning of Section~\ref{subsec:Resp} 
that, while for the two-body Green's function $G^{\text{ph}}$ 
one can introduce an integral equation, this is not in general 
possible for the polarization propagator. It becomes possible 
when one approximates the irreducible vertex function $\Gamma^{13}$ 
with the {\em direct} matrix elements of the interaction. In that 
case, in an infinite system one gets a simple algebraic equation 
whose solution, for the polarization propagators in 
Eq.~(\ref{eq:PiLT}) and the interaction in 
Eq.~(\ref{eq:pot}), is readily found to be
\begin{equation}
  \Pi^{\text{ring}}_{\text{X}}(q,\omega) = \frac{\Pi^{(0)}(q,\omega)}
    {1-\Pi^{(1)\text{d}}_{\text{X}}(q,\omega)\big/\Pi^{(0)}(q,\omega)}
    ,
\label{eq:Piring}
\end{equation}
where $\Pi^{(1)\text{d}}_{\text{X}}$ represents the first-order 
{\em direct} polarization propagator:
\begin{mathletters}
\label{eq:Pi1d}
\begin{eqnarray}
  \Pi^{(1)\text{d}}_{\text{L};I=0(1)}(q,\omega) &=& 
    \Pi^{(0)}(q,\omega)4V_{0(\tau)}(q)\Pi^{(0)}(q,\omega) , \\
  \Pi^{(1)\text{d}}_{\text{T};I=0(1)}(q,\omega) &=&
    \Pi^{(0)}(q,\omega)4[V_{\sigma(\sigma\tau)}(q)
      -V_{t(t\tau)}(q)]\Pi^{(0)}(q,\omega) .
\end{eqnarray}
\end{mathletters}
The effect of the exchange diagrams is often included through an 
effective zero-range interaction, calculated by taking the limit 
$q\to0$ of the first-order exchange contribution and rewriting it 
as an effective first-order direct term \cite{Ose82}. Exact 
calculations, however, show that extrapolating this approximation 
to finite transferred momenta is not always reliable \cite{Shi89}.

A more advanced approximation scheme is given by the {\em continued
fraction} (CF) {\em expansion} \cite{Len80,Del85,Del87,Fes92}. 
At infinite order the CF expansion exactly corresponds to the summation of
the perturbative series, so that it is not any easier to calculate 
than the exact expression. However, when truncated at finite order, 
not only does it reproduce the standard perturbative series at the 
same order, but in addition it yields an estimate for each one of the infinite 
number of higher-order contributions. Regrettably no general methods 
are available to predict the convergence of the CF expansion, the 
only reliable test being to compare the results at successive orders.

On the other hand, one should note that for zero-range forces the 
first-order CF expansion already gives the exact (albeit trivial) 
result, making one hope that the short-range nature of the nuclear 
interactions allows for a fast convergence. Indeed, all available 
calculations have been performed truncating the CF expansion at 
first order \cite{Del85,Del87,Alb93,Bar94,Bar96a,Bar96b}.
Here, as anticipated, we shall test the convergence up to second order.

The CF formalism for the polarization propagator is developed in 
Ref.~\cite{Del85} for the case of Tamm-Dancoff correlations and 
extended in Ref.~\cite{Del87} to the full RPA. Instead of following 
the rather involved formal derivation given there, here we shall briefly 
sketch a sort of heuristic derivation of the CF expansion.

Let us assume that we want to build a CF-like expansion for the 
polarization propagator, according to the pattern 
\begin{equation}
  \Pi^{\text{RPA}} = \frac{\Pi^{(0)}}
    {\displaystyle 1 - A - \frac{\strut B}{\displaystyle 1 - C - 
    \frac{\strut D}
    {\displaystyle 1 - ...}}} .
\end{equation}
We have said that the CF approach at $n$-th order exactly corresponds 
to the perturbative series at the same order and then it approximates 
the higher orders. Thus, if we want to approximate the exact RPA 
propagator at first order in CF (for sake of illustration we drop 
spin-isospin indices),
\begin{equation}
  \Pi^{\text{RPA}} = \sum_{n=0}^{\infty}\Pi^{(n)} ,
\end{equation}
we can rather naturally set
\begin{equation}
  \Pi^{(n)} \cong \Pi^{(0)} \left[\frac{\Pi^{(1)}}{\Pi^{(0)}}\right]^n .
\label{eq:Pin}
\end{equation}
In Eq.~(\ref{eq:Pin}) 
$\Pi^{(1)}\equiv\Pi^{(0)}4V\Pi^{(0)}+\Pi^{(1)\text{ex}}$ is the sum 
of the direct and exchange first-order terms of RPA --- since this 
yields the correct expression for the direct terms. With the 
approximation in Eq.~(\ref{eq:Pin}) the summation is trivial, yielding
\begin{equation}
  \Pi^{\text{RPA}}_{\text{CF1}} = 
    \frac{\Pi^{(0)}}{1-\Pi^{(1)}/\Pi^{(0)}} =
    \frac{\Pi^{(0)}}{1-4V\Pi^{(0)}-\Pi^{(1)\text{ex}}/\Pi^{(0)}} .
\end{equation}
We could then add in the denominator of the above expression the 
exact second-order term, $\Pi^{(2)}$, after subtracting its 
approximate estimate given by the first-order CF expansion, 
$[\Pi^{(1)}]^2/\Pi^{(0)}$. We would thus obtain
\begin{eqnarray}
  \Pi^{\text{RPA}}_{\text{CF2}} &=& 
    \frac{\Pi^{(0)}}{1-\Pi^{(1)}/\Pi^{(0)}
    -\{\Pi^{(2)}/\Pi^{(0)}-[\Pi^{(1)}/\Pi^{(0)}]^2\}} \nonumber \\
  &=& \frac{\Pi^{(0)}}{1-4V\Pi^{(0)}-\Pi^{(1)\text{ex}}/\Pi^{(0)}
    -\{\Pi^{(2)\text{ex}}/\Pi^{(0)}-[\Pi^{(1)\text{ex}}/\Pi^{(0)}]^2\}} .
\label{eq:PiCF2}
\end{eqnarray}
It is easily deduced from Eq.~(\ref{eq:PiCF2}) that the third-order 
term is approximated as 
$\Pi^{(3)}\cong\Pi^{(1)}\{2\Pi^{(2)}/\Pi^{(0)}-
[\Pi^{(1)}/\Pi^{(0)}]^2\}$.
Then, going ahead in a CF-style expansion we would guess for the 
exact RPA propagator the following expression:
\begin{equation}
  \Pi^{\text{RPA}} = \frac{\Pi^{(0)}}
    {\displaystyle 1 - \Pi^{(1)}/\Pi^{(0)}
    - \frac{\strut 
      \Pi^{(2)\text{ex}}/\Pi^{(0)}-[\Pi^{(1)\text{ex}}/\Pi^{(0)}]^2}
    {\displaystyle 1 - 
      \frac{\strut \Pi^{(3)\text{ex}}/\Pi^{(0)}
     +[\Pi^{(1)\text{ex}}/\Pi^{(0)}]^3
        -2[\Pi^{(1)\text{ex}}/\Pi^{(0)}][\Pi^{(2)\text{ex}}/\Pi^{(0)}]}
      {\displaystyle 
      \Pi^{(2)\text{ex}}/\Pi^{(0)}-[\Pi^{(1)\text{ex}}/\Pi^{(0)}]^2}
    - ...}} .
\end{equation}
This is the expression that one would get from the formalism of 
Refs.~\cite{Del85,Del87} when the expansion up to third order is 
worked out. Note that we did not assume any specific scheme (either
Tamm-Dancoff or RPA) in this heuristic derivation.

Thus, following Eq.~(\ref{eq:Piring}), we can write
\begin{equation}
  \Pi^{\text{RPA}}_{\text{X}} = \frac{\Pi^{(0)}}
    {\displaystyle 1 
    -\Pi^{(1)\text{d}}_{\text{X}}\big/\Pi^{(0)}
    -\Pi^{(1)\text{ex}}_{\text{X}}\big/\Pi^{(0)}
    -\frac{\strut 
      \Pi^{(2)\text{ex}}_{\text{X}}\big/\Pi^{(0)}
      -\left[\Pi^{(1)\text{ex}}_{\text{X}}\big/\Pi^{(0)}
      \right]^2}{\displaystyle 1-...}} ,
\end{equation}
where $\Pi^{(1)\text{d}}_{\text{X}}$ has been defined in 
Eq.~(\ref{eq:Pi1d}). Clearly, a truncation at $n$-th order would 
require the calculation of the exchange contributions up to that order.
Exploiting Eq.~(\ref{eq:PiLT}) these can be cast in the form
\begin{mathletters}
\begin{eqnarray}
  \Pi^{(n)\text{ex}}_{\text{L};I}(q,\omega) &=& 
    \text{tr}[\hat{O}^{I}_{\text{L}}\hat{\Pi}^{(n)\text{ex}}(q,\omega)
    \hat{O}^{I}_{\text{L}}] = 
    \sum_{\alpha_i}
   C^{\alpha_1...\alpha_n}_{\text{L};I}\Pi^{(n)\text{ex}}_{\alpha_1...\alpha_n}
    (q,\omega)
    , \\
  \Pi^{(n)\text{ex}}_{\text{T};I}(q,\omega) &=&
    \sum_{ij}\Lambda_{ji}
    \text{tr}[\hat{O}^{I}_{\text{T};i}\hat{\Pi}^{(n)\text{ex}}(q,\omega)
    \hat{O}^{I}_{\text{T};j}] = 
    \sum_{\alpha_i}
   C^{\alpha_1...\alpha_n}_{\text{T};I}\Pi^{(n)\text{ex}}_{\alpha_1...\alpha_n}
    (q,\omega)
    , 
\end{eqnarray}
\end{mathletters}
where the indices $\alpha_i$ run over all the spin-isospin channels 
and the spin-isospin factors are absorbed into the coefficients 
$C^{\alpha_1...\alpha_n}_{\text{X}} \equiv C^{(\alpha_1)}_{\text{X}} 
C^{(\alpha_2)}_{\text{X}}...C^{(\alpha_n)}_{\text{X}}$ (see 
Table~\ref{tab:I}).
\begin{table}
\begin{center}
\begin{tabular}{crrrrrr}
  $\strut\phantom{\Big|}X$ & $C^0_X$ & $C^\tau_X$ & $C^\sigma_X$ & 
  $C^{\sigma\tau}_X$ & $C^t_X$ & $C^{t\tau}_X$ \\ \tableline
  $L;I=0$ & 1  &  3  &  3  &  9  &  0  &  0  \\ 
  $L;I=1$ & 1  & -1  &  3  & -3  &  0  &  0  \\ 
  $T;I=0$ & 1  &  3  & -1  & -3  & -1  & -3  \\ 
  $T;I=1$ & 1  & -1  & -1  &  1  & -1  &  1  \\ 
\end{tabular}
\end{center}
\caption{ The spin-isospin coefficients $C^{\alpha}_X$ (see text) 
in the longitudinal and transverse isoscalar and isovector channels, 
for the interaction in Eq.~(\protect\ref{eq:pot}).
  }
\label{tab:I}
\end{table}
Moreover the ``elementary'' exchange contribution
$\Pi^{(n)\text{ex}}_{\alpha_1...\alpha_n}$ containing $n$ interaction 
lines $V_{\alpha_1}$...$V_{\alpha_n}$, namely\footnote{ The following 
formulae are valid for non-tensor interactions; the treatment of the 
tensor terms is slightly more complex and it is given in 
Appendix~\ref{app:C}.}
\begin{eqnarray}
  \Pi^{(n)\text{ex}}_{\alpha_1...\alpha_n}(q,\omega) &=&
    -i^{n+1}\int\frac{d^4K_1}{(2\pi)^4}\cdot\cdot\cdot
    \frac{d^4K_{n+1}}{(2\pi)^4}
    G^{(0)}(K_1)G^{(0)}(K_1+Q)V_{\alpha_1}(\bbox{k}_1-\bbox{k}_2)
    \cdot\cdot\cdot \nonumber \\
  &&\qquad \cdot\cdot\cdot V_{\alpha_n}(\bbox{k}_n-\bbox{k}_{n+1})
    G^{(0)}(K_{n+1})G^{(0)}(K_{n+1}+Q) \nonumber \\
  &=& (-1)^n\int\frac{d\bbox{k}_1}{(2\pi)^3}\cdot\cdot\cdot
    \frac{d\bbox{k}_{n+1}}{(2\pi)^3}
    G^{(0)}_{\text{ph}}(\bbox{k}_1,\bbox{q};\omega)
    V_{\alpha_1}(\bbox{k}_1-\bbox{k}_2)\cdot\cdot\cdot \nonumber \\
  &&\qquad \cdot\cdot\cdot V_{\alpha_n}(\bbox{k}_n-\bbox{k}_{n+1})
    G^{(0)}_{\text{ph}}(\bbox{k}_{n+1},\bbox{q};\omega) 
\end{eqnarray}
have been introduced. With the definition of $G^{(0)}_{\text{ph}}$ 
given in Eq.~(\ref{eq:G0phnew}) and by a suitable change of integration 
variables one can eliminate all of the $\theta$-functions that contain
angular integration variables, leaving a multiple integral with the 
following general structure:
\begin{eqnarray}
  && \Pi^{(n)\text{ex}}_{\alpha_1...\alpha_n}(q,\omega) =
    (-1)^n\int\frac{d\bbox{k}_1}{(2\pi)^3}\theta(k_F-k_1)
    \cdot\cdot\cdot
    \frac{d\bbox{k}_{n+1}}{(2\pi)^3}\theta(k_F-k_{n+1}) \nonumber \\
  &&\quad\times\Big[
    \frac{1}{\omega-\epsilon_{\bbox{k}_1+\bbox{q}}+\epsilon_{\bbox{k}_1}
    +i\eta_\omega} 
    V_{\alpha_1}(\bbox{k}_1-\bbox{k}_2)\cdot\cdot\cdot
    V_{\alpha_n}(\bbox{k}_n-\bbox{k}_{n+1})
   \frac{1}{\omega-\epsilon_{\bbox{k}_{n+1}+\bbox{q}}+\epsilon_{\bbox{k}_{n+1}}
    +i\eta_\omega} \nonumber \\
  &&\quad\quad+\sum(\omega\to-\omega) \Big] .
\label{eq:Pinex}
\end{eqnarray}
In Eq.~(\ref{eq:Pinex}), $\sum(\omega\to-\omega)$ stands for the 
sum of all the terms generated according to the following rules:
\begin{itemize}
  \item[i)] Take all of the terms obtained by substituting 
$\omega\to-\omega$ in one energy denominator in the second line 
of Eq.~(\ref{eq:Pinex}); then add the contribution obtained by 
performing the same substitution in two energy denominators and so 
on up to when the replacement $\omega\to-\omega$ has been performed 
in all the $n+1$ denominators;
  \item[ii)] Every time 
$(\omega-\epsilon_{\bbox{k}_i+\bbox{q}}+\epsilon_{\bbox{k}_i}
  +i\eta_\omega)^{-1}$ is replaced with
$(-\omega-\epsilon_{\bbox{k}_i+\bbox{q}}+\epsilon_{\bbox{k}_i}
  -i\eta_\omega)^{-1}$
then replace $\bbox{k}_i$ with $-\bbox{k}_i-\bbox{q}$ in the potential.
\end{itemize}

The number of integrations can be reduced by noticing that the 
azimuthal angles are contained only in the potential functions 
$V_{\alpha_i}$. For typical potentials this integration can be done 
analytically, hence it is convenient to introduce a new function 
representing the azimuthal integral of the potential. To this end, 
define the new variables:
\begin{eqnarray}
  |\bbox{k}-\bbox{k}'| &=& \sqrt{k^2+{k'}^2-2kk'[\cos\theta\,\cos\theta'+
    \sin\theta\,\sin\theta'\,\cos(\varphi-\varphi')]} \nonumber \\
  &=& \sqrt{k^2+{k'}^2-2[yy'+\sqrt{k^2-y^2}\sqrt{{k'}^2-{y'}^2}
    \cos(\varphi-\varphi')]} \\
  &=& \sqrt{x+x'-2\sqrt{x}\sqrt{x'}\cos(\varphi-\varphi')+(y-y')^2} , 
    \nonumber
\end{eqnarray}
where $y\equiv k\,\cos\theta$ and $x\equiv k^2-y^2$. Then, one can introduce
\begin{equation}
  W_{\alpha}(x,y;x',y') = \int_0^{2\pi}\frac{d\varphi}{2\pi}
    V_{\alpha}(\bbox{k}-\bbox{k}') = W_{\alpha}(x',y';x,y)
\label{eq:Walpha}
\end{equation}
and rewrite Eq.~(\ref{eq:Pinex}) as 
\begin{eqnarray}
  \Pi^{(n)\text{ex}}_{\alpha_1...\alpha_n}(q,\omega) &=&
    (-1)^n\left(\frac{m_N}{q}\right)^{n+1}\left(\frac{k_F}{2\pi}\right)^{2n+2}
    \int_{-1}^{1}dy_1\frac{1}{2}\int_{0}^{1-y_1^2}dx_1\cdot\cdot\cdot
    \int_{-1}^{1}dy_{n+1}\frac{1}{2}\int_{0}^{1-y_{n+1}^2}dx_{n+1} \nonumber \\
  && \times\frac{1}{\psi-y_1+i\eta_{\omega}}W_{\alpha_1}(x_1,y_1;x_2,y_2)
   \cdot\cdot\cdot W_{\alpha_n}(x_n,y_n;x_{n+1},y_{n+1})
   \frac{1}{\psi-y_{n+1}+i\eta_{\omega}} \nonumber \\
  && + \sum(\omega\to-\omega) .
\label{eq:Pinexpsi}
\end{eqnarray}
For $n=1$ one has
\begin{eqnarray}
  \Pi^{(1)\text{ex}}_{\alpha}(q,\omega) &=&
    -\left(\frac{m_N}{q}\right)^{2}\frac{k_F^4}{(2\pi)^4}
    \int_{-1}^{1}dy\frac{1}{2}\int_{0}^{1-y^2}dx
    \int_{-1}^{1}dy'\frac{1}{2}\int_{0}^{1-{y'}^2}dx' \nonumber \\
  && \times\frac{1}{\psi-y+i\eta_{\omega}}W_{\alpha}(x,y;x',y')
   \frac{1}{\psi-y'+i\eta_{\omega}} \nonumber \\
  && + \sum(\omega\to-\omega) \nonumber \\
  &=&  -\left(\frac{m_N}{q}\right)^{2}\frac{k_F^4}{(2\pi)^4}
    \left[{\cal Q}_\alpha^{(1)}(0,\psi) 
    - {\cal Q}_\alpha^{(1)}(\bar{q},\psi) 
    + {\cal Q}_\alpha^{(1)}(0,\psi+\bar{q})
    - {\cal Q}_\alpha^{(1)}(-\bar{q},\psi+\bar{q})\right]
    , \nonumber \\
\label{eq:Pi1ex}
\end{eqnarray}
where 
\begin{equation}
  {\cal Q}_\alpha^{(1)}(\bar{q},\psi) = 2 \int_{-1}^1 dy
   \frac{1}{\psi-y+i\eta_{\omega}}\int_{-1}^1 dy' \, {W_\alpha}''(y,y';\bar{q})
    \frac{1}{y-y'+\bar{q}}
\label{eq:Q1alpha}
\end{equation}
and
\begin{equation}
  W''_\alpha(y,y';\bar{q}) = \frac{1}{2}\int_0^{1-y^2}dx\,
    \frac{1}{2}\int_0^{1-{y'}^2}dx' \, W_\alpha(x,y+\bar{q};x',y') .
\label{eq:Wppalpha}
\end{equation}
Note that in getting to Eq.~(\ref{eq:Q1alpha}) use has been made of the 
Poincar\'e--Bertrand theorem \cite{Bal63}. For the potential in 
Eq.~(\ref{eq:mes-exch}) $W''_\alpha$ can be calculated analytically 
(see Appendix~\ref{app:D}), so that the calculation of the first-order
exchange contribution to the polarization propagator is reduced to the 
numerical evaluation of two-dimensional integrals for the real part 
and of one-dimensional integrals for the imaginary part.

For $n=2$ one has
\begin{eqnarray}
  &&\Pi^{(2)\text{ex}}_{\alpha\alpha'}(q,\omega) =
    \left(\frac{m_N}{q}\right)^{3}\frac{k_F^6}{(2\pi)^6}
    \int_{-1}^{1}dy_1\frac{1}{2}\int_{0}^{1-y_1^2}dx_1
    \int_{-1}^{1}dy_2\frac{1}{2}\int_{0}^{1-y_2^2}dx_2
    \int_{-1}^{1}dy_3\frac{1}{2}\int_{0}^{1-y_3^2}dx_3 \nonumber \\
  && \phantom{\Pi^{(2)\text{ex}}_{\alpha\alpha'}(q,\omega)=}
    \times\frac{1}{\psi-y_1+i\eta_{\omega}}W_{\alpha}(x_1,y_1;x_2,y_2)
    \frac{1}{\psi-y_2+i\eta_{\omega}}W_{\alpha}(x_2,y_2;x_3,y_3)
    \frac{1}{\psi-y_3+i\eta_{\omega}} \nonumber \\
  && \phantom{\Pi^{(2)\text{ex}}_{\alpha\alpha'}(q,\omega)=}  
    + \sum(\omega\to-\omega) \nonumber \\
  && \phantom{\Pi^{(2)\text{ex}}_{\alpha\alpha'}(q,\omega)} =
    \left(\frac{m_N}{q}\right)^{3}\frac{k_F^6}{(2\pi)^6}
    \left[{\cal Q}_{\alpha\alpha'}^{(2)}(0,0;\psi) 
        - {\cal Q}_{\alpha\alpha'}^{(2)}(0,\bar{q};\psi) 
        - {\cal Q}_{\alpha\alpha'}^{(2)}(\bar{q},0;\psi) 
        + {\cal Q}_{\alpha\alpha'}^{(2)}(\bar{q},\bar{q};\psi) \right.
    \nonumber \\
  &&\qquad \left.
        - {\cal Q}_{\alpha\alpha'}^{(2)}(0,0;\psi+\bar{q})
        + {\cal Q}_{\alpha\alpha'}^{(2)}(0,-\bar{q};\psi+\bar{q})
        + {\cal Q}_{\alpha\alpha'}^{(2)}(-\bar{q},0;\psi+\bar{q})
        - {\cal Q}_{\alpha\alpha'}^{(2)}(-\bar{q},-\bar{q};\psi+\bar{q})
    \right] , \nonumber \\
\label{eq:Pi2ex}
\end{eqnarray}
where 
\begin{equation}
  {\cal Q}_{\alpha\alpha'}^{(2)}(\bar{q}_1,\bar{q}_2;\psi) = \int_{-1}^1 dy
    \frac{1}{2}\int_{0}^{1-y^2} dx \, 
    {\cal G}_{\alpha}(x,y+\bar{q}_1;\psi+\bar{q}_1)
    \frac{1}{\psi-y+i\eta_{\omega}}
    {\cal G}_{\alpha'}(x,y+\bar{q}_2;\psi+\bar{q}_2)
\label{eq:Q2aap}
\end{equation}
and
\begin{mathletters}
\label{eq:GWp}
\begin{eqnarray}
  {\cal G}_{\alpha}(x,y;\psi) &=& \int_{-1}^1 dy'
    \frac{1}{\psi-y'+i\eta_{\omega}}W'_\alpha(x,y;y') ,
\label{eq:Galpha} \\
  W'_\alpha(x,y;y') &=& \frac{1}{2}\int_0^{1-{y'}^2}dx' \, W_\alpha(x,y;x',y')
    .
\label{eq:Wpalpha}
\end{eqnarray}
\end{mathletters}
For the potential in Eq.~(\ref{eq:mes-exch}) $W'_\alpha$ can be 
calculated analytically (see Appendix~\ref{app:D}) and one is left 
with the numerical integration of Eqs.~(\ref{eq:Q2aap}) and 
(\ref{eq:Galpha}), so that the calculation of the second-order 
exchange contribution to the polarization propagator is effectively 
reduced to the numerical evaluation of at most three-dimensional integrals.
Higher orders add a numerical two-dimensional integration for each
additional interaction line, since, for a potential of the form 
in Eq.~(\ref{eq:mes-exch}), only the azimuthal integration can be 
performed analytically for the interaction lines that do not close 
on the external vertices.

Finally we recall that the nucleon propagators can be dressed by the 
HF field, as explained in Section~\ref{subsec:HF-resp}, by replacing 
$\psi\to\psi^*$ and $m_N\to m_N^*$, where $\psi^*$ and $m_N^*$ have 
been defined in Eqs.~(\ref{eq:mNstar}) and (\ref{eq:mNstarrel}), 
multiplying by the appropriate power of the normalization factor 
$1+\omega/m_N+\Delta^{(1)}/m_N$ when relativistic kinematics are 
employed (see Eqs.~(\ref{eq:Pinrelnonrel}) and (\ref{eq:JHF})).

\subsection{ Effective particle-hole interaction }
\label{subsec:eff-int}

In order to assess the contributions to the nuclear responses arising 
from the various approximation schemes introduced so far and to show 
typical results, first of all we have to choose an effective
interaction. This choice can be rather delicate as it may introduce 
uncontrolled uncertainties in the calculation. Here, however, we are 
not interested so much in comparisons with data, but rather with setting up 
working many-body schemes. For this purpose, we shall use the 
$G$-matrix based on the Bonn potential of Ref.~\cite{Nak84}, adapted 
to the quasielastic regime as in Ref.~\cite{DeP97}. Although the 
attraction provided in the scalar-isoscalar channel by this
interaction is definitely too strong \cite{DeP97}, it will serve 
our illustrative needs.

Two approaches to determine the effective ph interaction in the 
nuclear medium appear to be possible: one can either directly fix an effective 
potential by fitting some phenomenological properties or start with 
a bare nucleon-nucleon interaction and calculate the related $G$-matrix. 
Parameterizations of the ph interaction based upon the first 
procedure are generally only available at very low momentum 
transfers (in terms of Migdal-Landau parameters), and since we 
are probing relatively high momenta, we have resorted to using a 
$G$-matrix. We have chosen the one of Ref.~\cite{Nak84}, that, 
in our view, has the following appealing features: it is based 
upon a realistic boson-exchange potential; it accounts for the 
density dependence; and it includes (nonlocal) exchange contributions 
in the effective interaction, which are conveniently parameterized 
in terms of Yukawa functions.

A feature related to the effective inclusion of antisymmetrization
effects is particularly interesting in order to test a specific 
widely employed approximation scheme, the so-called {\em ring 
approximation},  in which the exchange diagrams of the RPA series 
are dropped and their effect mimicked by adding to the direct 
interaction matrix elements an effective exchange contribution 
(see, e.~g., Ref.~\cite{Ose82}). Indeed, below we shall compare 
calculations employing the fully antisymmetrized formalism developed 
in the previous subsections using the direct part of the $G$-matrix, 
to those employing the ring approximation using the antisymmetrized 
effective interaction.

To facilitate the comparison with the original parameterization of
Ref.~\cite{Nak84}, the potential is given here using the standard 
representation of Eq.~(\ref{eq:pot}) in spin and isospin (no spin-orbit 
contribution will be considered in the following), but employing 
different symbols for the momentum space potentials (and adding 
the tensor contributions in the exchange channel):
\begin{eqnarray}
  V(\bbox{k}_f,\bbox{k}_i;k_F) &=& F + 
     F' \bbox{\tau}_1\cdot\bbox{\tau}_2 +
     G  \bbox{\sigma}_1\cdot\bbox{\sigma}_2 +
     G' \bbox{\sigma}_1\cdot\bbox{\sigma}_2 \ \bbox{\tau}_1\cdot\bbox{\tau}_2
     \nonumber \\
  && +
     T  S_{12}(\hat{\bbox{q}}) +
     T' S_{12}(\hat{\bbox{q}}) \bbox{\tau}_1\cdot\bbox{\tau}_2 +
     H  S_{12}(\hat{\bbox{Q}}) +
     H' S_{12}(\hat{\bbox{Q}}) \bbox{\tau}_1\cdot\bbox{\tau}_2 \ ,
\end{eqnarray}
where $\bbox{q}=\bbox{k}_i-\bbox{k}_f$, 
$\bbox{Q}=\bbox{k}_i+\bbox{k}_f$ ($\bbox{k}_i$, $\bbox{k}_f$ being 
the relative momenta in the initial and final states, respectively) 
and the coefficients are density and momentum dependent.

\begin{figure}[p]
\begin{center}
\mbox{\epsfig{file=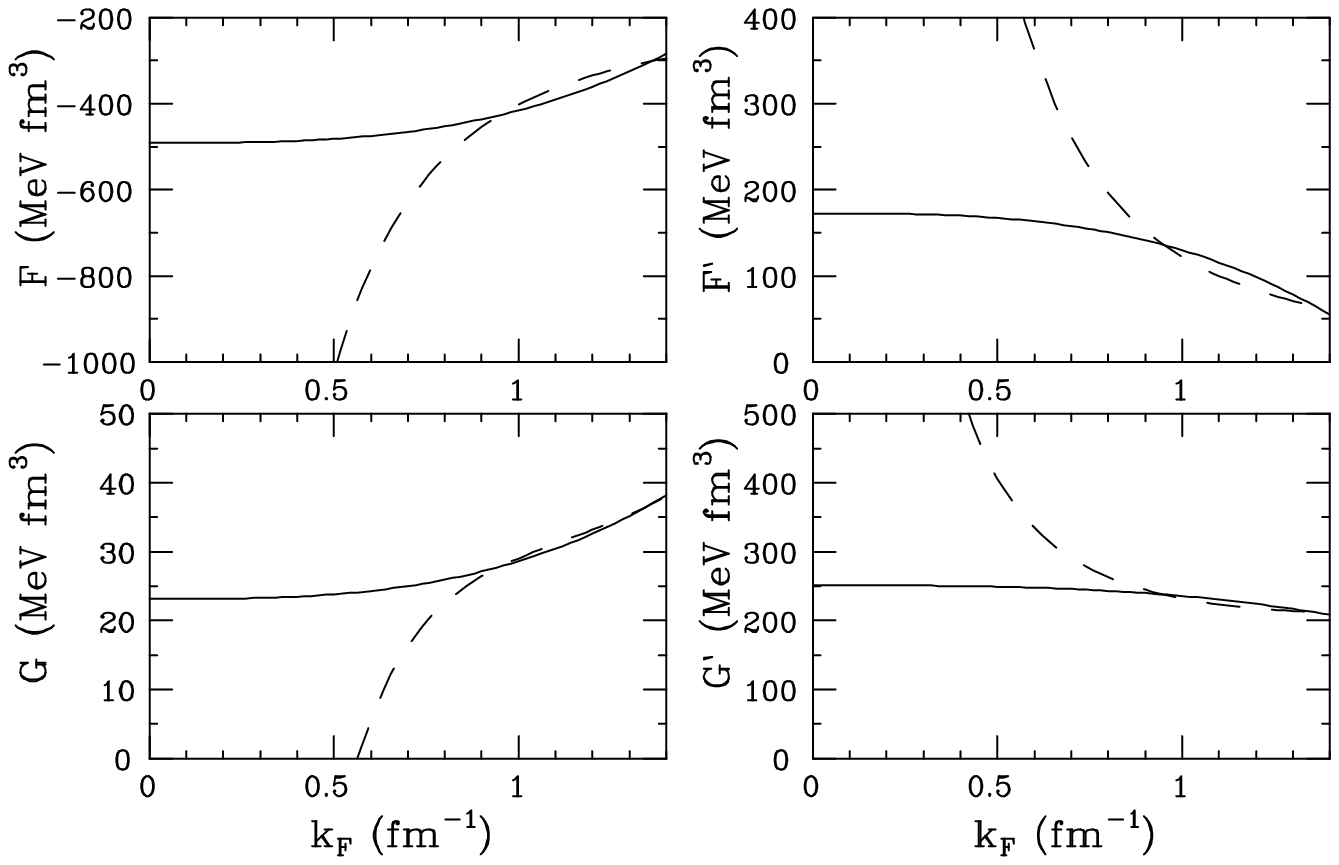,width=.85\textwidth}}
\caption{ Effective interaction in the non-tensor channels as a function of
$k_F$ at $q=0$; linear (solid) and from Ref.~\protect\cite{Nak84} (dashed)
density dependence.
  }
\label{fig:Veff-kF}
\vskip 3mm
\mbox{\epsfig{file=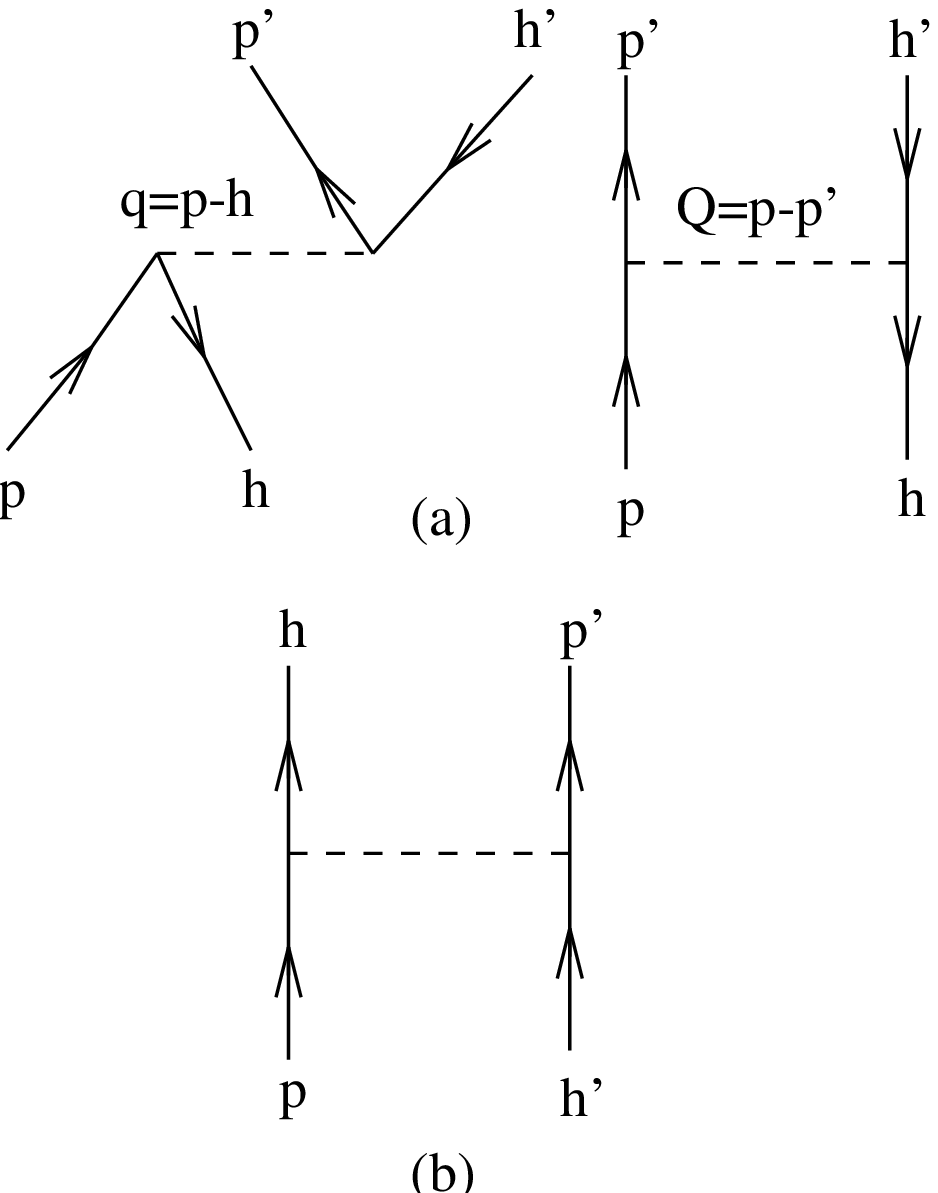,width=.50\textwidth}}
\caption{ (a) Direct and exchange ph matrix elements; (b) direct pp matrix 
element.
}
\label{fig:Veff-diag}
\end{center}
\end{figure}

Before utilizing the interaction of Ref.~\cite{Nak84}
in a calculation of quasielastic responses, a few issues have to
be addressed \cite{DeP97}.

a) The density dependence of the $G$-matrix is given in terms of 
density-dependent coupling constants, which is not very useful 
for applications to finite nuclei. Furthermore, the parameterization 
is fitted for 0.95 fm$^{-1} < k_F <$ 1.36 fm$^{-1}$, which spans 
a range of densities down to roughly 1/3 of the central density. 
Extrapolation of this parameterization to lower densities (which 
is crucial for application to hadron scattering) gives unreasonable 
results. Thus, we have chosen to employ a linear $\rho$-dependence 
($V=V^{\text{ex}}+V^\rho \rho$), which is considered a reasonable
choice (see, e.~g., Ref.~\cite{Spe77}). In Fig.~\ref{fig:Veff-kF} 
one can see a comparison of the two parameterizations for the 
$k_F$-dependence of the effective interaction. It should be noted 
that most of the contribution to the quasielastic responses comes 
from densities where the two descriptions differ by only a few percent.

\begin{figure}[p]
\begin{center}
\mbox{\epsfig{file=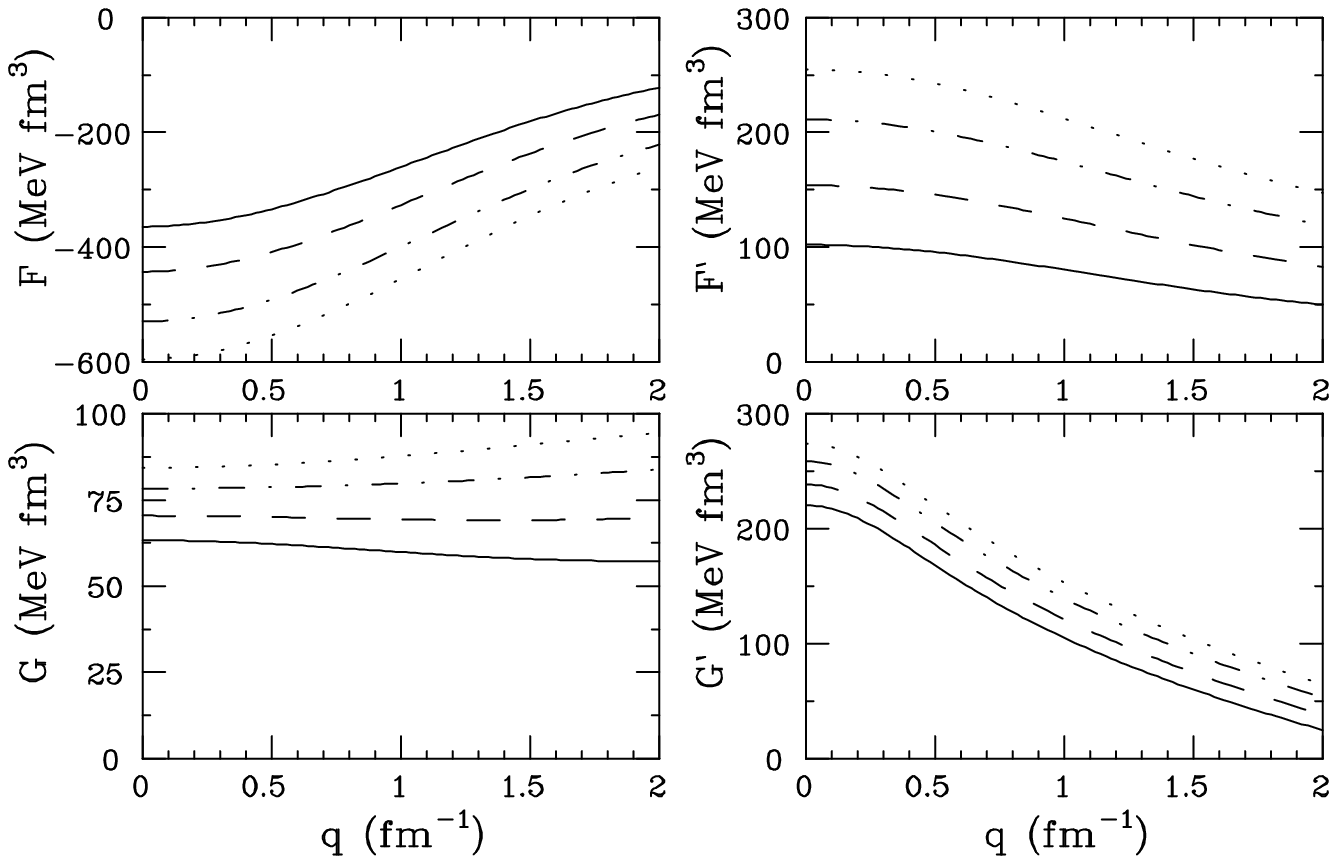,width=.85\textwidth}}
\caption{ Effective ph interaction in the non-tensor channels as a function of
$q$ at $k_F=1.36$ (solid), 1.25 (dashed), 1.10 (dot-dashed) and 0.95 fm$^{-1}$ 
(dotted).
}
\label{fig:Veff-nontensor}
\vskip 3mm
\mbox{\epsfig{file=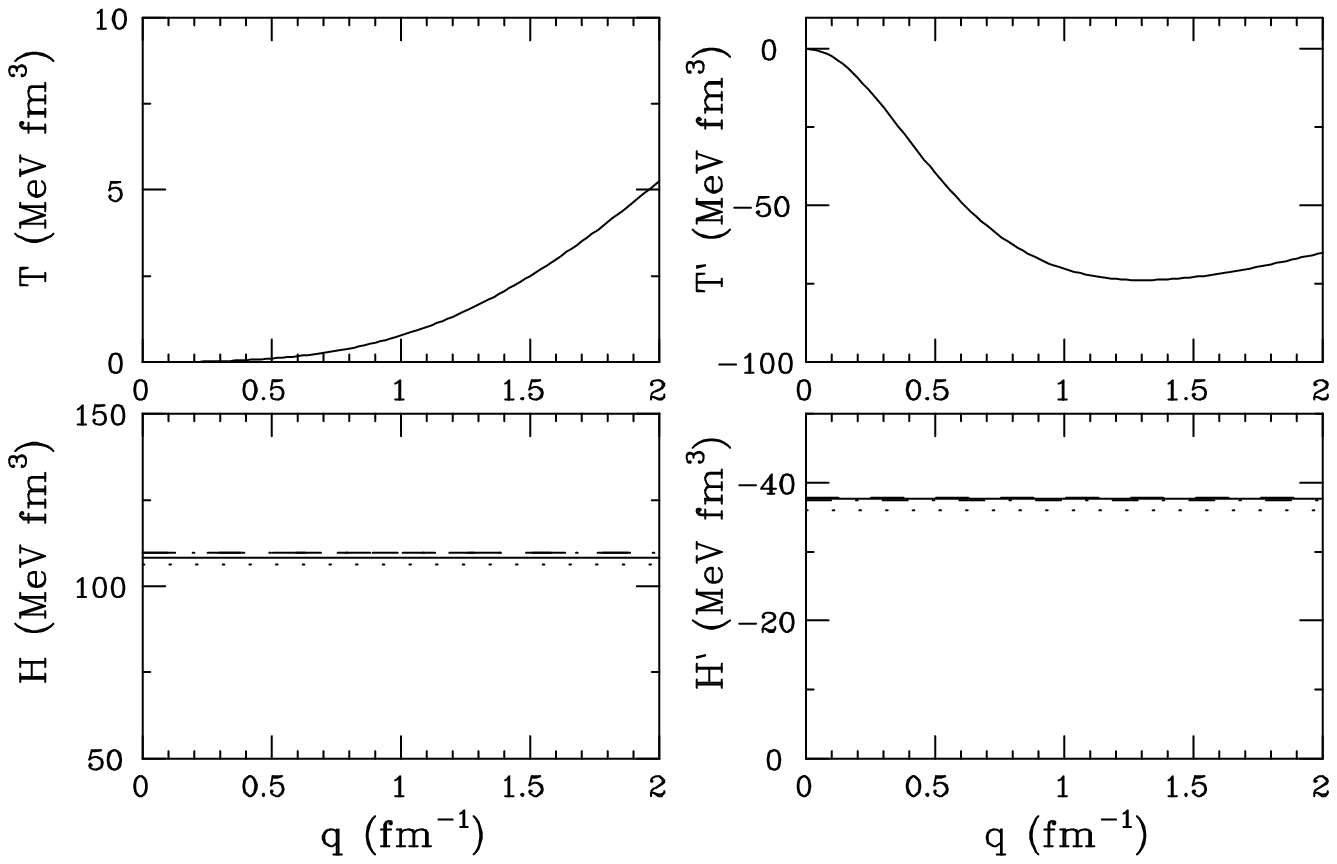,width=.85\textwidth}}
\caption{ As in Fig.~\protect\ref{fig:Veff-nontensor}, but for the tensor
channels; $T$ and $T'$ do not depend on the density.
}
\label{fig:Veff-tensor}
\end{center}
\end{figure}

b) In order to obtain a local interaction at a fixed density, one 
can use the relation between $q$, $Q$ and $k_i$, i.~e., 
$Q=\sqrt{4k_i^2-q^2}$ and take for $k_i$ a suitably chosen average 
value, $\langle k_i \rangle$; then, the only independent momentum is 
$q$. The authors of Ref.~\cite{Nak84} were interested in a 
potential for low excitation energy nuclear structure 
calculations and hence they 
assumed that the two nucleons in the initial state lie on the 
Fermi surface and so averaged over the relative angle, getting 
$\langle k_i \rangle \approx 0.7 k_F$. Clearly, in this case one 
has the constraint $0<q\lesssim 1.4k_F$. On the other hand, we are 
interested in the ph interaction in the quasielastic region where 
one nucleon in the initial state is below the Fermi sea, while the 
other can be well above it. A look at Fig.~\ref{fig:Veff-diag} shows 
that $\bbox{k}_i$ is defined in terms of the particle and hole momenta as 
$\bbox{k}_i=(\bbox{p}-\bbox{h}')/2=(\bbox{h}-\bbox{h}'+\bbox{q})/2$. 
Thus, at fixed $\bbox{q}$ one should average $k_i$ over $\bbox{h}$ and 
$\bbox{h}'$, getting $\langle k_i \rangle \approx
\sqrt{6k_F^2/5+q^2}/2$. Now $k_i$ grows with $q$, so that there are 
no longer constraints on $q$ and the exchange momentum turns out 
to be constant, $Q=\sqrt{6/5}k_F$. In Fig.~\ref{fig:Veff-nontensor} 
one can see the resulting interaction in the non-tensor channels.

c) The tensor channels are simpler, since in the parameterization 
of Ref.~\cite{Nak84} there is no explicit density dependence
(Fig.~\ref{fig:Veff-tensor}). The coefficients of the exchange 
tensor operator, $H$ and $H'$, display a very mild density 
dependence induced by $Q$, which is completely negligible. The 
only drawback concerns the treatment of $S_{12}(\bbox{Q})$: assuming 
that $\bbox{q}$ and $\bbox{Q}$ are orthogonal, with some algebra 
one can show that 
$S_{12}(\hat{\bbox{Q}})=-S_{12}(\hat{\bbox{q}})/2$. 

\subsection{ Testing the model }
\label{subsec:test}

First of all we have to choose the Fermi momentum. Of course, one could
easily perform a local density calculation to achieve a better 
description of finite nuclei. Here, for sake of illustration, we 
prefer to use the pure Fermi gas. The choice of $k_F$ can be made 
in several ways --- here we shall choose an average value according 
to the formula (see, e.~g., \cite{Cen97b}) 
\begin{equation}
  \bar{k}_F = \frac{1}{A}\int d\bbox{r} k_F(r) \rho(r) ,
\end{equation}
where $\rho(r)$ is the empirical Fermi density distribution normalized
to the number of nucleons and $k_F(r)=[(3\pi/2)\rho(r)]^{1/3}$. For 
$^{12}$C one gets $\bar{k}_F\approx195$ MeV/c and this is the value 
used in the calculations that follow.

\begin{figure}[t]
\begin{center}
\mbox{\epsfig{file=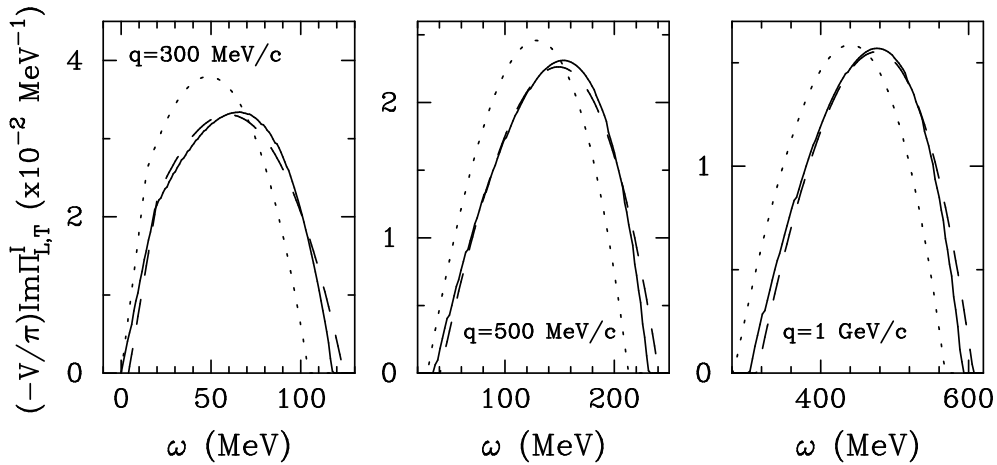,width=.8\textwidth}}
\vskip 2mm
\caption{ Fermi gas responses for $k_F=195$ MeV/c at $q=300$, 500 and 
1000 MeV/c: Free (dotted), exact HF (solid) and HF approximated
according to the prescription of Section~\protect\ref{subsec:HF-resp} 
(dashed). The ph interaction is the $G$-matrix discussed in the text; 
the kinematics are relativistic.
 }
\label{fig:HFspeth}
\end{center}
\end{figure}

Let us start with the HF response. In Fig.~\ref{fig:HFspeth} we
display the HF response of $^{12}$C at $q=300$, 500 and 1000 MeV/c. 
As anticipated, Eq.~(\ref{eq:HFappr}) turns out to be a good 
approximation to the exact expression (\ref{eq:HFexact}) (except 
on the borders of the response region, where the Fermi gas is anyway 
unrealistic). The HF correlations widen the response region and 
quench and harden the position of the quasielastic peak, as is 
well known. Note however that the short-range correlations, which 
are embodied in the effective interaction based on a $G$-matrix, 
reduce the amount of hardening that is observed in calculations 
based on the bare Bonn potential \cite{Bar96a}. Note also that the 
same level of accuracy is obtained using either non-relativistic 
or relativistic kinematics.

\begin{figure}[p]
\begin{center}
\mbox{\epsfig{file=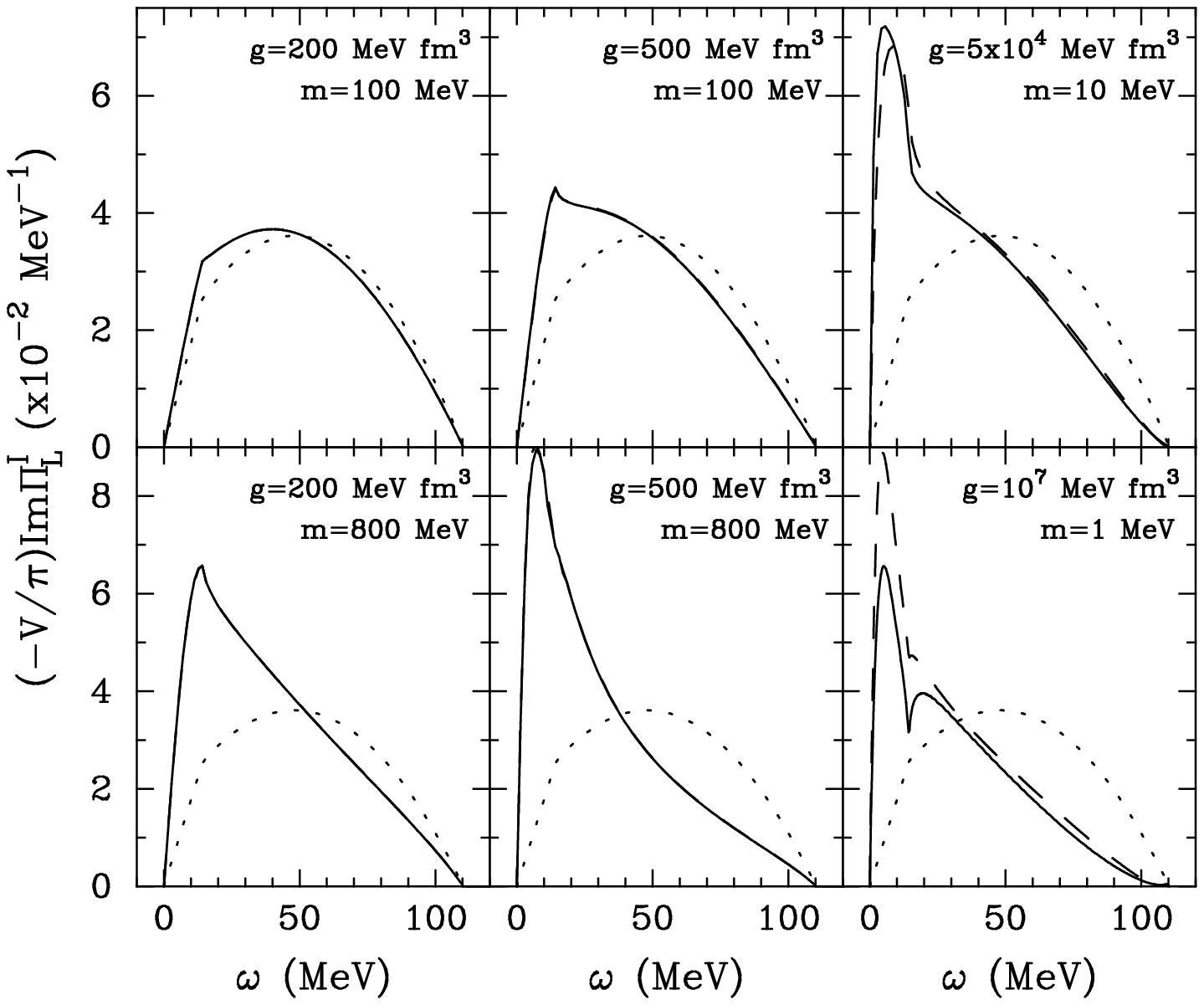,width=.62\textwidth}}
\vskip 2mm
\caption{ Fermi gas longitudinal responses for $k_F=195$ MeV/c at 
$q=300$ MeV/c, with a spin-spin one-boson-exchange interaction, 
for various values of the coupling constant and of the boson mass: 
Free response (dotted), RPA with the first-order CF expansion (dashed)
and RPA with the second-order CF expansion (solid). Note that in 
the left and middle panels the dashed and solid lines are 
indistinguishable. The kinematics are non-relativistic.
 }
\label{fig:cf2sigma}
\vfill
\mbox{\epsfig{file=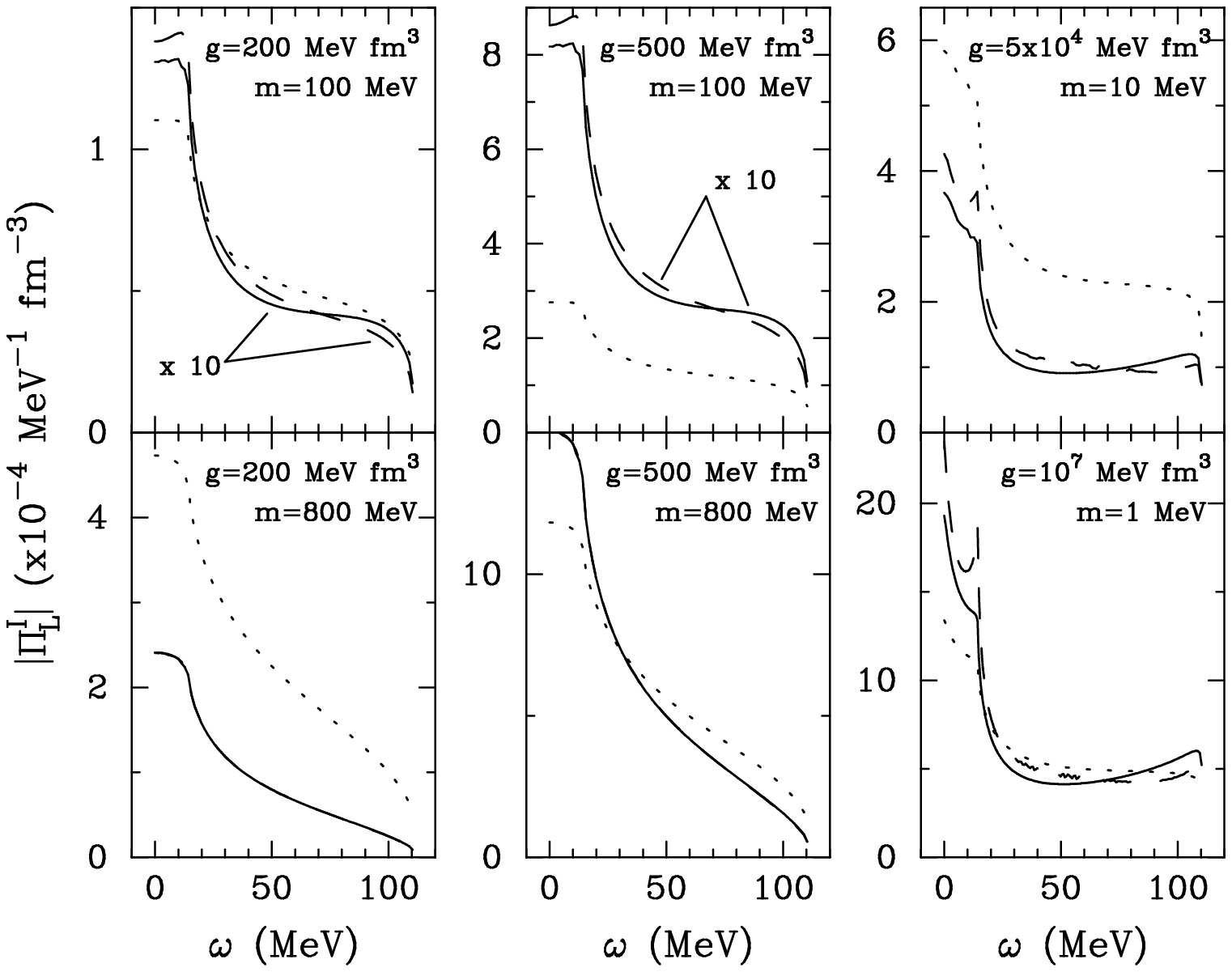,width=.62\textwidth}}
\vskip 2mm
\caption{ Modulus of the longitudinal polarization propagator for 
$k_F=195$ MeV/c at $q=300$ MeV/c, with a spin-spin one-boson-exchange 
interaction, for various values of the coupling constant and of the 
boson mass: First-order, $\Pi^{\text{(1)}}$ (dotted); exact 
second-order, $\Pi^{\text{(2)}}$ (dashed); CF approximation to second order,
$\Pi^{\text{(2)appr}}={\Pi^{\text{(1)}}}^2/\Pi^{\text{(0)}}$ (solid).
 }
\label{fig:r1vs2sigma}
\end{center}
\end{figure}

Before discussing the RPA results, we would like to test the
convergence of the CF expansion. For this purpose, in 
Fig.~\ref{fig:cf2sigma} we compare the longitudinal RPA responses at 
first and second order in the CF expansion using a model
one-boson-exchange interaction,
$V_\sigma(k)=\bbox{\sigma}_1\cdot\bbox{\sigma}_2 g [m^2/(m^2+k^2)]$ 
(the spin operators having the purpose of killing the direct (ring) 
contribution). For values of the coupling constant $g$ and of the 
boson mass $m$ typical of realistic nucleon-nucleon potentials one 
finds that the first- and second-order results match at the level 
of a few percent (in the left and middle panels of 
Fig.~\ref{fig:cf2sigma}, the solid and dashed curves are actually 
indistinguishable). One has to go to very low boson masses (a few 
MeV) and, consequently to very high values of $g$ in order to find 
some discrepancies. To understand these results better,  
in Fig.~\ref{fig:r1vs2sigma} we display the modulus of the polarization
propagator at first order, $\Pi^{(1)}$ (dotted), at second order, 
$\Pi^{(2)}$ (dashed) and the approximation to $\Pi^{(2)}$ generated 
by the first-order CF expansion (see Section \ref{subsec:RPA-resp}), 
$\Pi^{(2)\text{appr}}\equiv{\Pi^{(1)}}^2/\Pi^{(0)}$ (solid). From 
inspection of the curves, it is clear that the first important 
element to guarantee good convergence is the range of the 
interaction. Indeed, for $m=800$ MeV (short-range), $\Pi^{(2)}$ 
and $\Pi^{(2)\text{appr}}$ practically coincide independent of 
the strength of the interaction. This, of course, should be expected, 
since for zero-range interactions the first-order CF expansion gives 
the exact result. For masses of the order of the pion mass one starts 
finding discrepancies between $\Pi^{(2)}$ and $\Pi^{(2)\text{appr}}$. 
However, for realistic values of the interaction strength the 
second-order contribution turns out to be one order of magnitude 
smaller than the first-order one, and thus these discrepancies have 
little effect on the full response functions (Fig.~\ref{fig:cf2sigma}).

To understand these results it may be useful to compare the strength 
of the interactions employed here to that of one-pion-exchange, 
$g_\pi m_\pi^2\equiv f_\pi^2/3\cong0.33$ (in natural units). With 
the same units, the cases with $m=100$ MeV correspond to 
$g\,m^2=0.26$ and 0.65; those with $m=800$ MeV to $g\,m^2=16.7$ 
and 41.7; for $m=1$ and 10 MeV one has $g\,m^2=1.3$ and 0.65, respectively.

To summarize, from the left and middle panels of 
Fig.~\ref{fig:r1vs2sigma} one can understand that the validity of 
the CF expansion originates from the interplay between range and 
strength of the interaction. For short-range potentials where the 
conventional perturbative expansion may not converge, the CF technique
yields a good approximation for the propagators at all orders; for 
long-range (on the nuclear scale) forces, the CF approximation is 
less accurate, but the relative weakness of the interaction already 
guarantees the convergence of the conventional perturbative
expansion. One has to go to unreasonably low masses to find a 
situation where the interaction range is very long and $\Pi^{(1)}$ 
and $\Pi^{(2)}$ are of the same order (right panels in 
Fig.~\ref{fig:r1vs2sigma}).

\begin{figure}[p]
\begin{center}
\vskip 2mm
\mbox{\epsfig{file=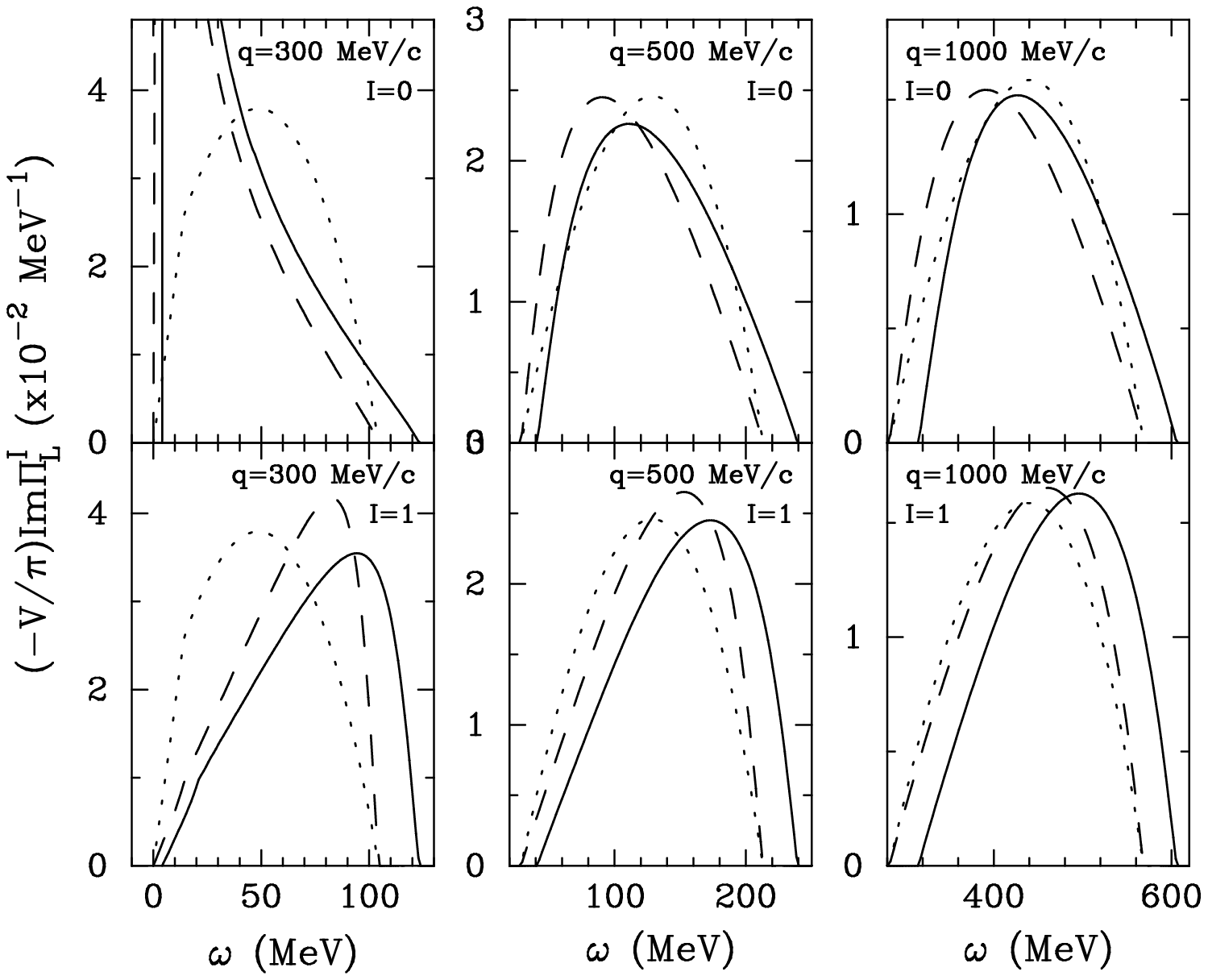,width=.72\textwidth}}
\caption{ Fermi gas longitudinal responses for $k_F=195$ MeV/c at 
$q=300$, 500 and 1000 MeV/c, with the $G$-matrix discussed in the 
text: Free response (dotted), RPA (dashed) and BHF-RPA (solid). 
The kinematics are relativistic.
 }
\label{fig:CF1HFLspeth}
\vfill
\mbox{\epsfig{file=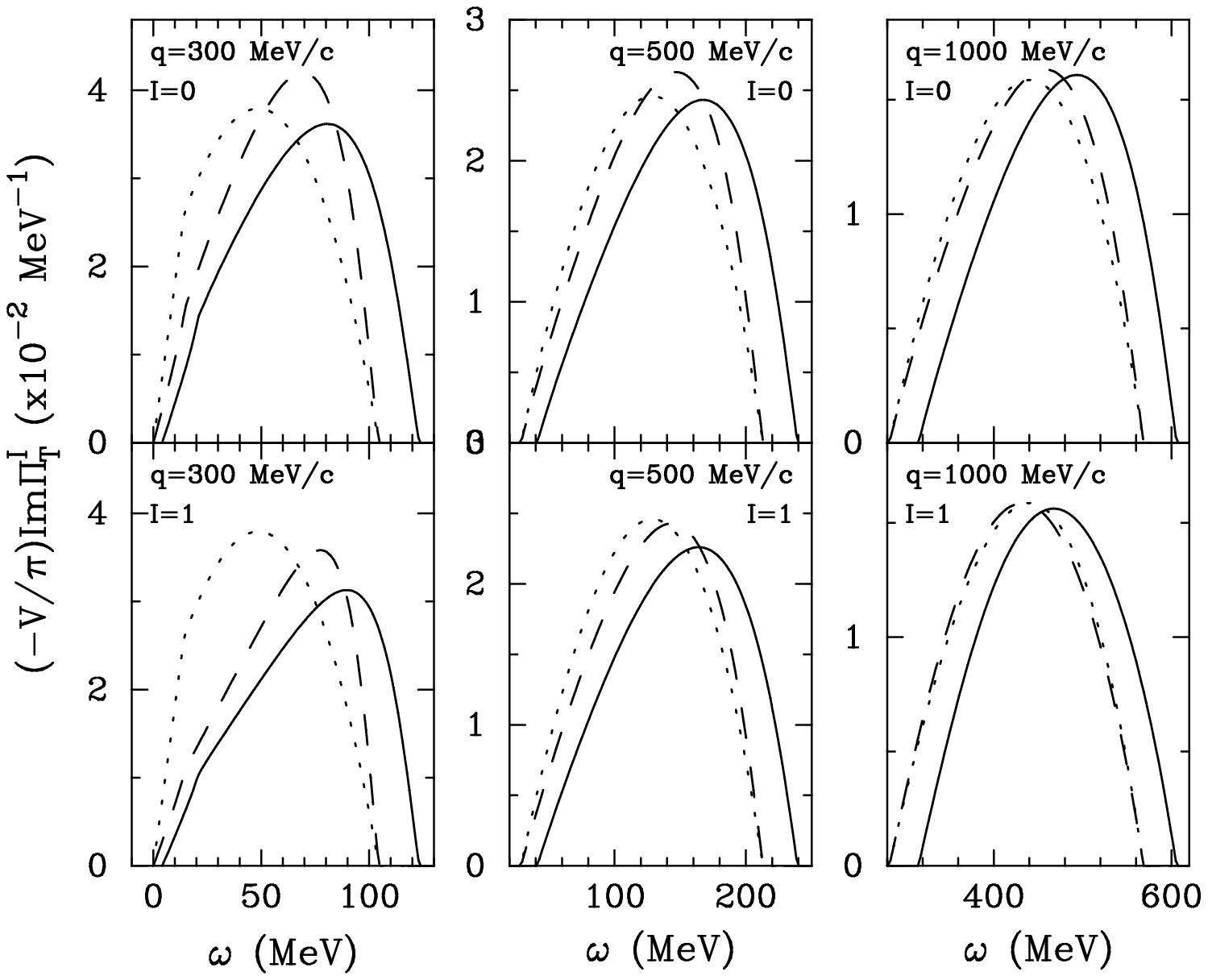,width=.72\textwidth}}
\caption{ The same as in Fig.~\protect\ref{fig:CF1HFLspeth}, but for the
transverse channel.
 }
\label{fig:CF1HFTspeth}
\end{center}
\end{figure}

We can thus conclude that the calculations of nuclear response
functions in the antisymmetrized RPA performed at first order 
in the CF expansion are indeed quite accurate. The same conclusion 
is also supported by calculations with a realistic effective 
interaction --- such as the $G$-matrix parameterization discussed 
above --- and including HF and relativistic kinematical effects. In fact, in 
Fig.~\ref{fig:CF1HFLspeth} we show the RPA and BHF-RPA longitudinal
responses of $^{12}$C at $q=300$, 500 and 1000 MeV/c, using the full 
$G$-matrix introduced at the beginning of this section. Also for the 
full interaction, the discrepancies between the first- and
second-order CF responses are too small to be displayed. They are at 
the level of fractions of percent everywhere, except for the case of 
the isoscalar channel at 300 MeV/c, where they rise to a few percent 
due to the closeness of a singularity in the propagator induced by the
strongly attractive interaction. Indeed, as already mentioned, the 
scalar-isoscalar channel is (too) attractive \footnote{As indicated 
by the energy position of the breathing modes; in other words, nuclear
matter with such an interaction becomes unstable.} and softens the 
quasielastic peak; the scalar-isovector one is repulsive and
gives rise to a hardening. The effect of the HF correlations is the 
same as in the discussion of Fig.~\ref{fig:HFspeth}. In 
Fig.~\ref{fig:CF1HFTspeth} the transverse response is displayed 
for the same conditions.

Finally, it is interesting and important to test the validity of the 
ring approximation --- where exchange diagrams are not included --- 
since this approximation has been widely used in the literature
because of its simplicity. In this scheme, the effect of 
antisymmetrization is simulated by adding to the direct interaction 
matrix elements an effective exchange contribution (see, e.~g., 
Ref.~\cite{Ose82}). For details see also Ref.~\cite{DeP97}, where 
a prescription to determine the effective exchange momentum designed 
for use in the quasifree region has been given.

\begin{figure}[t]
\begin{center}
\mbox{\epsfig{file=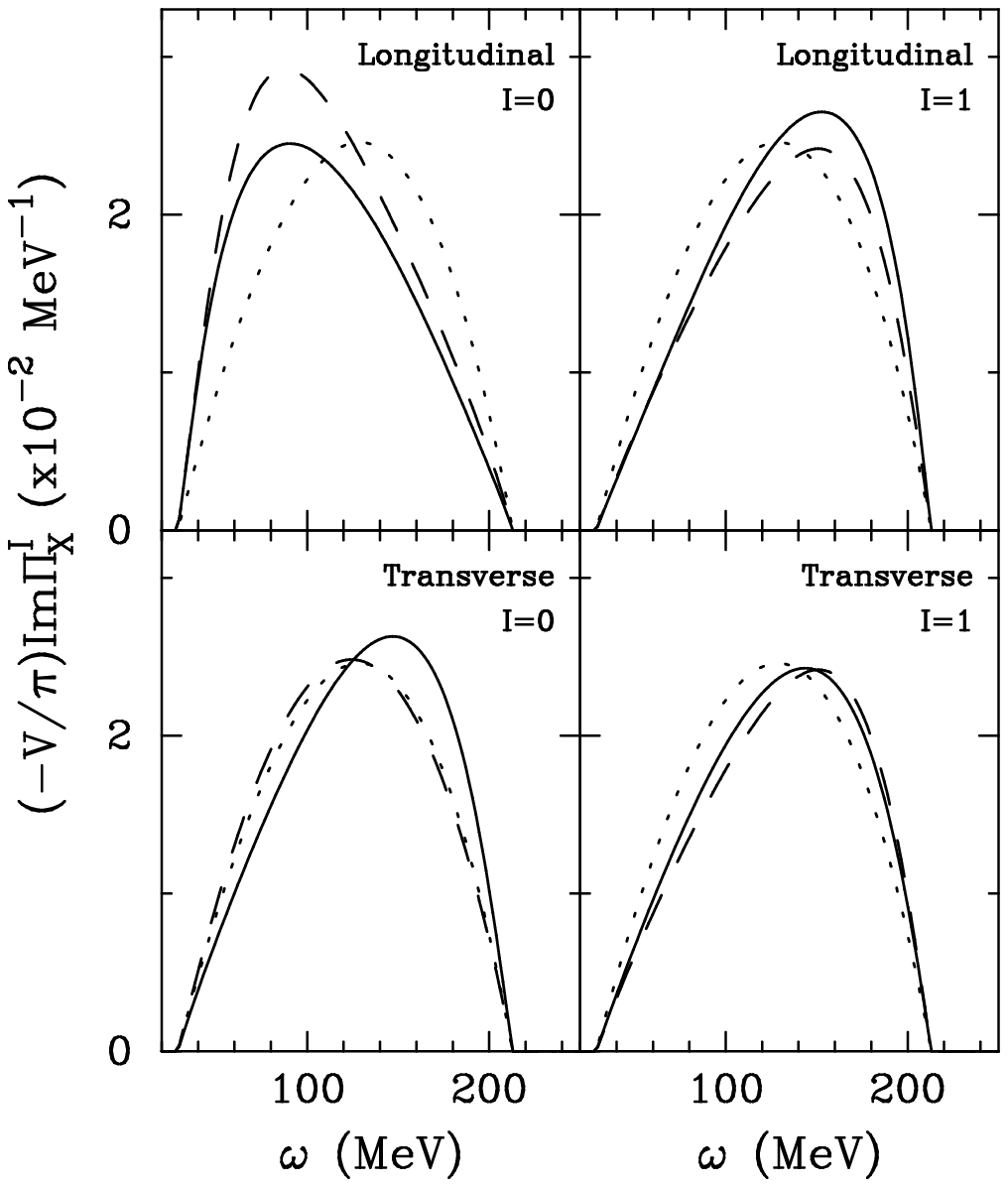,width=.6\textwidth}}
\vskip 2mm
\caption{ Fermi gas responses for $k_F=195$ MeV/c at $q=500$ MeV/c, 
with the $G$-matrix parameterization discussed in the text: Free 
response (dotted), ring approximation (dashed) and RPA (solid). The 
kinematics are relativistic.
 }
\label{fig:rpavsring}
\end{center}
\end{figure}
In Fig.~\ref{fig:rpavsring} we display the ring and RPA responses of 
$^{12}$C at $q=500$ MeV/c, using the $G$-matrix parameterization. 
It is apparent that the only channel where the ring approximation works 
reasonably well is the spin-isovector one, which, incidentally, is the
dominant one in ($e$,$e'$) magnetic scattering; it is less accurate in
all other channels, especially in the scalar-isoscalar one. The 
same considerations also apply when the HF mean field is included in 
the ring and RPA responses. Note that these results confirm those of 
Ref.~\cite{Shi89}, where a comparison of ring and RPA calculations had
been done using a numerically rather involved finite nucleus
formalism. Also in that calculation the $G$-matrix of
Ref.~\cite{Nak84} had been employed.

\section{ Parity-violating electron scattering and axial responses }
\label{sec:pv}

\subsection{ The asymmetry, the currents and the RFG responses}
\label{sec:asym}

A new window on the inclusive nuclear responses that allows us to 
unravel aspects of nuclear and nucleon structure that are otherwise 
inaccessible to unpolarized probes is offered by parity-violating
electron scattering from nuclei. See Ref.~\cite{MJMrev} for a 
general review of the subject. Experiments of this type exploit 
longitudinally polarized electrons to measure the helicity asymmetry 
${\cal A}$, defined as the difference between the inclusive nuclear 
scattering of right- and left-handed electrons divided by their sum, namely
\begin{equation}
{\cal A} = \frac{d^2\sigma^{+}-d^2\sigma^{-}}{d^2\sigma^{+}+d^2\sigma^{-}} .
\label{eq:Asym}
\end{equation}

\begin{figure}[t]
\begin{center}
\mbox{\epsfig{file=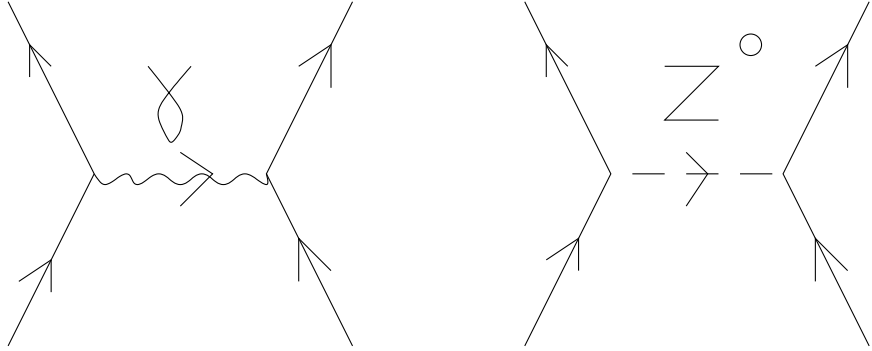,width=.7\textwidth}}
\vskip 2mm
\caption{Single-photon-exchange and $Z^0$-exchange diagrams for
polarized electron scattering from nuclei.}
\label{fig:ampl}
\end{center}
\end{figure}

${\cal A}$ arises from the interference between the electromagnetic 
current that is purely vector (V) and the weak neutral current 
that has both vector and axial-vector (A) components. Diagrams 
for the associated amplitudes in leading order of the bosons 
exchanged (the photon and the $Z^0$) are displayed in 
Fig.~\ref{fig:ampl}. In this approximation Eq.~(\ref{eq:Asym}) 
can be cast in the form
\begin{equation}
{\cal A} = {\cal A}_0
\frac{v_{\text L} R_{\text L}^{\text AV}(q,\omega)
    + v_{\text T} R_{\text T}^{\text AV}(q,\omega)
    + v_{\text T'}R_{\text T'}^{\text VA}(q,\omega)}
{     v_{\text L} R_{\text L}(q,\omega)
    + v_{\text T} R_{\text T}(q,\omega)} .
\label{eq:AsR}
\end{equation}
The numerator is parity-violating (PV) and the denominator 
parity-conserving (PC). Here
\begin{equation}
v_{\text L} = \left(\frac{Q^2}{\bbox{q}^2}\right)^2
\end{equation}
\begin{equation}
v_{\text T} = \frac{1}{2}\left|\frac{Q^2}{\bbox{q}^2}\right|
                + \tan^2\frac{\theta}{2}
\end{equation}
and
\begin{equation}
v_{\text T'} = \tan\frac{\theta}{2}
 \sqrt{\left|\frac{Q^2}{\bbox{q}^2}\right| + \tan^2\frac{\theta}{2}}
\end{equation}
are the usual leptonic kinematical factors, $\theta$ is the electron 
scattering angle and, as before, $Q^2=\omega^2-\bbox{q}^2$ is the 
spacelike four-momentum transferred from the electron to the nucleus.

In Eq.~(\ref{eq:Asym}) the nuclear (and nucleon's) structure are
embedded in the electromagnetic PC nuclear responses $R_{\text L}$ and 
$R_{\text T}$ discussed in the previous sections, while their PV analogs 
$R_{\text L}^{\text AV}$, $R_{\text T}^{\text AV}$ and 
$R_{\text T'}^{\text VA}$ are discussed in this section. Here the 
first (second) index in the superscript refers to the vector (V) or 
axial (A) nature of the leptonic (hadronic) WNC. For brevity we shall 
often simply refer to these by their hadronic character, i.e., the L and 
T PV responses are called ``vector'' and the $T'$ response
``axial''. Finally the scale of the asymmetry is set by the factor
\begin{equation}
{\cal A}_0 = \frac{\sqrt{2} G_F m_N^2}{\pi\alpha} \frac{|Q^2|}{4m_N^2}
\approx 6.5\times 10^{-4} \tau ,
\label{eq:A0}
\end{equation}
which is defined in terms of the EM $(\alpha)$ and Fermi $(G_F)$ 
coupling constants. If there were no additional dependence on $q$ 
and $\omega$, then the expression for ${\cal A}_0$ would imply that the
asymmetry grows with $\tau=|Q^2|/4m_N^2$ and hence it is not
surprising that the first parity violation in electron scattering 
was observed at high energies at SLAC \cite{Pre78,Pre79}. On the other
hand, it is also clear that selective processes such as elastic 
scattering do contain additional dependences on $(q,\omega)$ via 
form factors that may make measurements at large $\tau$ extremely 
difficult. In fact, only a very few have been performed to date. 
Even more challenging, but not impossible, are experiments whose 
goal is to disentangle in Eq.~(\ref{eq:AsR}) the separate
contributions of the PV responses
$R_{\text L}^{\text AV}$, $R_{\text T}^{\text AV}$ and
$R_{\text T'}^{\text VA}$. 

In the investigation of the PV nuclear responses --- these 
can assume positive as well as negative values --- of 
central importance is the isospin decomposition of the hadronic four-current
\begin{equation}
\left(J_\mu\right)_{I,I_z} = \beta^{(0)}\left(J_\mu\right)_{0,0} 
                                 + \beta^{(1)}\left(J_\mu\right)_{1,0} 
\label{eq:Jmu}
\end{equation}
into isoscalar ($I$=0) and isovector ($I$=1) components ($\mu$ is
the Lorentz index). In the Standard Model at tree level the coefficients in
Eq.~(\ref{eq:Jmu}) in the EM sector read 
\begin{equation}
\beta^{(0)}_{\text V,EM} =\beta^{(1)}_{\text V,EM}=\frac{1}{2} ,
\end{equation}
i.e. one has the usual responses $R_{\text L}$ and $R_{\text T}$.
In the WNC sector, instead, they read 
\begin{eqnarray}
\beta^{(0)}_{\text V,WNC} &=& -2\sin^2\theta_W \simeq -0.461
\label{eq:b0vwn}
\\
\beta^{(1)}_{\text V,WNC} &=& 1-2\sin^2\theta_W \simeq 0.539
\label{eq:b1vwn}
\end{eqnarray}
for the vector coupling and
\begin{eqnarray}
\beta^{(0)}_{\text A} &=& 0
\\
\beta^{(1)}_{\text A} &=& 1
\end{eqnarray}
for the axial one. The above results obtain with the following value
of the weak mixing angle
\begin{equation}
\sin^2\theta_W = 0.23055\pm 0.00041 .
\label{eq:sin2t}
\end{equation}

The importance of isospin becomes clear if one assumes it to be an 
exact symmetry in nuclei (which is of course only approximately 
true). Then for PV {\em elastic} scattering on spin zero, isospin 
zero nuclei, Eq.~(\ref{eq:AsR}) reduces to 
\begin{equation}
\frac{{\cal A}}{{\cal A}_0} = 2\sin^2\theta_W 
\label{eq:As00}
\end{equation}
(see Eq.~(\ref{eq:aVaA}) for the coefficient of the axial leptonic 
current), which suggests using PV experiments as a tool for testing 
the Standard Model in the low-energy regime (see, for example, 
Ref.~\cite{MJMrev}). 

To this point we have been assuming that the strangeness content in 
the nucleon or nucleus is negligible. If this is not the case, then 
it has been realized that PV electron scattering can be invaluable in 
exploring this aspect of hadronic structure (again, see 
Ref.~\cite{MJMrev} for a review that contains discussion of this 
issue). Indeed, when strangeness content is taken into account 
Eq.~(\ref{eq:As00}) is modified as follows
\begin{equation}
\frac{{\cal A}}{{\cal A}_0} = 2\sin^2\theta_W + \frac{G_E^{(s)}}{G_E^{(0)}} . 
\label{eq:As00s}
\end{equation}
In the above $G_E^{(0)}=G_{E_p}+G_{E_n}$ and $G_E^{(s)}$ are the 
electric isoscalar EM and strange form factors of the nucleon (the 
indices $p$ and $n$ refer to the proton and the neutron, respectively).
Formula (\ref{eq:As00s}) will be exploited to extract $G_E^{(s)}$ in an
experiment planned at CEBAF involving elastic PV scattering from 
$^4$He. 

A further clue to the strangeness content may be seen in studying PV 
elastic scattering from the proton: in this case Eq.~(\ref{eq:AsR}) becomes
\begin{equation}
\frac{{\cal A}}{{\cal A}_0} = - \frac{\varepsilon G_{E_p}\tilde G_{E_p}+
      \tau G_{M_p}\tilde G_{M_p}+
      \delta G_{M_p}\tilde G_{A_p}}
     {\varepsilon G_{E_p}^2+\tau G_{M_p}^2} ,
\label{eq:Ap}
\end{equation}
where
\begin{equation}
\varepsilon = \frac{1}{1+2(1+\tau)\tan^2(\theta/2)}
\label{eq:epsilon}
\end{equation}
and
\begin{equation}
\delta = (1-4\sin^2\theta_W)\sqrt{\tau(1+\tau)(1-\varepsilon^2)} .
\label{eq:delta}
\end{equation}
In Eq.~(\ref{eq:Ap}) $\tilde G_{E_p}$, $\tilde G_{M_p}$ and 
$\tilde G_{A_p}$ are the electric, magnetic and axial weak form 
factors of the proton, that read
\begin{eqnarray}
\tilde G_{E_p,M_p} &=& \frac{1}{2} 
 \left[ \beta^{(0)}_{\text V,WNC} + \beta^{(1)}_{\text V,WNC}\right] 
 G_{E_p,M_p}
+\left[ \beta^{(0)}_{\text V,WNC} - \beta^{(1)}_{\text V,WNC}\right] 
 G_{E_n,M_n}
\nonumber\\
&=& \frac{1}{2} (1-4\sin^2\theta_W)G_{E_p,M_p}-\frac{1}{2}G_{E_n,M_n}
   -\frac{1}{2} G_{E,M}^{(s)}  
\end{eqnarray}
and
\begin{equation}
\tilde G_{A_p} = \frac{1}{2} G_{A_p}-\frac{1}{2}G_{A_n}
                -\frac{1}{2} G_{A}^{(s)} .
\end{equation}

\begin{figure}[tde]
\begin{center}
\mbox{\epsfig{file=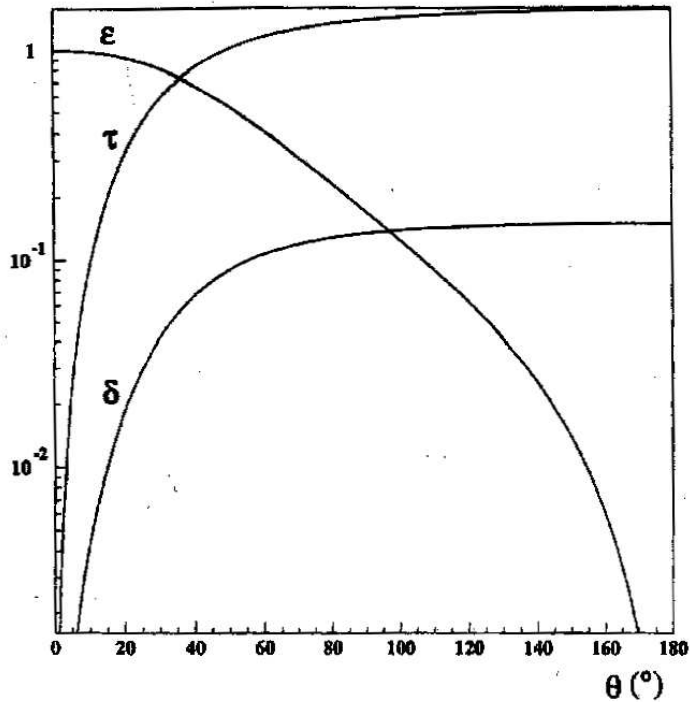,width=.6\textwidth}}
\vskip 2mm
\caption{The kinematical coefficients $\tau$, $\varepsilon$ and 
$\delta$ versus the scattering angle $\theta$; the energy of the 
incoming electron is 3.355 GeV.}
\label{fig:tde}
\end{center}
\end{figure}

Equation~(\ref{eq:Ap}) has already been exploited in experiments
performed at Bates (SAMPLE) and CEBAF (HAPPEX) (see 
Refs.~\cite{SAMPLE} and \cite{HAPPEX}, respectively) for unraveling 
the strangeness content of the proton. Figure~\ref{fig:tde} may 
help in understanding the magnitudes of the quantities given above. 
There one infers that at backward scattering angles (SAMPLE) it is 
mainly the magnetic strangeness that is measured, however with some 
contamination arising from the axial contribution, whereas at forward angles
(HAPPEX) it is a mixture of electric and magnetic strangeness that is
observed, the axial contribution being totally negligible in that case.

Although the above discussions of strangeness content are focused on 
the responses of the nucleon rather than of the nucleus, in fact it is
also relevant for the latter, since the nuclear responses measured in 
the quasielastic regime are indeed affected by the nucleon's isoscalar
form factors and these in turn are affected by strangeness.
In fact this is an excellent example of how the theoretical
predictions in nuclear many-body and particle physics are interrelated.

In particular, with regard to the responses of the nucleus, PV
experiments offer the opportunity of

\begin{itemize}
\item[i)] disentangling the isoscalar and isovector contributions to
$R_{\text L}^{\text AV}$, $R_{\text T}^{\text AV}$ and
$R_{\text T'}^{\text VA}$,
\item[ii)] exploring the Coulomb sum rule separately in the isoscalar
and isovector channels (see Section \ref{sec:sumrules}),
\item[iii)] measuring the neutron distribution in nuclei (see 
Ref.~\cite{DDS}; not discussed here),
\item[iv)] investigating the nuclear axial response especially in the
$\Delta$ region (also not discussed in the present work)

and
\item[v)] unambiguously revealing the role of the pion in nuclear
excitations through the (possible) existence of a zero in the 
frequency behavior of the asymmetry.
\end{itemize}

\begin{figure}[p]
\begin{center}
\mbox{\epsfig{file=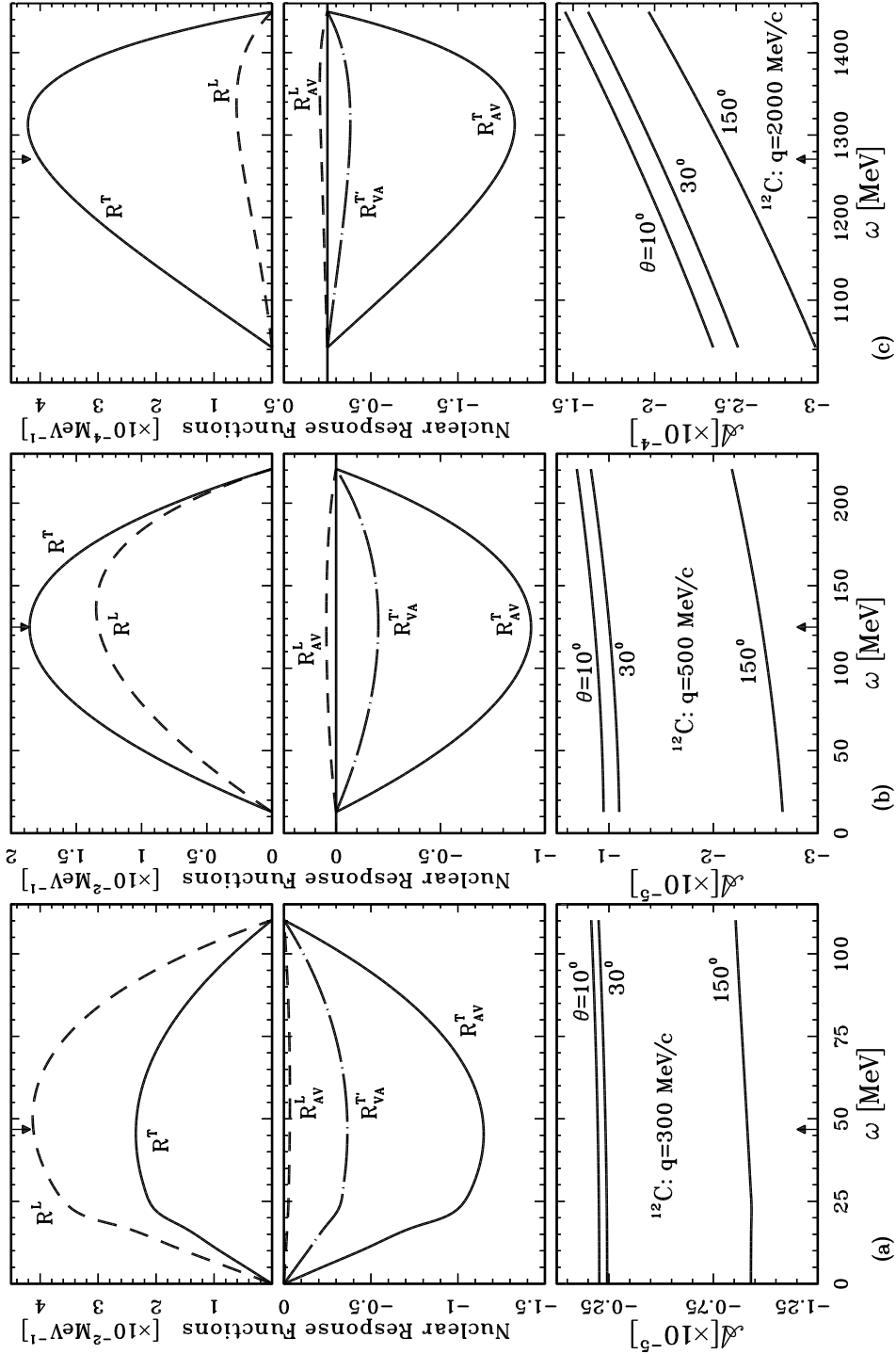,width=.82\textwidth}}
\caption{ 
Relativistic Fermi gas response functions and asymmetry for $^{12}C$ at 
$q$=300 (a), 500 (b) and 2000 (c) MeV/c shown as functions of $\omega$. 
The location of $\omega=\frac{1}{2}|Q^2|/m_N$ is indicated by the 
arrow. {\em Upper}: the EM responses $R_{\text L}$ and $R_{\text T}$ 
are displayed. {\em Middle}: the parity-violating responses 
$R^{\text AV}_{\text L}$, $R^{\text AV}_{\text T}$ and 
$R^{\text VA}_{\text T'}$ are shown. {\em Lower}: the asymmetry 
${\cal A}$ is given for three values of the scattering angle $\theta$. 
}
\label{fig:resp}
\end{center}
\end{figure}

To see how this occurs let us first split the PC and PV responses into
their isospin components (we are neglecting the strangeness content 
at this point) according to
\begin{eqnarray}
R_{\text L,\text T} =  \beta^{(0)}_{\text V,EM}R^{I=0}_{\text L,\text T}
              +  \beta^{(1)}_{\text V,EM}R^{I=1}_{\text L,\text T} \\ 
R_{\text L,\text T}^{\text AV} =  
              \beta^{(0)}_{\text V,WNC}R^{I=0}_{\text L,\text T}
              +  \beta^{(1)}_{\text V,WNC}R^{I=1}_{\text L,\text T}  
\end{eqnarray}
for the vector channel. The PV axial channel is purely isovector at 
tree level. Next let us focus on the RFG model where the longitudinal 
isoscalar response is essentially proportional to 
$\left(G_E^{(0)}\right)^2 = \left(G_{E_p}+G_{E_n}\right)^2$
and the isovector one to 
$\left(G_E^{(1)}\right)^2 = \left(G_{E_p}-G_{E_n}\right)^2$.
Since $G_{E_n}$ is small, especially at low $\tau$, it follows that
\begin{equation}
R^{I=0}_{\text L} \simeq R^{I=1}_{\text L} .
\end{equation}
But then it is clear that the RFG PV longitudinal response is 
{\em almost vanishing} because of the opposite sign and approximately 
equal magnitude of the coefficients in Eqs.~(\ref{eq:b0vwn}) and 
(\ref{eq:b1vwn}). This dramatic consequence of the Standard Model 
is displayed in Fig.~\ref{fig:resp}, where the five responses entering
in the definition of the asymmetry in Eq.~(\ref{eq:AsR}) are shown 
for $|\bbox{q}|$=300, 500 and 2000 MeV/c together with ${\cal A}$.
Beyond the fact that $R^{\text AV}_{\text L}$ is very small, one 
also observes in the figure that

\begin{itemize}
\item[i)] a similar cancellation {\it does not} occur in the transverse 
channel, where in fact the isoscalar and the isovector responses are 
quite different because now 
$(G_M^{(0)})^2\propto (\mu_p+\mu_n)^2\simeq 0.77$ and 
$(G_M^{(1)})^2\propto (\mu_p-\mu_n)^2\simeq 22.1$;
\item[ii)] $R^{\text AV}_{\text T}$ and $R^{\text VA}_{\text T'}$ are 
{\em negative}, this being related to the sign of the axial 
coefficient of the leptonic WNC
\begin{equation}
j_\mu(k',s';k,s)^{WNC} = \overline{u}(k',s') (a_V\gamma_\mu +
a_A \gamma_5 \gamma_\mu) u(k,s) ,
\end{equation}
where $u(k,s)$ is the Dirac spinor of the electron. Indeed, according 
to the Standard Model,
\begin{equation}
a_V = - (1-4\sin^2\theta_W)\simeq -0.092
\ \ \ \mbox{and}\ \ \ 
a_A=-1 ;
\label{eq:aVaA}
\end{equation}
\item[iii)] as a consequence the asymmetry is negative as well, reflecting 
the left-handed nature of the weak interaction;
\item[iv)] the asymmetry, as previously mentioned, grows with 
$\tau$ and $\theta$.

\end{itemize}

How do interactions among the constituents of the RFG modify 
the above predictions?

\subsection{ The role of the pion and other mesons}
\label{sec:pion}

In answering the last question of the previous section we extend our 
model from a strict RFG framework to one where pions are also 
included \cite{Alb93}, because then we can at least approximately 
preserve the two major requirements of {\it Lorentz covariance} and 
{\it gauge invariance}. Indeed the RFG cross section is built (apart 
from overall kinematical factors) from the contraction of the 
leptonic and hadronic Lorentz tensors and is therefore a relativistic 
invariant, although the partition into longitudinal and transverse 
responses depends, of course, upon the reference system. Moreover the 
single-nucleon four-current entering into the RFG nuclear tensor
is conserved and hence the non-interacting RFG is gauge invariant.

When the nucleon-nucleon interaction carried by the pion is switched
on it is not obvious that the two above mentioned properties are retained.
Indeed, when the correlations and meson exchange currents (MEC)
associated with the one pion exchange potential (OPEP) are introduced, 
as shown in Ref.~\cite{Alb93} the Lorentz covariance and gauge 
invariance are violated (however, only slightly so) due to the 
following approximations that are usually made:
\vspace{0.2in}

i) the pion propagator is assumed to be static,

ii) and a non-relativistic expansion of the two-body currents is 
performed~\cite{Alb90}.
\vspace{0.2in}

In order to achieve a treatment of forces and currents that is
as consistent as possible we first limit ourselves to the study of 
diagrams with {\it only one pionic line}, namely, we work in the 
first perturbative order in the N-N interaction. The correlation
diagrams to be evaluated in this scheme are the so-called 
self-energy and exchange contributions, that when iterated to infinite order 
generate the previously discussed HF and RPA series, respectively. Note
that the tadpole and ring diagrams vanish due to the spin and isospin 
structure of the OPEP. Concerning MEC, three contributions occur, the 
pion-in-flight term, the contact term and the one associated with the 
$\Delta$. A direct comparison with the exact relativistic 
calculation \cite{Ama98} shows that the non-relativistic expansion 
of Ref.~\cite{Alb90} is indeed quite accurate up to momentum transfers
of the order of 1 GeV/c.

The outcome of this is that sizable pionic contributions
to the EM longitudinal (spin scalar, $\sigma=0$) and transverse
(spin vector, $\sigma=1$) nuclear responses are found. In both cases 
the correlation effects produce a hardening of the responses, that is,
a shift of the strength to higher $\omega$. The PV longitudinal and 
transverse correlated responses are simply obtained from the EM ones 
through the isospin rotations implied by the structure of the WNC 
discussed in the previous subsection; the axial-vector response
will be treated separately in Section \ref{subsec:axas}.

\begin{figure}[t]
\begin{center}
\mbox{\epsfig{file=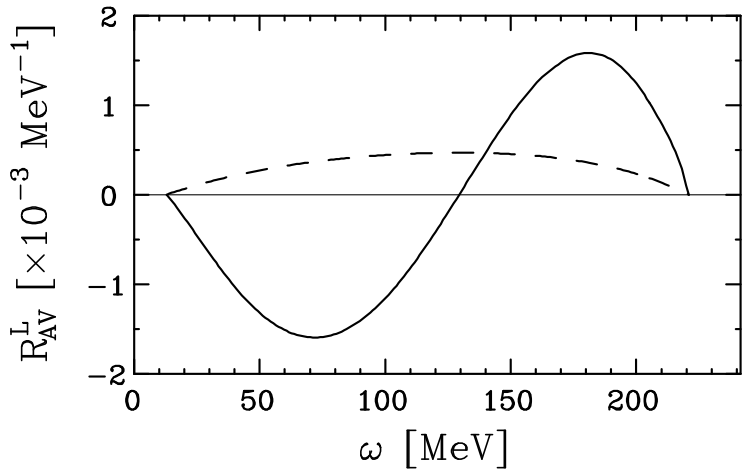,width=.6\textwidth}}
\vskip 2mm
\caption{The longitudinal weak neutral current response 
$R_{\text L}^{\text AV}$
at $q$=500 MeV/c, showing the free RFG (dashed) and correlated (solid) 
results.}
\label{fig:RLAV}
\end{center}
\end{figure}

The main points emerging from this analysis are:
\begin{itemize}
\item[a)] In isospace the contribution of the self-energy diagram to
the charge response is almost equally split between isoscalar ($I=0$) 
and isovector ($I=1$) components. The latter, on the other hand, is 
of course overwhelming in the transverse response, due to the
dominance of the isovector magnetic moment. In contrast, in the case 
of the pionic force the $I=0$ part of the exchange diagram turns out to be
{\it three times as large} as the $I=1$ one in the charge response and
this imbalance, that becomes even stronger in higher orders of 
perturbation theory, is further strengthened by the difference between
the isoscalar and isovector form factors. The isoscalar dominance of 
the pionic exchange correlations has dramatic consequences for the 
PV longitudinal response function, as may be seen in 
Fig.~\ref{fig:RLAV}, where this response is displayed as a function 
of $\omega$ with and without pionic correlations.

\begin{figure}[t]
\begin{center}
\mbox{\epsfig{file=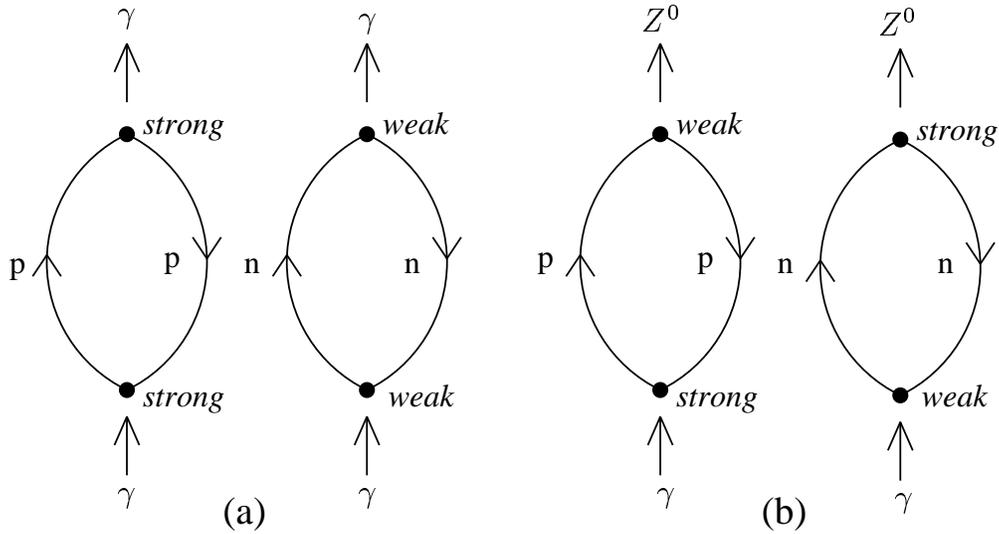,width=.8\textwidth}}
\vskip 2mm
\caption{Feynman diagrams representing the free particle-hole 
polarization propagator for the EM (a) and PV (b) longitudinal 
response. The excitation of proton (p) and neutron (n) particle-hole 
pairs is shown separately. The labels {\it strong} and {\it weak} 
refer to the strength of the nucleon coupling to the photon $\gamma$ 
or to the vector boson $Z^0$.
}
\label{fig:diagrams-free}
\end{center}
\end{figure}

A physical interpretation of why $R_{\text L}^{\text AV}$ is small in 
the independent-particle model and why isospin-correlations are so 
important in determining its ultimate size follows from the expression
for the observables in a language that explicitly refers to neutrons 
and protons rather than employing isospin labeling. Indeed, by
inspecting Fig.~\ref{fig:diagrams-free}, where the diagrams describing
both the EM and PV longitudinal responses for a free system are 
displayed, one easily understands why in the non-interacting case the 
EM longitudinal response turns out to be substantial: both of its 
vertices can in fact be large, as they involve the coupling of a 
longitudinal photon to a proton. In contrast, one of the vertices 
entering in the non-interacting PV response is {\em always} small, 
since either the longitudinal coupling of a photon to a neutron or of 
a $Z^0$ to a proton is involved. This last fact is often phrased by 
saying that the $\gamma$ is blind to neutrons and the $Z^0$ is 
blind to protons (i.e., in the longitudinal 
channel). To quantify the meaning of ``large'' and ``small'' we note 
that typically the $\gamma -n$ coupling is about 1/10 that of the 
$\gamma -p$, and likewise the $Z^0 -p$ coupling is about 1/10 that 
of the $Z^0 -n$.

The exchange correlations corresponding to the exchange of an
isovector charged meson between the particle and hole convert 
a neutron (proton) into a proton (neutron), and thus give rise to 
a diagram where {\it both} couplings are {\it large} --- hence the 
crucial role of such isovector correlations in determining 
$R_{\text L}^{\text AV}$.

The above arguments clearly do not apply to the transverse response 
both because in this case the $I=0$ channel is much weaker than the
$I=1$ channel, being essentially proportional to the squares of the 
very different isoscalar and isovector magnetic moments of the
nucleon, and furthermore because protons and neutrons can both
couple strongly to photons via their (comparable) magnetic moments. 
Being also essentially isovector, the axial-vector response likewise 
does not display the sensitivity expected for the longitudinal response.

\item[b)] While the tensor component of the OPEP never contributes to 
the self-energy in a translationally invariant system, the exchange 
diagram gets a tensor contribution, {\it but only in the transverse 
channel} and mostly via the backward-going graphs. This implies a 
different role for the pionic force in the two EM responses, a finding
that should be tested against experiment.

\item[c)] When we extend our analysis from first to infinite order of 
perturbation theory, thus generating the HF and RPA responses, the
results do not change substantially. Although the pionic interaction 
is strong, it nevertheless therefore appears that for
quasielastic kinematics its effects are reasonably small at not too 
small $q$, thus rendering perturbation theory quite accurate already 
at the lowest order, at least for the classes of diagrams studied here.

\item[d)] The contribution of the central part of the pionic
interaction to the exchange diagram stems largely from the
$\delta$-force and not from the finite-range one. Notably for 
pointlike nucleons the continuity equation is obeyed by the OPEP 
and by the related pionic MEC, but is not affected by its $\delta$ component.
However in keeping with the usual approach taken in studies of pionic 
effects, we include a $\pi$NN vertex function $\Gamma_{\pi}$, whose 
scale is set by a mass parameter $\Lambda_\pi$, to smear out the
$\delta$-piece of OPEP.  As a consequence the continuity equation 
is modified by the presence of $\Gamma_{\pi}$ and to restore its 
validity additional MEC should be introduced. Significantly these counterterms
affect only the pion-in-flight current, which is tiny, and therefore
are quantitatively negligible.

\item[e)] In contrast to the exchange diagram, the self-energy term 
gets a contribution only from momentum-dependent forces, and therefore
an unmodified pionic $\delta$-interaction does not contribute in this 
channel. In fact, the contribution coming from the $\delta$-function 
in OPEP, when modified by the $\pi$NN vertex form factor, contains an 
effective momentum-dependence and so is nonzero.

\item[f)] Most remarkably when the interactions carried by heavier 
mesons (namely $\rho$, $\sigma$ and $\omega$) are switched on, the 
previous conclusions on the PV responses (in particular the dramatic 
enhancement of $R_{\text L}^{\text AV}$) are not substantially 
changed, as illustrated in Ref.~\cite{Bar96a}, thus showing the 
central role played by the pion in the nuclear dynamics for 
quasielastic kinematics. This is accomplished, however, through 
rather subtle aspects of the nuclear many-body problem, where the 
interference between the pion and the other mesons turns out to be crucial. 

\end{itemize}


\subsection{The axial response and the asymmetry}
\label{subsec:axas}

Let us now turn to a discussion of the asymmetry in the 
pion-correlated Fermi gas model. For its evaluation the axial-vector 
response function $R_{\text T'}^{\text VA}$ is needed. In a 
non-relativistic context the latter is related to the isovector 
component of the EM transverse response, 
$R^{\text T}_{{\text AV},I=1}$, through the simple formula \cite{Bar94}
\begin{equation}
  R_{\text T'}^{\text VA}(q,\omega) = a_V {G_A^{(1)}\over 
  G_M^{(1)}}{1\over\kappa}
  R_{\text T}^{I=1}(q,\omega)\ ,
\label{eq:RTprimenonrel}
\end{equation}
which can be extended into the relativistic regime via the prescription
\begin{equation}
  R_{\text T'}^{\text VA}(q,\omega) \cong a_V
   {G_A^{(1)}\over G_M^{(1)}}\sqrt{\tau+1\over\tau}
   R_{\text T}^{\text I=1}(q,\omega) \ ;
\label{eq:RTprimeapp}
\end{equation}
these agree to better than 2\% in the momentum range 300 MeV/c $<q<1$ 
GeV/c for the RFG. Although not immediately apparent, 
Eq.~(\ref{eq:RTprimeapp}) has been demonstrated to be preserved 
even in presence of pionic correlations.
The effect of pionic correlations and MEC effects in this response
is a ``hardening'' (a shift to higher $\omega$) of the peak of the 
response at intermediate values of $q$, which then fades away and 
even leads to a slight ``softening'' of the response at the highest 
momentum transfers considered.

\begin{figure}[p]
\begin{center}
\mbox{\epsfig{file=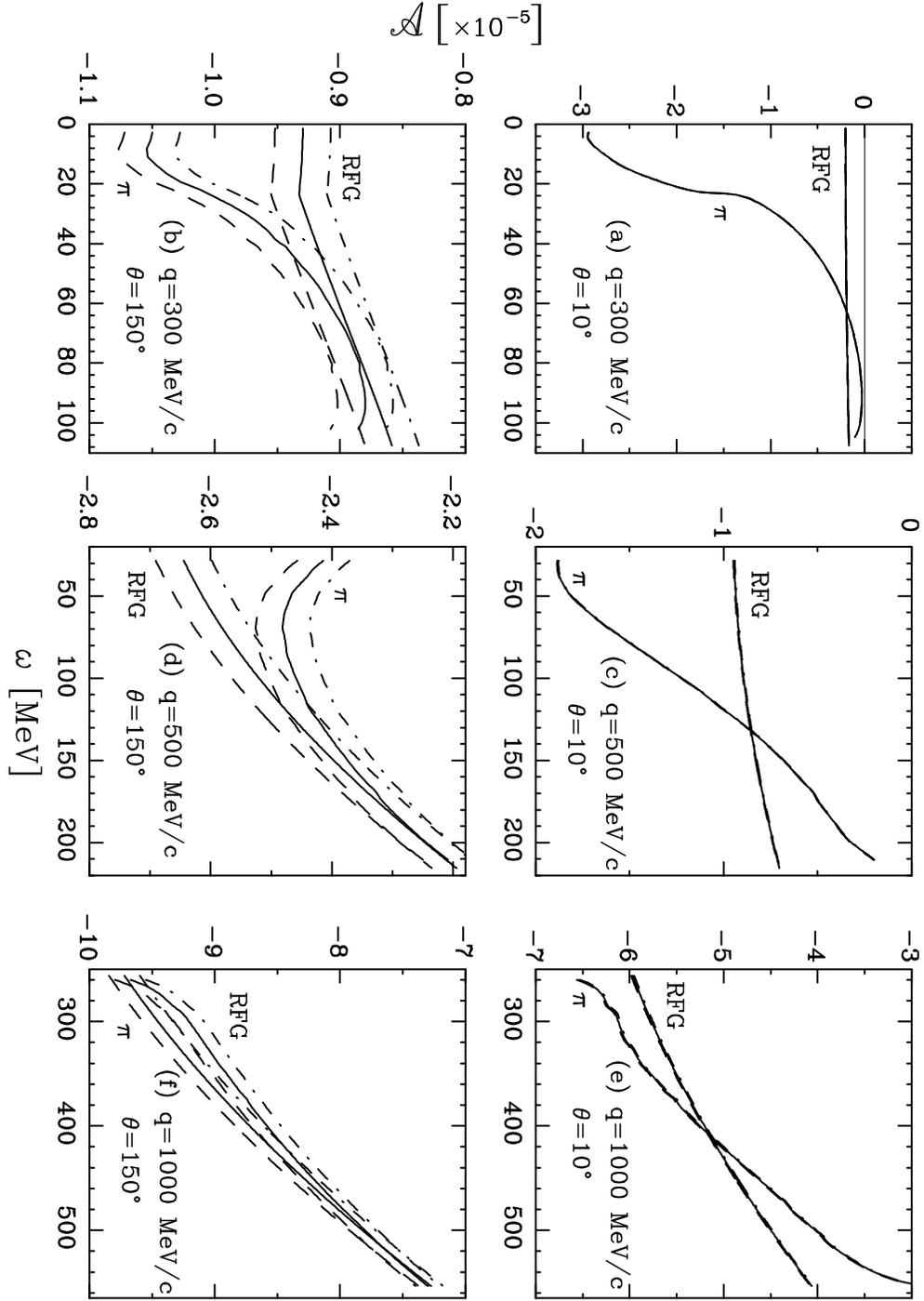,width=.8\textwidth}}
\vskip 2mm
\caption{The $\omega$-dependence of ${\cal A}$ for different
kinematical conditions. The relativistic Fermi gas model results 
are labeled RFG, while the ones with pionic effects included are 
labeled $\pi$. The two families of curves have $g_A^{(1)}$=1.26 (solid), 
1.26+10\% (dashed) and 1.26-10\% (dash-dotted). 
}
\label{fig:Asym}
\end{center}
\end{figure}

We are now in the position to calculate the asymmetry ${\cal A}$,
which is displayed in Fig.~\ref{fig:Asym} as a function of $\omega$ 
for three values of $q$ and for forward ($\theta=10^0$) and backward 
($\theta=170^0$) scattering angles.

Upon examining Fig.~\ref{fig:Asym}, we notice the significant effect
occurring at moderate $q$ (say 300--500 MeV/c), small $\omega$ and 
forward angles. As previously discussed, it is related to the large 
negative value assumed by the correlated $R_{\text L}^{\text AV}$, 
which leads to a pionic asymmetry that is an order-of-magnitude 
larger than the free RFG one. However, as $\omega$ increases
$R_{\text L}^{\text AV}$ rapidly decreases until it changes sign, 
while $R_{\text T}^{\text AV}$ stays negative; accordingly, they 
largely cancel in the numerator of the ratio expressing ${\cal A}$ 
and this becomes substantially lowered.

Interestingly, an energy is reached (about 60 MeV for $q=300$ MeV/c)
where the correlated and free RFG values of ${\cal A}$ coincide. At 
still larger $\omega$ a further reduction of ${\cal A}$ is seen to 
occur until at about 90 MeV it nearly vanishes. This constitutes 
an example of a dynamical {\em restoration} of a symmetry (here the 
left-right parity symmetry) and reflects the complex nature of the 
PV longitudinal response. The near-vanishing of the asymmetry at 
$\omega\approx90$ MeV in Fig.~\ref{fig:Asym} stems from the 
cancellation between the {\em positive} contribution it gets from 
$R_{\text L}^{\text AV}$ and the {\em negative} one it gets from 
$R_{\text T}^{\text AV}$. This trend of course fades away at larger 
$\theta$, where the role of the longitudinal PV response gradually 
becomes irrelevant. Finally, we see that at larger momenta, where 
the impact of correlations is no longer so strongly felt, the nearly 
perfect restoration of the left-right symmetry does not show up 
anymore. It is, however, still true (even at 1 GeV/c) that an energy 
exists where the free and the correlated values of the asymmetry coincide.

From the lower panels of Fig.~\ref{fig:Asym} it clearly appears that at
backward angles ${\cal A}$ is almost equally sensitive to the nuclear 
dynamical effects and to the uncertainties in the nucleonic form 
factors. This raises the question of whether or not the asymmetry 
itself is a suitable observable for extracting information on either 
the nuclear correlations or the single-nucleon form factors. As a 
consequence we are led to consider three different integrated 
observables that have been introduced to emphasize one of the two 
aspects of the problem. They are:

\begin{itemize}
\item[]
 \begin{equation}
  \Delta{\cal A}(q,\theta) \equiv {1 \over{\Delta\omega}}\Bigl[
  \int_{\omega_{min}}^{\omega_{QEP}}\!\!d\omega
   {\cal A}(\theta;q,\omega) -
  \int_{\omega_{QEP}}^{\omega_{max}}\!\!d\omega
   {\cal A}(\theta;q,\omega)\Bigr]
\label{eq:sigmin}
\end{equation}

\item[]
\begin{equation}
  {\overline{\cal A}}(q,\theta) \equiv {1\over{\Delta\omega}}
    \int_{\omega_{min}}^{\omega_{max}}\!\!d\omega\,
    {\cal A}(\theta;q,\omega) 
\label{eq:abar}
\end{equation}
and

\item[]
\begin{equation}
  {\cal R}(q,\theta) \equiv {
    \int_{\omega_{min}}^{\omega_{max}}\!\!d\omega\ W^{PV}(q,\omega)
    \big/\widetilde{X}_{\text T}(\theta,\tau,\psi;\eta_F) \over
    \int_{\omega_{min}}^{\omega_{max}}\!\!d\omega\ W^{EM}(q,\omega)
    \big/X_{\text T}(\theta,\tau,\psi;\eta_F)
  } \,.
\label{eq:sigmatilde}
\end{equation}
\end{itemize}
Here $\omega_{min}$ and $\omega_{max}$ are the RFG response boundaries
for a fixed $q$
\begin{eqnarray}
  \omega_{min} &=& \sqrt{(k_F-q)^2+m_N^2}-\sqrt{k_F^2+m_N^2}
    \label{eq:omegamin}\\
  \omega_{max} &=& \sqrt{(k_F+q)^2+m_N^2}-\sqrt{k_F^2+m_N^2}\ ,
    \label{eq:omegamax}
\end{eqnarray}
the energy interval $\Delta\omega$ is
\begin{eqnarray}
  \Delta\omega &\equiv& \omega_{max} - \omega_{min}
    \label{eq:domega}\\
  &=& \sqrt{(k_F+q)^2+m_N^2} - \sqrt{(k_F-q)^2+m_N^2}
    \label{eq:domegb}\\
  &\cong& 2 k_F q/\sqrt{q^2+m_N^2} 
    \label{eq:domegc}
\end{eqnarray}
and the hadronic functions 
\begin{eqnarray}
W^{EM} &=& v_{\text L} R_{\text L} +v_{\text T} R_{\text T} \\
W^{PV} &=& v_{\text L} R_{\text L}^{\text AV} 
     +v_{\text T} R_{\text T}^{\text AV} 
     +v_{\text T'}R_{\text T'}^{\text VA}
\end{eqnarray}
are divided by
\begin{eqnarray}
  X_{\text T}(\theta,\tau,\psi;\eta_F) &=& v_{\text T} \left( 2\tau G_M^2
+\frac{G_E^2+\tau G_M^2}{1+\tau}\Delta \right) \\
  {\widetilde X}_{\text T}(\theta,\tau,\psi;\eta_F) &=& a_A v_{\text T} \left(
2 \tau G_M{\widetilde G_M} 
+ \frac{ G_E{\widetilde G}_E+\tau G_M{\widetilde G}_M}
{1+\tau} \Delta \right)
\label{eq:XT}
\end{eqnarray}
in order to extract the single-nucleon content from the many-body one
\cite{Alb88}. The quantity $\Delta$ is given later in
Eq.~(\ref{eq:Delta}): since $\Delta\sim\eta_F^2\ll 1$, the contributions
containing $\Delta$ may often safely be neglected. Removing the
single-nucleon content in this way has the following advantages:

\begin{figure}[p]
\begin{center}
\mbox{\epsfig{file=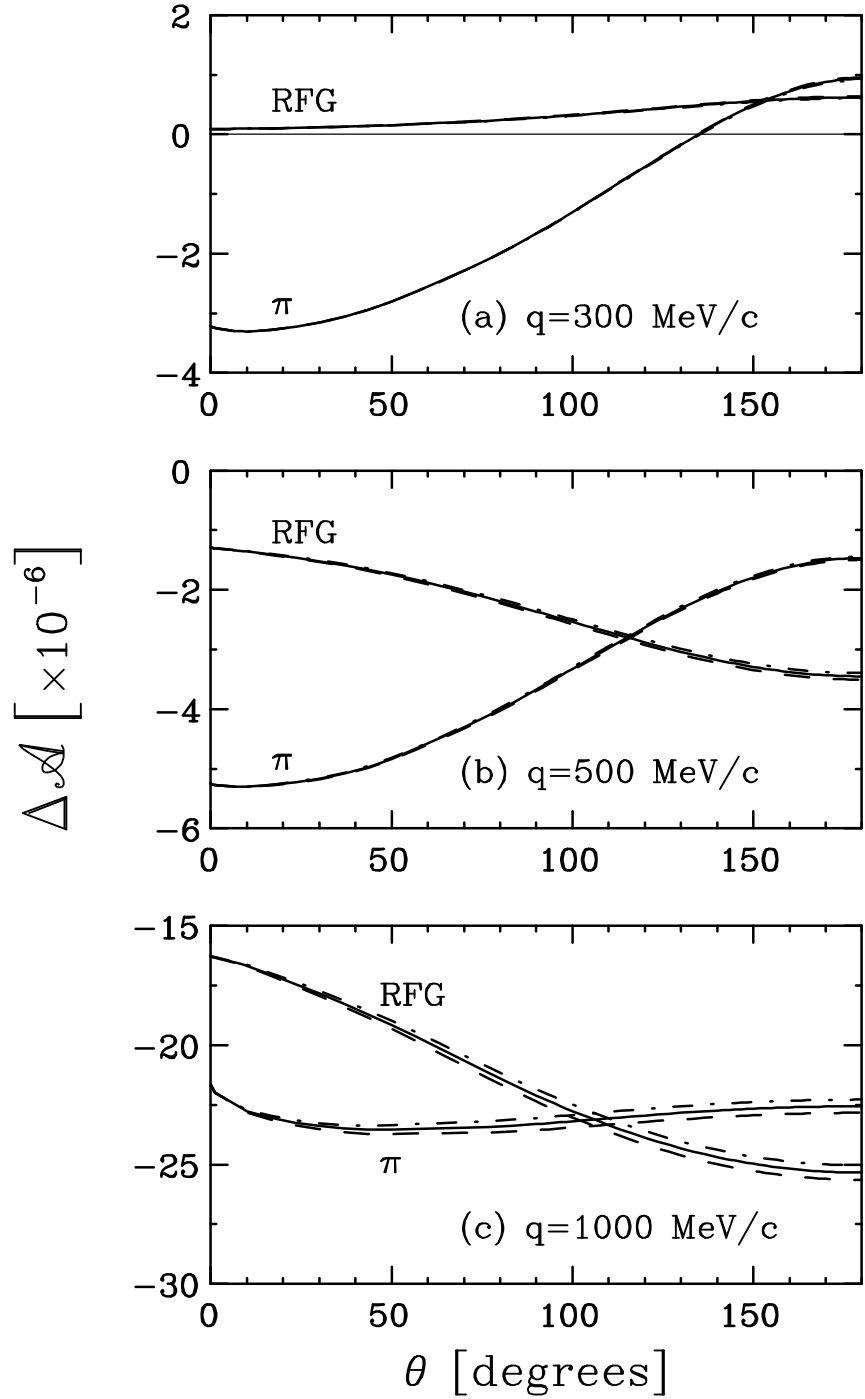,width=.7\textwidth}}
\vskip 2mm
\caption{The quantity $\Delta {\cal A}$ shown as a function of $\theta$
for $q$=300 (a), 500 (b) and 1000 MeV/c (c). The labeling of the curves is
as in Fig.~\ref{fig:Asym}.
}
\label{fig:DeltaA}
\end{center}
\end{figure}

\begin{itemize}
\item[i)] The pionic correlations are particularly felt by 
$\Delta{\cal A}$, as is clearly apparent from Fig.~\ref{fig:DeltaA}. 
Indeed, there we first observe that at $q=300$ MeV/c the results 
obtained with the free RFG model almost vanish because of the 
nearly perfect cancellation between the contributions where 
$\omega<\omega_{QEP}$ and those where $\omega>\omega_{QEP}$;
however, at larger $q$ this cancellation becomes less complete, 
owing partly to the role played by the nucleonic form factors and 
partly to the RFG model itself, whose responses (in contrast to the 
non-relativistic case) become less and less symmetric as $q$ increases.
It is also clear that the correlations, in particular the exchange 
diagram, dramatically alter the prediction of the free RFG, yielding 
a huge $\Delta{\cal A}_{\pi}$ at small $\theta$. This result is 
simply interpreted by observing that of the nuclear responses that 
enter in the asymmetry the pion has its greatest effect on
$R_{\text L}^{\text AV}$  and although in the RFG model the latter 
accounts only for at most about 10\% of the total asymmetry (and 
this only in the forward direction), nevertheless the impact of 
the pionic correlations is violent enough to induce a large negative value of 
$R_{\text L}^{\text AV}$ at small $\omega$, which is in turn reflected
in the large negative value of $\Delta{\cal A}_{\pi}$ at small 
$\theta$ displayed in Fig.~\ref{fig:DeltaA}. We deduce from this 
that the characteristic behavior of $\Delta{\cal A}$ with $\theta$ 
shown in the figure represents one of the most transparent signatures of
pion-induced isoscalar correlations in nuclei (we recall that 
$R_{\text T'}^{\text VA}$ is purely isovector and that in 
$R_{\text T}^{\text AV}$ the isoscalar contribution is strongly 
suppressed --- see Ref.~\cite{Alb93}). 

The observable $\Delta{\cal A}$ has been specifically devised to
enhance the signal for nuclear correlations and to minimize the 
sensitivity to the single-nucleon form factors. This can be inferred 
by observing the three (almost overlapping) lines for each family 
of curves in Fig.~\ref{fig:DeltaA}, corresponding to a variation of 
the strength of the effective axial-vector coupling, $g^{(1)}_A$,  
of $\pm10$\% around the canonical value $g^{(1)}_A=1.26$. The impact 
on $\Delta{\cal A}$ of pionic correlations is seen to be more than 
an order-of-magnitude larger than that arising from variations in 
the axial-vector form factor. Similar results are found for the 
magnetic and electric strangeness form factors (see Ref.~\cite{Bar94}).

\item[ii)] 
The energy-averaged asymmetry ${\overline{\cal A}}$, on the other 
hand, has been devised to minimize the sensitivity to the pionic 
correlations, as these tend to cancel out in the symmetrical integral 
in Eq.~(\ref{eq:abar}). The MEC contribution, on the other hand, 
does not average out in ${\overline{\cal A}}$, although in the $ph$ 
sector of the nuclear excitations it turns out to have a rather
insignificant effect. Typical results are shown in Fig.~\ref{fig:Abar}
for $q=$ 500 MeV/c.

\begin{figure}[t]
\begin{center}
\mbox{\epsfig{file=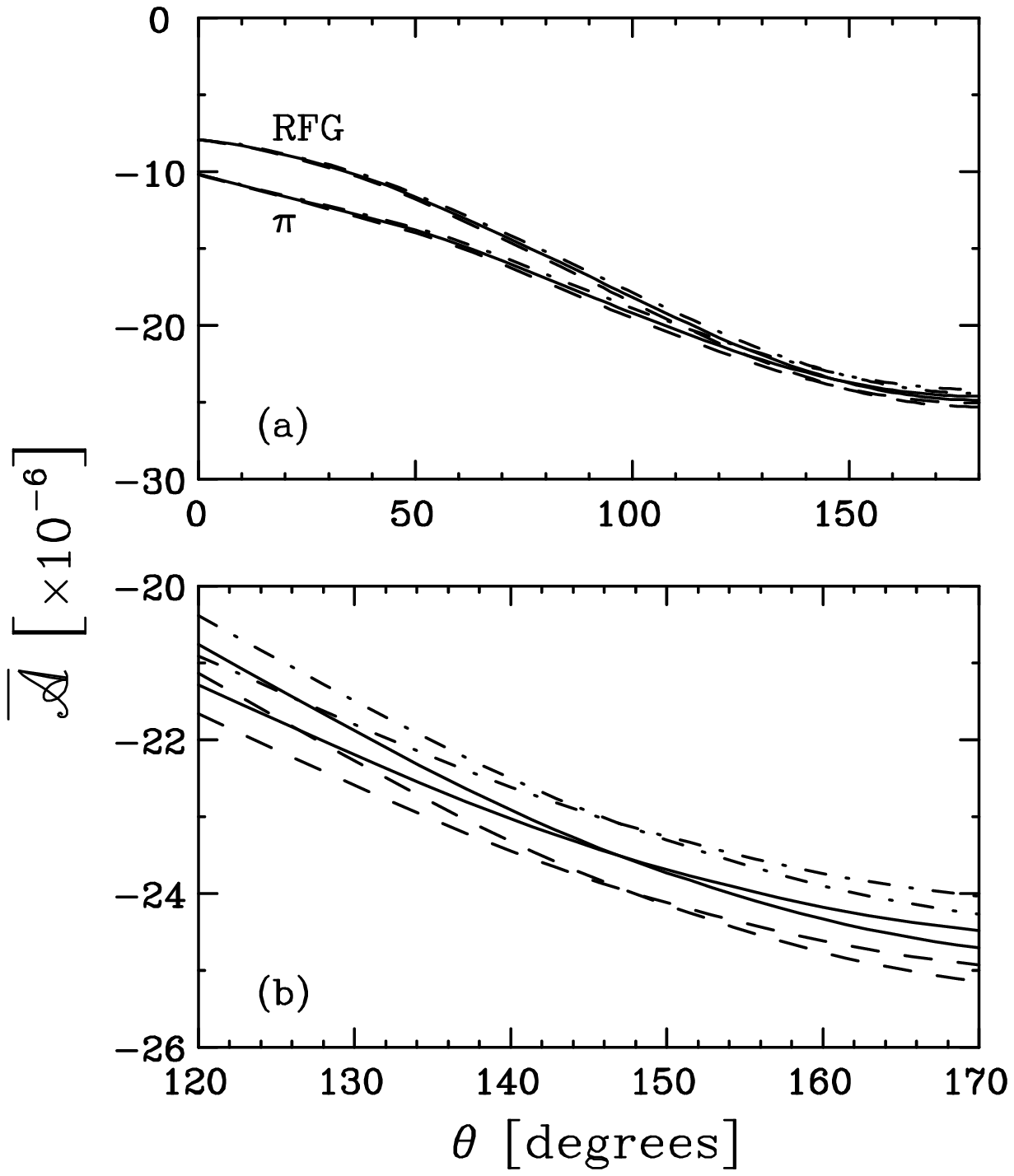,width=.6\textwidth}}
\vskip 2mm
\caption{The energy-averaged asymmetry ${\overline{\cal A}}$ 
displayed as a function of $\theta$ for $q$=500 MeV/c, with the same
labeling of the curves as in Fig.~\ref{fig:Asym}. Panel (a) shows the
entire angular range, while (b) only shows the backward-angle region in
greater detail. 
}
\label{fig:Abar}
\end{center}
\end{figure}

In particular, as seen in the expanded view in the figure, at very
backward scattering angles where one might hope to determine the
effective axial-vector coupling $g_A^{(1)}$ (again variations of
$\pm 10$\% around 1.26 are shown in the figure) the free RFG and pionic
correlated results for the energy-averaged asymmetry come together
(for $q$=500 MeV/c at $\theta\approx 147^o$, which however varies with
$q$). Since this special condition is presumably model-dependent,
it is unlikely that one can count on using such particular kinematics
to effect a determination of $g_A^{(1)}$ through the variations shown
in the figure. 

In contrast to the backward-angle situation, at forward scattering 
angles where the pionic correlations induce drastic modifications in 
$R_{\text L}^{\text AV}$, as we have seen, here the two families of 
curves differ although certainly not as much as in the case of 
$\Delta{\cal A}$. In other words, the observable ${\overline{\cal A}}$
has some of the properties that we are looking for when we construct 
quantities that suppress the effects of correlations while bringing 
out the dependences on the single-nucleon form factors; however, this 
particular observable appears not to be entirely optimal. Since the 
PV longitudinal response is so strongly affected by the presence of 
pionic correlations, it is necessary to adopt an alternative approach 
to minimize these effects, leading to the introduction of the 
observable in Eq.~(\ref{eq:sigmatilde}).

\begin{figure}[t]
\begin{center}
\mbox{\epsfig{file=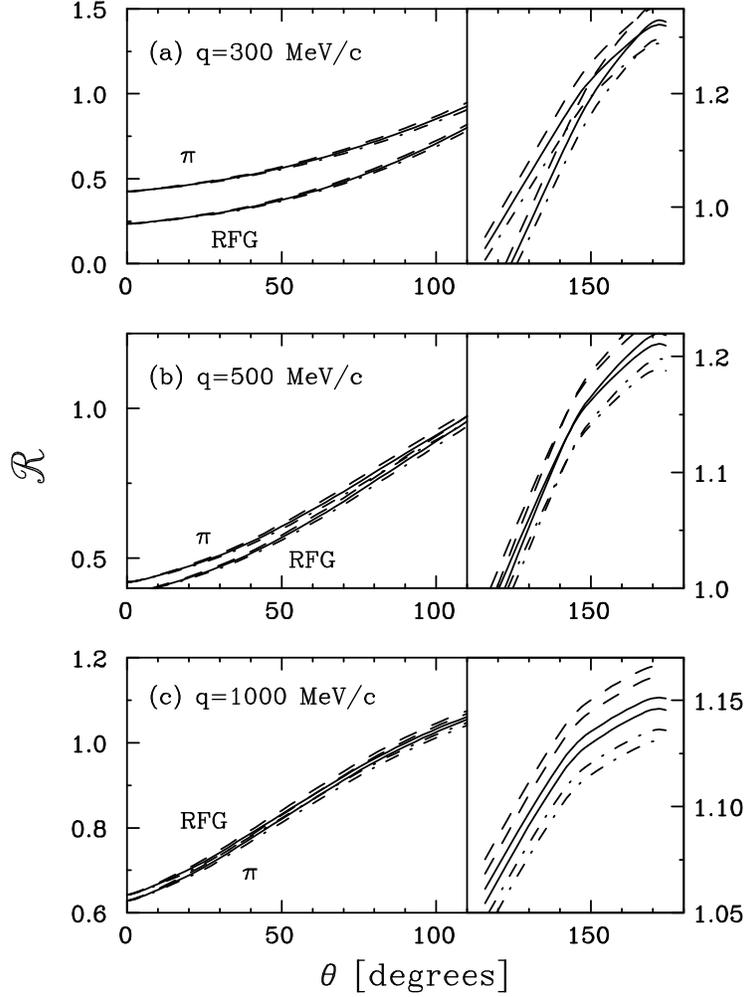,width=.6\textwidth}}
\vskip 2mm
\caption{The quantity ${\cal R}$ versus $\theta$ for $q$=300 (a), 
500 (b) and 1000 MeV/c (c). The curves are labeled as in 
Fig.~\ref{fig:Asym}.
}
\label{fig:R}
\end{center}
\end{figure}

\item[iii)]
The procedure proposed in Refs.~\cite{Bar94,Alb88} for the definition of 
the quantity ${\cal R}$ has the goal of scaling the results (see 
the next section) by dividing out most of the single-nucleon content 
through the use of the dividing factors $X_{\text T}$ and 
${\widetilde X}_{\text T}$.
In Fig.~\ref{fig:R} we show ${\cal R}$ as a function of $\theta$
for three values of $q$.  Again two families of curves are displayed, 
one for the free RFG and one for the model with pionic effects
included (labeled $\pi$), and each family has three curves 
($g_A^{(1)}=1.26$, $1.26+10$\% and $1.26-10$\%). In panel b  we see 
a significant range of angles over which the pionic effects provide 
negligible modifications with respect to the RFG results and where 
the (merged) curves with the three values of the 
isovector/axial-vector strength can clearly be discerned. This 
behavior is similar at $q=1$ GeV/c, although not quite as nicely 
separated. Even so, the difference between the RFG and pionic families
at backward scattering angles amounts to an effective change in 
$g_A^{(1)}$ of only about 4\%.  

It thus appears that the observable ${\cal R}$ is sufficiently 
uncontaminated by correlation effects for favorable kinematics and 
better suited than ${\overline{\cal A}}$ to disentangling the 
nucleonic form factors  from the nuclear dynamics. It has been 
specifically designed to integrate out the anomalous 
$\omega$-dependence, leaving quantities that for the most part only 
retain sensitivities to variations in the single-nucleon form factors. 

\end{itemize}

\section{ Scaling and sum rules}
\label{sec:sumrules}

In this section we briefly address several issues that arise in 
discussing scaling and sum rules, showing how these two properties 
are interrelated and how they constrain the nuclear models. At 
sufficiently large three-momentum transfer $q$ the so-called 
$y$-scaling region occurs when the energy transfer $\omega$ is 
lower than its value at the quasielastic peak and when 
non-quasielastic processes such as meson production do not 
affect the nuclear responses. The $y$-scaling approach attempts 
to find some function of $q$ and $\omega$, here denoted $G$, such 
that when divided into the inclusive electron scattering cross
section, the result is a reduced response $F(q,y)$ that scales as a function
of $y$. Here $y$ is an appropriately chosen scaling variable (see 
below), is a function of $(q,\omega)$ and replaces $\omega$ (i.e., 
one uses the variables $q$ and $y$, rather than $q$ and $\omega$). 
``Scaling'' means that for sufficiently large momentum transfers the 
function $F$ becomes universal, namely a function only of $y$, but not of $q$:
\begin{equation}
F(q,y) \stackrel{q \to \infty}\longrightarrow F(y)\equiv F(\infty,y) .
\end{equation}
The choice of the dividing function and scaling variable must be such 
as to remove the single-nucleon content from the nuclear responses 
in as model-independent a manner as possible while still retaining 
essential relativistic effects whenever feasible. 

A parallel strategy concerns the Coulomb Sum Rule (CSR) 
\cite{Cen97b,Amo97}: a dividing function $H_{\text L}$ can be devised such 
that the corresponding reduced longitudinal response 
$r_{\text L} = R_{\text L}/H_{\text L}$ fulfills the CSR.

In past work medium- and high-energy data have been tested with both 
the usual $y$-scaling approach (for a review, see Ref.~\cite{Day90}), 
while more recently the RFG-motivated approach has been applied and 
seen to scale successfully \cite{Wil97,DonS99,Don99}. Actually these 
data appear to support not only scaling, but also superscaling, 
namely the existence of a function related to $F$ that is the same 
for all nuclei \cite{Alb88,DonS99,Don99}. Additionally, it now appears
that the experimental CSR is reasonably well saturated at high 
momentum transfers \cite{Jou96}. As a consequence, any reliable 
nuclear model should simultaneously fulfill the two major requirements of
\vspace{0.2in}

1) {\it scaling}

2) {\it fulfilling the Coulomb Sum Rule}.
\vspace{0.2in}

We now show that the RFG model simultaneously satisfies the above 
properties. In the context of the PWIA for $(e,e'N)$ reactions 
the $y$-scaling variable is defined to be equal and opposite to the smallest 
value of the missing momentum, $p_{min}$, attained in the 
$y$-scaling region: It turns out to read
\begin{eqnarray}
  y &=& -p_{min} = \frac{1}{2W^2} \left\{\left(M^0_A + \omega\right) 
    \sqrt{W^2-\left(M^0_{A-1} + m_N\right)^2} 
    \sqrt{W^2-\left(M^0_{A-1}-m_N\right)^2}\right. \nonumber \\
    && -\left.q\left[W^2+\left(M^0_{A-1}\right)^2 - m_N^2\right]\right\} ,
\label{eq:y-var}
\end{eqnarray}
where
\begin{equation}
W= \sqrt{\left(M^0_A + \omega\right)^2 - q^2} ,
\label{eq:defW}
\end{equation}
$M^0_A$ and $M^0_{A-1}$ being the masses of the initial and daughter 
nuclei (in their ground states), respectively. The energy transfer 
can, of course, be expressed in terms of $q$ and $y$. In particular, 
it must lie in the range \cite{Cen97b} $\omega_t \leq \omega \leq q $, where
\begin{equation}
  \omega_t = E_S + \sqrt{(M^0_{A-1}+m_N)^2+q^2}-(M^0_{A-1}+m_N)
\end{equation}
is the threshold energy, with $E_S=m_N+M^0_{A-1}-M^0_A$ being the 
nuclear separation energy. The scaling variable in 
Eq.~(\ref{eq:y-var}) vanishes when
\begin{equation}
\omega = \omega_0 = E_S + \sqrt{m_N^2 + q^2} - m_N ,
\end{equation}
which is roughly the position of the quasielastic peak, and hence the 
scaling region is characterized by having $y$ negative. 

In order to study the scaling behavior of the {\it inclusive} $(e,e')$
process one should be able to remove the effective $eN$ cross section 
from under the integrals involved in going from coincidence to inclusive 
scattering. These integrals extend over the missing momentum $(p_m)$ 
and over an energy that characterizes the degree of excitation of 
the daughter nucleus, ${\cal E}$:
\begin{equation}
{\cal E} = E_{A-1}-E_{A-1}^0\geq 0 ,
\label{eq:calE}
\end{equation} 
where $E_{A-1}$ is the energy of the unobserved daughter system (in
general in an excited state) and $E_{A-1}^0$ is that energy when this
system is in its ground state, i.e., has mass $M_{A-1}^0$. Naturally
${\cal E}$ can be re-expressed in terms of the missing energy $E_m$; 
one has roughly that $E_m\cong {\cal E}+E_S$.
If one assumes that the proton and neutron distributions 
inside the nucleus are equal, which is a reasonable approximation for 
$N=Z$ nuclei, and that the most important contributions to the 
nuclear spectral function arise from the lowest values of 
$(p,{\cal E})$ that can be reached for given values of $q$ and 
$y$ (in the scaling region these are ${\cal E}=0$ and $p=-y$), 
then the function one hopes will scale as a function of $y$ when 
$q\rightarrow\infty$ is
\begin{equation}
  F(q,y) \equiv \frac{ d^2\sigma / d\Omega_e d\omega }
    { {\tilde\sigma}_{eN}(q,y;p=-y,{\cal E}=0) } .
\label{eq:Fqy}
\end{equation}
In the PWIA this is indeed found to be the case, since in the limit 
$q\to\infty$ Eq.~(\ref{eq:Fqy}) becomes a function only of $y$, 
namely it scales \cite{Day90}.

Turning now to the RFG model, its spectral function \cite{Cen97b} is
\begin{equation}
  \tilde S^{\text{RFG}} (p, {\cal E}) = \frac{3 A}{8\pi k^3_F}
    \theta (k_F - p) \delta \left [{\cal E} (p) 
    - {\cal E}^{\text{RFG}} (p) \right] ,
\label{eq:support}
\end{equation}
where the excitation energy is
\begin{equation}
{\cal E}^{\text{RFG}} (p) = \left( \sqrt{k^2_F + m_N^2}
    - \sqrt{p^2 + m_N^2} \right) \ .
\label{eq:EFG}
\end{equation}
Defining the RFG scaling variable through the intercept of the support
of the RFG spectral function given in Eq.~(\ref{eq:support}) and 
the kinematical boundaries in the missing energy-missing momentum 
plane one obtains the $y$-scaling variable of RFG
\begin{equation}
  y_{\text{RFG}} = m_N \zeta = m_N \left(\lambda \sqrt{1+\frac{1}{\tau}} - 
    \kappa\right) ,
\label{eq:yRFG}
\end{equation}
where the dimensionless variables $\lambda = \omega/2m_N$, 
$\kappa = q/2m_N$ and $\tau = \kappa^2 - \lambda^2$ have been introduced. 

A different scaling variable was originally proposed for the RFG in 
Ref.~\cite{Alb88}, namely 
\begin{equation}
  \psi = \frac{1}{\sqrt{\xi_F}} \frac{\lambda - \tau}
   {\sqrt{(1+\lambda)\tau + \kappa \sqrt{\tau (1+\tau)}}} ,
\label{eq:psi}
\end {equation}
where $\xi_F = \epsilon_F - 1 = \sqrt{1 + \eta_F^2} - 1$ and 
$\eta_F =k_F/m_N$ are the dimensionless Fermi kinetic energy and 
momentum, respectively. With some algebra it can be shown that 
the relation between the two scaling variables is
\begin{equation}
 \xi_F \psi^2 = \sqrt{1 + (y_{\text RFG}/m_N)^2} - 1 .
\label{eq:psi2}
\end{equation}
The physical significance of $\psi$ is then immediately apparent: 
among the nucleons responding to an external probe one has the 
smallest kinetic energy and this is given by $\psi^2$ (in units 
of the dimensionless Fermi kinetic energy $\xi_F$). Instead of 
working from the unseparated inclusive cross section towards a 
reduced response that, if successful, would scale as 
$q \to \infty$, one can work directly with the separated 
longitudinal and transverse responses, $R_{\text L}$ and $R_{\text T}$, since 

\begin{itemize}
\item[1)] we are most interested in model-to-model comparisons and 
the same procedures may be followed in each case (i.e., focusing 
on L or T responses directly), 
\item[2)]  
a few cases exist where L/T separations have been performed experimentally,
\item[3)] we wish to draw comparisons with studies of the CSR where 
only the $L$ response is relevant.
\end{itemize}

In this spirit we seek reduced responses denoted 
$F_{\text L,T} (\kappa,\psi)$ that scale. These are to be obtained from 
the inclusive response functions $R_{\text L,T} (\kappa,\lambda)$ by 
dividing through by specific functions, denoted 
$G_{\text L,T} (\kappa,\lambda)$:
\begin{equation}
F_{\text L,T} (\kappa,\psi)\equiv R_{\text L,T} (\kappa, \lambda)
  /G_{\text L,T} (\kappa, \lambda) .
\label{eq:defG}
\end{equation}
If the dividing functions are chosen appropriately, then as above the 
reduced responses defined in Eq.~(\ref{eq:defG}) will scale, namely, 
become functions only of a single scaling variable such as $\psi$ 
defined above when $\kappa\rightarrow\infty$,
\begin{equation}
F_{\text L,T} (\kappa,\psi) 
\stackrel{\kappa \to \infty}\longrightarrow F_{\text L,T} 
(\psi)\equiv F_{\text L,T} (\infty,\psi) .
\end{equation}
Such dividing functions, derived in Ref.~\cite{Bar98,Cen97b}, are
\begin{eqnarray}
G_{\text L} (\kappa,\lambda) &=& \frac{Z U_{\text Lp}+N U_{\text Ln}}
    {2\kappa [1+\xi_F (1+\psi^2)/2]} \\
  &=& \frac{1}{2\kappa} (Z U_{\text Lp}+N U_{\text Ln}) + {\cal O}(\xi_F) 
\label{eq:RFG-G}
\end{eqnarray}
and
\begin{eqnarray}
G_{\text T} (\kappa,\lambda) &=& \frac{Z U_{\text Tp}+N U_{\text Tn}}
    {2\kappa [1+\xi_F (1+\psi^2)/2]} \\
 &=& \frac{1}{2\kappa} (Z U_{\text Tp}+N U_{\text Tn}) + {\cal O}(\xi_F) ,
\label{eq:RFG-GT}
\end{eqnarray}
where (see Ref.~\cite{Don91})
\begin{equation}
  U_{\text Lp,n} = \frac{\kappa^2}{\tau} \left[G_{\text Ep,n}^2 (\tau) 
   + W_{2{\text p,n}}(\tau) \Delta\right] 
\label{eq:URFG}
\end{equation}
\begin{equation}
  U_{\text Tp,n} = 2\tau G_{\text Mp,n}^2 (\tau) 
   + W_{2\text{p,n}}(\tau) \Delta \ ,
\label{eq:URFGT}
\end{equation}
with 
\begin {equation}
  W_{2{\text p,n}}(\tau) = \frac{1}{1 + \tau} \left[G_{\text Ep,n}^2 (\tau) 
   + \tau G^2_{\text Mp,n} (\tau)\right] 
\label{eq:WRFG}
\end{equation}
and
\begin{equation}
  \Delta = \frac{\tau}{\kappa^2} \left[ \frac{1}{3} \left( \epsilon^2_F 
    + \epsilon_F \sqrt{1 + \zeta^2} + 1 + \zeta^2\right) 
    + \lambda \left(\epsilon_F + \sqrt{1 + \zeta^2}\right) + \lambda^2\right]
    - (1 + \tau) .
\label{eq:Delta}
\end{equation}
Dividing the longitudinal and transverse RFG response functions by 
$G_{\text L}$ and $G_{\text T}$  yields the reduced responses
\begin{equation}
  F_{\text L}^{\text{RFG}}(\psi) = 
  F_{\text T}^{\text{RFG}}(\psi) = 
 \frac{3\xi_F}{2m_N \eta^3_F} 
    (1 - \psi^2)\theta(1 - \psi^2)\left[1+\frac{1}{2} \xi_F(1+\psi^2)\right] ,
\label{eq:RFGscale}
\end {equation}
that, by construction, scale with $\psi$, $\zeta$ or $y_{\text RFG}$.

In parallel with the scaling behavior of the RFG one can study the 
CSR and the various energy-weighted moments of another reduced 
response denoted $r_{\text L} (\kappa, \lambda)$, introduced in 
Refs.~\cite{Alb88,Cen97b}. Here the longitudinal response 
$R_{\text L} (\kappa,\lambda)$ is divided by a function 
$H_{\text L} (\kappa, \lambda)$ to yield
\begin{equation}
r_{\text L} (\kappa,\psi)\equiv 
R_{\text L} (\kappa, \lambda)/H_{\text L} (\kappa, \lambda) 
\label{eq:defH}
\end{equation}
and the n$^{th}$ moment of the longitudinal response of the nucleus is
given by
\begin{equation}
  \Xi^{(n)} = \int\limits_0^\kappa d\lambda \, \lambda^n \, 
   r_{\text L} (\kappa, \lambda) .
\end{equation}
In particular, the $n=0$ moment, $\Xi^{(0)}$, is the CSR. In the 
case of the RFG the dividing function is
\begin{equation}
  H_{\text L} (\kappa, \lambda) = \frac{\kappa\eta_F^3}{2\xi_F}
  \left( Z U_{\text Lp}+N U_{\text Ln} \right)
 /\left(\partial\psi/\partial\lambda\right) \ .
\label{eq:HL}
\end{equation}

Thus, upon dividing the charge response of the RFG by Eq.~(\ref{eq:HL}) 
one obtains the following reduced longitudinal response
\begin{equation}
  r^{\text{RFG}}_{\text L} (\kappa, \lambda) = \frac{3}{8 m_N } 
    (1 - \psi^2) \theta 
    (1 - \psi^2) \frac{{\partial\psi}}{\partial\lambda} ,
\end{equation}
which, by construction, fulfills the CSR in the non-Pauli-blocked 
domain, as can easily be verified. 

The two dividing functions $G_{\text L}$ (related to the scaling) and 
$H_{\text L}$ 
(related to the CSR) are linked  according to \cite{Cen97b}
\begin {eqnarray}
  G_{\text L} (\kappa, \lambda) &=& \left( \frac{\eta^3_F}{4\xi_F} \right)
    \frac{\partial \psi}{\partial\lambda}
    \frac{1}{1 + \xi_F (1 + \psi^2)/2 } H_{\text L} (\kappa, \lambda) \\
&=& \frac{1}{2} \left( \frac{\kappa}{\tau} \right) 
    \left( \frac{1+2\lambda}{1+\lambda} \right) 
    H_{\text L} (\kappa, \lambda)  + {\cal O}(\xi_F) .
\end{eqnarray}

When experimental data are reduced using the above dividing functions,
they are seen to yield a CSR at high-$q$ and to scale when plotted 
versus any of the scaling variables previously introduced. 

Finally we explore the scaling and CSR properties of nuclear models 
other than the RFG, specifically:

\begin{itemize}

\item[1)] the hybrid model (HM), introduced in Ref.~\cite{Cen97b};

\item[2)] the quantum hadrodynamical (QHD) model \cite{Ser86}.

\end{itemize}

The hybrid model is designed to account for the binding of the 
nucleons inside the nucleus, thus curing a flaw of the RFG related 
to its negative separation energy. The HM has continuum states that are plane
waves, as in the RFG model, but has bound states described by 
shell-model wave functions obtained by solving the Schr\"odinger 
equation with some choice of potential well (in \cite{Cen97b} 
harmonic oscillator bound-state wave functions were used to 
simplify the analysis in the limit where $A\to\infty$). In the HM 
the scaling variable cannot be obtained analytically, since the model 
can only be treated numerically. However, scaling variables exist that 
incorporate the shift result from using Eq.~(\ref{eq:yRFG}) for 
$\zeta'$ (and a corresponding dimensionful variable $y'$) or Eq.~(\ref{eq:psi})
for $\psi'$ by making the replacements $\lambda \to \lambda^\prime$ and 
$\tau \to \tau^\prime = \kappa^2 - {\lambda^\prime}^2$, where
\begin{equation}
  \lambda^\prime = \lambda - \lambda_{\text{shift}} ,
\label{eq:shift}
\end {equation}
with
\begin{equation}
\lambda_{\text{shift}} = \frac{1}{2m_N} (T_F + E_S) 
\label{eq:shiftlam}
\end{equation}
and $T_F= m_N\xi_F $ the Fermi kinetic energy. The HM turns out to 
have the width of its reduced response identical to a RFG computed 
with a Fermi momentum that is somewhat larger than the usual one 
(237 MeV/c for the HM, versus the 230 MeV/c value for $k_F$ used 
for the RFG to correspond to nuclei near $^{40}$Ca or $^{56}$Fe).

Scaling may then be examined for the HM by computing
\begin{equation}
F^{\text{HM}}_{\text L} (\kappa,\psi)\equiv 
  R^{\text{HM}}_{\text L} (\kappa, \lambda)/G_L (\kappa, \lambda)
\label{eq:F-HM}
\end{equation}
and the various energy-weighted moments of the longitudinal response, 
including the zeroth moment or CSR, may be computed using 
\begin{equation}
r^{\text{HM}}_{\text L} (\kappa,\psi)\equiv 
  R^{\text{HM}}_{\text L} (\kappa, \lambda)/H_{\text L} (\kappa, \lambda) .
\label{eq:r-HM}
\end{equation}
Note that the {\em same} dividing factors $G_{\text L}$ and $H_{\text L}$ that
were developed from our discussions of the RFG are used.

\begin{figure}[t]
\begin{center}
\mbox{\epsfig{file=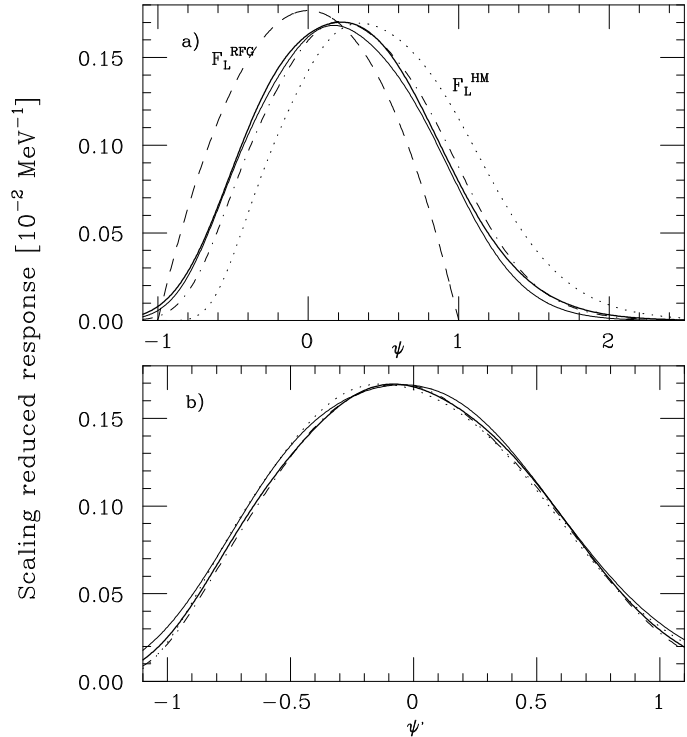,width=.55\textwidth}}
\caption{The reduced response $F^{\text{HM}}_{\text L}$ (the hybrid model 
scaling function) is shown as a function of $\psi$ in panel (a) and 
$\psi'$ in panel (b) for 4 values of $q$ (fine dotted --- 0.5, 
dot-dashed --- 1.0, heavy solid --- 1.5 and solid --- 2.0 GeV/c). 
The RFG result, which scales exactly as a function of $\psi$, is also 
shown for reference as a dashed curve in panel (a).
}
\label{fig:scaHM}
\end{center}
\end{figure}

In Fig.~\ref{fig:scaHM} the scaling function $F^{\text{HM}}_{\text L}$ for 
$^{40}$Ca is displayed versus $\psi$ and $\psi'$ for four different 
values of $q$ (the RFG result is also shown for reference): the HM 
scales either with $\psi$ or with $\psi'$ as $q$ becomes
large. Indeed, only the $q=$ 500 MeV/c plot versus $\psi$ shows 
any appreciable violation of scaling, whereas the scaling versus 
$\psi'$ is excellent. 

In the QHD model \cite{Ser86} protons and neutrons in the nucleus are
described by Dirac spinors and move in strong Lorentz scalar and
vector mean fields. These in turn arise self-consistently from the exchange
of $\sigma$ and $\omega$ mesons between the same nucleons on which
they act. The scalar field dresses the bare mass of the nucleon, 
considerably lowering its value; the vector field uniformly shifts
the fermion spectrum. As a consequence the QHD charge response of
nuclear matter in Hartree approximation is unaffected by the vector
field, while it turns out to be quite sensitive to the effective mass
$m^\ast_N$ induced by the scalar field. This is, of course, true in 
the simple approximation of constant relativistic mean fields. An 
improved description allows for an energy-dependence of the latter, 
which helps to account for the data of proton-nucleus elastic scattering.

As for the HM model, scaling can be examined in the QHD model by 
computing
\begin{equation}
F^{\text{QHD}}_{\text L} (\kappa,\psi)\equiv 
  R^{\text{QHD}}_{\text L} (\kappa, \lambda)/G_{\text L} (\kappa, \lambda)
\label{eq:F-QHD}
\end{equation}
and likewise the various energy-weighted moments of the 
longitudinal response computed using 
\begin{equation}
r^{\text{QHD}}_{\text L} (\kappa,\psi)\equiv 
  R^{\text{QHD}}_{\text L} (\kappa, \lambda)/H_{\text L} (\kappa, \lambda) .
\label{eq:r-QHD}
\end{equation}
Since the dividing factors are (at least to a very good level 
of approximation) {\it universal}, accordingly we use the 
{\em same} dividing factors $G_{\text L}$ and $H_{\text L}$ 
that were developed 
from our discussions of the RFG.

\begin{figure}[t]
\begin{center}
\mbox{\epsfig{file=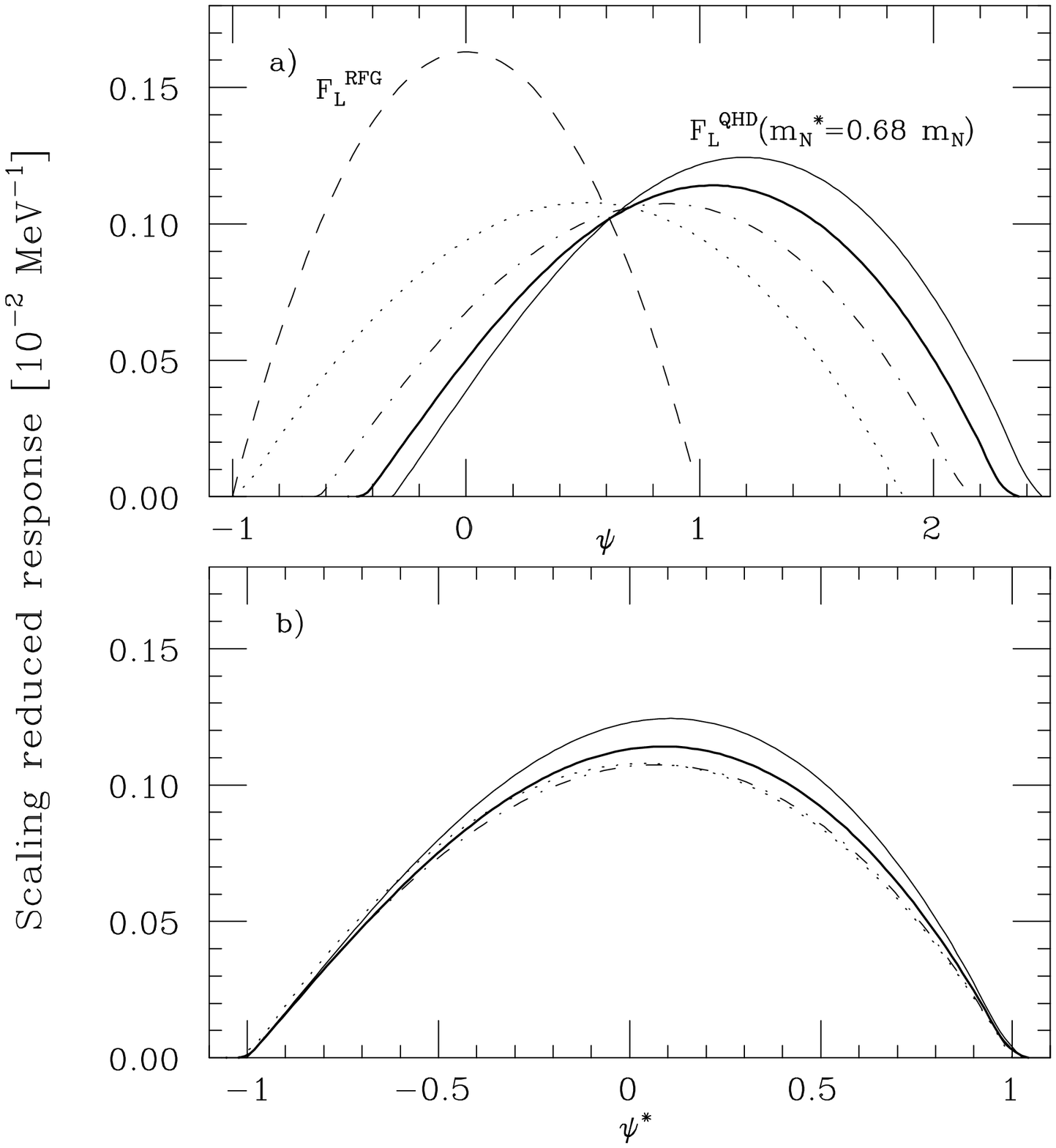,width=0.55\textwidth}}
\caption{The reduced response $F^{\text{QHD}}_{\text L}$ with $m_N^\ast =$ 
0.68 $m_N$ is shown as a function of $\psi$ in panel (a) and 
$\psi^\ast$ in panel (b) for 4 values of $q$ (fine dotted --- 0.5, 
dot-dashed --- 1.0, heavy solid --- 1.5 and solid --- 2.0 GeV/c). 
The RFG result, which scales exactly as a function of $\psi$, is 
also shown for reference as a dashed curve in panel (a).
}
\label{fig:scaQHD}
\end{center}
\end{figure}

In Fig.~\ref{fig:scaQHD} we display the reduced responses in 
Eqs.~(\ref{eq:F-QHD}) and (\ref{eq:r-QHD}) as functions both of 
$\psi$ and also $\psi^{\ast}$, namely, the RFG scaling variable given in 
Eq.~(\ref{eq:psi}) with $m_N$ replaced by $m_N^\ast$ (two different 
values of the effective mass, $m_N^\ast = 0.68\,  m_N$ and 0.8 
$m_N$ are used). It is clearly seen that $F^{\text{QHD}}_{\text L}$ does 
not scale versus $\psi$ when the effective mass is constant and differs from 
$m_N$. As $q$ continues to grow beyond the range of values shown in 
the figures, the results continue to shift to higher $\omega$ and 
never coalesce into a universal curve. When plotted versus 
$\psi^\ast$ the behavior, while better, still does not scale. This 
is in contrast with the RFG and HM results displayed above and, 
importantly, is not what is seen experimentally where the world data 
do appear to scale in $\psi$ \cite{Don99}. The fact that 
experimentally the scaling is observed to occur successfully for 
$q$ greater than about 1 GeV/c suggests that $m_N^\ast/m_N$ should 
not deviate appreciably from unity for such kinematics.

\begin{figure}[t]
\begin{center}
\mbox{\epsfig{file=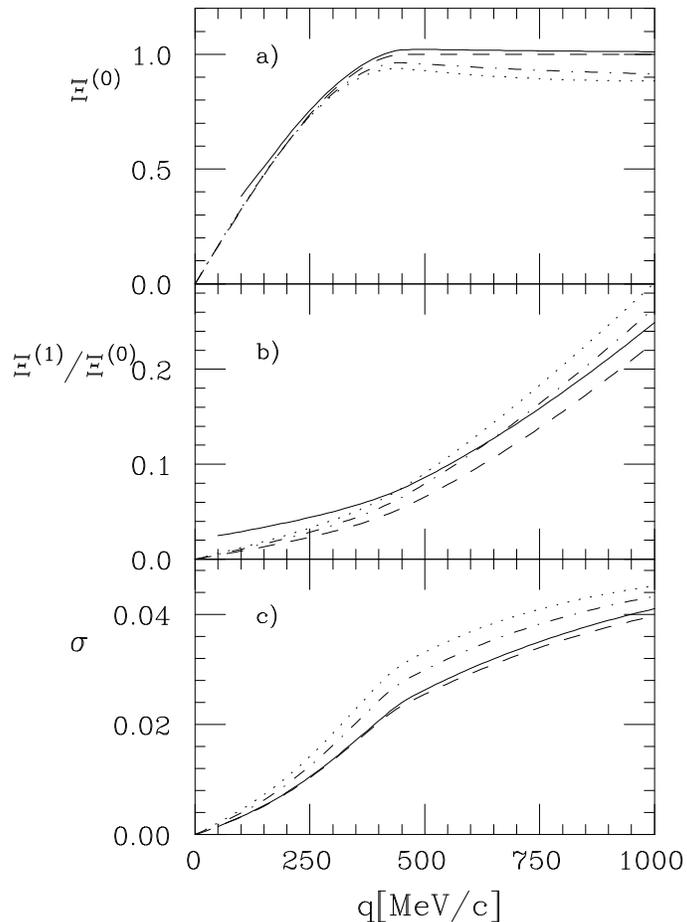,width=0.55\textwidth}}
\caption{The Coulomb sum rule (a), energy-weighted sum rule (b) 
and variance (c) are shown as functions of $q$ for three models:
dashed --- the RFG, solid --- the HM and the QHD model with 
$m_N^\ast$=0.8 $m_N$ (dot-dashed) and 0.68 $m_N$ (dotted). 
}
\label{fig:CSR}
\end{center}
\end{figure}

Finally, the CSR ($\Xi^{(0)}$), the energy-weighted sum rule 
($\Xi^{(1)}/\Xi^{(0)}$) and the variance 
$\sigma=\sqrt{\Xi^{(2)}-(\Xi^{(1)})^2}$ of $^{40}$Ca corresponding to
the HM, QHD (with $m_N^\ast = 0.68$ $m_N$ and 0.8 $m_N$) and RFG
models are displayed in Fig.~\ref{fig:CSR}. The RFG model and the 
HM both saturate the CSR at high-$q$. In contrast, the QHD model 
does so only if the effective value of $m_N^\ast/m_N$ evolves with 
increasing $q$ towards unity, as suggested by the latest version of the model. 

\newpage

\section{ Outlook and perspectives }
\label{sec:outlook}

In this paper we have discussed the parity-conserving and violating- inclusive
nuclear responses, addressing the following issues

\begin{itemize}

\item[a)] Where do they occur?

\item[b)] How can we describe them?

\item[c)] What is their input?

\end{itemize}

With respect to the first item we have limited our focus to the quasielastic 
peak (QEP), since at larger excitation energies where for example the 
$\Delta$ plays a role we do not expect a description of the nucleus only
in terms of nucleonic and mesonic degrees of freedom to be adequate.
At some point QCD degrees of freedom should become the more appropriate
ones to describe inclusive scattering. On the other hand at energies 
significantly lower than those characterizing the QEP where discrete
excitations, giant resonances, etc. are seen the theoretical 
many-body framework is different from the one used here.

Concerning the second item let us again stress that {\it any} theoretical 
framework should first fulfill (as much as possible) Lorentz
covariance and gauge invariance. For this the RFG appears to be a 
good starting point: It is a covariant model, since its ingredients 
are the fully relativistic nucleon propagators and EM (or WNC) vertices,
and it respects gauge invariance, since the vector currents of the 
nucleon are conserved. When mesons are added to the picture then 
Lorentz covariance and gauge invariance are fulfilled to the extent 
that one allows for a dynamical propagation of the mesons and treats 
the forces and the currents consistently. This turns out to be 
possible for the case of the pion.

Of course the RFG misses surface and finite-size effects; however, 
first of all, these are of minor relevance for the scattering of 
electrons in the QEP and $\Delta$ peak domains, and secondly, they 
can be satisfactorily accounted for within the semiclassical approach, 
which exploits the advantages offered by the translational invariance 
of the RFG and accommodates these advantages to fit the physics of finite 
systems \cite{Alb98}.

In the framework of the RFG we have treated nucleon-nucleon correlations 
in a perturbative scheme. Alternatives are represented by

\begin{itemize}

\item[i)] variational approaches

\item[ii)] loop expansions in the path integral framework,
which however represent an alternative regrouping of perturbation theory.

\end{itemize}

Concerning perturbation theory we have seen that basic ingredients are
the diagrams associated with the HF mean field and those related to 
the {\it antisymmetrized} RPA. The emphasis on the antisymmetrization 
arises from the recognition that very important carriers of the
nuclear force, notably $\pi$ and $\rho$ mesons, only act through 
the exchange diagrams.

The short-range correlations (SRC) induced by the violent repulsion present in
the N-N force at small distances are then inserted via the ladder diagrams.
It is, however, doubtful whether ladders can be covariantly computed, 
especially at high densities (or Fermi momenta $k_F$) where they become
increasingly important and the role of relativity cannot be ignored.
In addition, it may be that searching for a totally deterministic 
account of the N-N correlations is not the optimal way to proceed ---
should, for example, quantum chaos be at work in atomic nuclei, then 
costly efforts to compute SRC would not represent the most efficient 
way of interpreting the nuclear response functions.

A further serious challenge facing the perturbative approach relates to
the estimate of the size of diagrams that are not considered. In this 
connection it appears that only the loop expansion offers a consistent
criterion (the number of loops) to organize the perturbative series in classes
of homogeneous diagrams allowing at the same time an estimate of the rate of 
convergence of the resulting new expansion. However, although this is 
theoretically established, its practical implementation is far from trivial.

Finally a few words on the input to be fed into the perturbative scheme
are appropriate. In general the Bonn potential appears as a 
well-founded representation of the N-N force. Indeed it provides an 
excellent representation of the energy behavior of the N-N phase 
shifts in free space and of the deuteron's properties. In
addition, it can be derived from a Lagrangian defined in terms of
nucleonic and mesonic fields, and therefore cast in the framework of an
effective field theory that, from the theoretical viewpoint, represents a
highly desirable feature. In particular, it allows for a test of the 
gauge invariance of the theory, i.e. the consistency between the
forces and currents. In this connection it is worthwhile to 
emphasize that the most direct way to assess the validity of the
mesons-plus-nucleons model of the nucleus ultimately rests on a deeper
understanding of the role of meson exchange currents (MEC) in the 
nuclear responses. Much has been done in this field, although much 
remains to be done.

It seems to us that at present the analysis of the 
nuclear responses is best performed on the basis of a hadronic
approach such as the one we have pursued using the Bonn potential.
It is important to remember, however, that such potentials are being
used for nucleons within the nucleus and that these are differently
off-shell from the conditions found in studying N-N scattering or the
ground state of the deuteron. 

In summary, in Section II of the present article we have analyzed the
quasielastic response functions for inclusive electron scattering in the
standard framework of HF plus RPA. In addition, the impact of 
short-range correlations on the effective particle-hole force has 
been explored within the context of the G-matrix approach.
The analysis has then been extended in Section III to include the 
observables that occur in studying parity-violating electron 
scattering, with emphasis placed on the ``new'' nuclear
axial response. In Section IV the attention has been directed towards 
several general, important properties of the response functions, in 
particular their Coulomb sum rule and scaling/superscaling behavior.
Finally, a short account of what lies ahead has been provided in Section V.

\appendix

\section{ Electromagnetic form factors of the nucleon }
\label{app:A}

Here we give the formulae for the EM form factors introduced in 
the definition of the response functions in Eq.~(\ref{eq:RILT}).

\subsection{ Non-relativistic Fermi gas }

In a non-relativistic calculation one defines, in terms of Sach's form factors,
\begin{mathletters}
\begin{eqnarray}
  {f_{L}^{(I)}}^2 &=& {G_{E}^{(I)}}^2 \\
  {f_{T}^{(I)}}^2 &=& 2\tau{G_{M}^{(I)}}^2 , \qquad I=0,1 ,
\end{eqnarray}
\end{mathletters}
where $\tau=|Q^2|/4m_N^2=(q^2-\omega^2)/4m_N^2$,
$G^{(I)}_X=G_{X_p}+(-1)^I G_{X_n}$ ($X=E,M$). A typical 
parameterization (although many others are possible) is the 
dipolar-plus-Galster one, namely
\begin{equation}
  \begin{array}{lcl}
    G_{E_p}(\tau) &=& G_D^V(\tau) \\
    G_{M_p}(\tau) &=& \mu_p G_D^V(\tau) \\
    G_{M_n}(\tau) &=& \mu_n G_D^V(\tau) \\
    G_{E_n}(\tau) &=& -\mu_n\tau G_D^V(\tau)\xi_n(\tau) .
  \end{array}
\label{eq:GXpn}
\end{equation}
Here $G_D^V(\tau)=(1+\lambda_D^V\tau)^{-2}$ is the vector dipole 
form factor, with $\lambda_D^V\cong4.97$, whereas 
$\mu_p\cong2.793$ and $\mu_n\cong-1.913$ are the proton and neutron 
magnetic moments, respectively. For $G_{E_n}$ we have adopted the 
Galster parameterization \cite{Gal71} with 
$\xi_n(\tau)=(1+\lambda_n\tau)^{-1}$ and $\lambda_n\cong5.6$.

\subsection{ Relativistic Fermi gas }

In Ref.~\cite{Alb90} it was shown that in an RFG calculation a 
very good approximation to the exact treatment of the EM vertices 
can be obtained with the following definitions:
\begin{mathletters}
\begin{eqnarray}
  {f_{\text L}^{(I)}}^2 &=& 
\left[\frac{1}{1+\tau}{G_E^{(I)}}^2+\tau {G_M^{(I)}}^2
    \frac{k_F^2}{2m_N^2}(1-\psi_r)^2\right] , \\
  {f_{\text T}^{(I)}}^2 &=& \frac{2\tau}{(1+\omega/2m_N)^2} {G_M^{(I)}}^2 ,
\end{eqnarray}
\end{mathletters}
where $\psi_r$ is the relativistic scaling variable in
Eq.~(\ref{eq:psiR}), $k_F$ is the Fermi momentum and the other 
quantities have been defined in Eqs.~(\ref{eq:GXpn}).

\section{ First-order self-energy }
\label{app:B}

Here we give the analytic expressions for the first-order self-energy 
based on the potential in Eqs.~(\ref{eq:pot})--(\ref{eq:mes-exch}).
$\Sigma^{(1)}(k)$ is the sum of direct (``Hartree'') and exchange
(``Fock'') terms, namely
\begin{equation}
  \Sigma^{(1)}(k) \equiv \Sigma^{\text{H}}(k) + \Sigma^{\text{F}}(k) ,
\end{equation}
where
\begin{equation}
  \Sigma^{\text{H}}(k) = \rho V_0(0)
\end{equation}
and
\begin{equation}
  \Sigma^{\text{F}}(k) = -\frac{3}{8}\rho\sum_\alpha C^{(\alpha)}_{\text{F}} 
    {\cal S}^{\text{F}}_\alpha(k) ,
\end{equation}
$\rho=2 k_F^3/3\pi^2$ being the nuclear density. In the last equation 
we have introduced the spin-isospin coefficients (note that the 
tensor channels do not contribute)
\begin{eqnarray}
  C^{(0)}_{\text{F}}         &=& 1, \quad
  C^{(\tau)}_{\text{F}}       =  3, \quad
  C^{(\sigma)}_{\text{F}}     =  3, \quad
  C^{(\sigma\tau)}_{\text{F}} =  9, \nonumber \\
  C^{(t)}_{\text{F}} &=& C^{(t\tau)}_{\text{F}} =0 ,
\end{eqnarray}
and defined 
\begin{equation}
  {\cal S}^{\text{F}}_\alpha(k) = \frac{1}{2\pi}\int d\bbox{k}'\,
    \theta(k_F-k') V_\alpha(\bbox{k}-\bbox{k}') .
\end{equation}
In any non-tensor channel $\alpha$ the potential is expressed as 
a combination of the ``$\delta$'' and ``momentum-dependent'' pieces in 
Eq.~(\ref{eq:mes-exch}), for which one finds
\begin{equation}
  {\cal S}^{\text{F}}_{\delta}(k)=\left\{
    \begin{array}{ll}
      g_\delta \frac{2}{3} , & \quad \ell=0 \\
      g_\delta (\lambda^2-\mu^2) w^{\text{F}}_a(\lambda|k)  , &\quad \ell=1 \\
      g_\delta (\lambda^2-\mu^2)^2 w^{\text{F}}_b(\lambda|k), &\quad \ell=2 
    \end{array}
    \right.
\end{equation}
and
\begin{equation}
  {\cal S}^{\text{F}}_{\text{MD}}(k)=\left\{
    \begin{array}{ll}
      g_{\text{MD}}\, \mu^2 w^{\text{F}}_a(\lambda|k) , & \quad \ell=0 \\
      g_{\text{MD}}\, \mu^2 
        [w^{\text{F}}_a(\mu|k)-w^{\text{F}}_a(\lambda|k)], & \quad \ell=1 \\
      g_{\text{MD}}\, \mu^2
        [w^{\text{F}}_a(\mu|k)-w^{\text{F}}_a(\lambda|k)
        -(\lambda^2-\mu^2)w^{\text{F}}_b(\lambda|k)] , & \quad \ell=2 ,
    \end{array}
    \right. 
\end{equation}
where $\ell$ represents the power of the form factors (see 
Eq.~(\ref{eq:mes-exch})). Here we have introduced the dimensionless 
form factor cut-off, $\lambda=\Lambda/k_F$, and meson mass, 
$\mu=m/k_F$, and we have defined
\begin{mathletters}
\begin{eqnarray}
  w_a^{\text{F}}(\lambda|k) &=& 1-\lambda\left[
    \arctan\left(\frac{1-k}{\lambda}\right)
    +\arctan\left(\frac{1+k}{\lambda}\right) \right] 
    -\frac{\lambda^2-k^2+1}{4k}
    \ln\left|\frac{\lambda^2+(k-1)^2}{\lambda^2+(k+1)^2} \right| 
    , \nonumber \\ 
  \\
  w_b^{\text{F}}(\lambda|k) &=& \frac{1}{2\lambda}
    \left[\arctan\left(\frac{1-k}{\lambda}\right)
    +\arctan\left(\frac{1+k}{\lambda}\right) \right] 
    +\frac{1}{4k} \ln\left|\frac{\lambda^2+(k-1)^2}{\lambda^2+(k+1)^2} \right|
    .
\end{eqnarray}
\end{mathletters}

\section{ Tensor interaction in the exchange diagrams }
\label{app:C}

The $n$-th order exchange polarization propagator in presence of tensor
interactions has an expression that is slightly more complicated than 
that in Eq.~(\ref{eq:Pinexpsi}), because the tensor operators do not 
in general allow for a factorization of the azimuthal integrations. 
A generic diagram with $m$ non-tensor and $n-m$ tensor interaction 
lines can instead be written as 
\begin{eqnarray}
 \Pi^{(n)\text{ex}}_{\alpha_1...\alpha_m,\alpha_{m+1}...\alpha_n}(q,\omega) &=&
    (-1)^n\left(\frac{m_N}{q}\right)^{n+1}\left(\frac{k_F}{2\pi}\right)^{2n+2}
    \nonumber \\ && \times
    \int_{-1}^{1}dy_1\frac{1}{2}\int_{0}^{1-y_1^2}dx_1\cdot\cdot\cdot
    \int_{-1}^{1}dy_{n+1}\frac{1}{2}\int_{0}^{1-y_{n+1}^2}dx_{n+1} \nonumber \\
  && \times\frac{1}{\psi-y_1+i\eta_{\omega}}W_{\alpha_1}(x_1,y_1;x_2,y_2)
   \cdot\cdot\cdot W_{\alpha_m}(x_m,y_m;x_{m+1},y_{m+1}) \nonumber \\
  && \times W_{\alpha_{m+1}...\alpha_n}(x_{m+1},y_{m+1};...;x_{n+1},y_{n+1})
   \frac{1}{\psi-y_{n+1}+i\eta_{\omega}} \nonumber \\
  && + \sum(\omega\to-\omega) ,
\end{eqnarray}
where $W_{\alpha_i}$ has been defined for the non-tensor channels 
in Eq.~(\ref{eq:Walpha}) and
\begin{eqnarray}
  && W_{\alpha_{m+1}...\alpha_n}(x_{m+1},y_{m+1};...;x_{n+1},y_{n+1}) =
    2^{n-m}\sum_{ij}\sum_{l_1...l_{n-m}}\Lambda_{ji}
    \int_0^{2\pi}\frac{d\varphi_{m+1}}{2\pi}...
    \int_0^{2\pi}\frac{d\varphi_{n+1}}{2\pi} \nonumber \\
  && \qquad\times 
    V_{\alpha_{m+1}}(\bbox{k}_{m+1}-\bbox{k}_{m+2})
      S_{il_1}(\widehat{\bbox{k}_{m+1}-\bbox{k}_{m+2}}) ...
    V_{\alpha_{n}}(\bbox{k}_{n}-\bbox{k}_{n+1})
      S_{l_{n-m}j}(\widehat{\bbox{k}_{n}-\bbox{k}_{n+1}}) .
\end{eqnarray}
In the last expression we have introduced the tensors
\begin{equation}
  S_{ij}(\hat{\bbox{k}}) = 3\hat{\bbox{k}}_i \hat{\bbox{k}}j-\delta_{ij} ,
\end{equation}
such that 
$\sum_{ij}\sigma_i\sigma_j S_{ij}(\hat{\bbox{k}}) = S_{12}(\hat{\bbox{k}})$.

The first-order case is rather simple, since one again gets 
Eqs.~(\ref{eq:Pi1ex})--(\ref{eq:Wppalpha}) with
\begin{equation}
  W_{\alpha}(x,y;x',y') = \int_0^{2\pi}\frac{d\varphi}{2\pi}
    V_{\alpha}(\bbox{k}-\bbox{k}') S_{zz}(\widehat{\bbox{k}-\bbox{k}'}) .
\label{eq:WalphaTN}
\end{equation}
At second order, however, one can use 
Eqs.~(\ref{eq:Pi2ex})--(\ref{eq:GWp}) only when just one tensor 
interaction is present.

\section{ First- and second-order exchange diagrams }
\label{app:D}

Here we give the explicit expressions for the first- and 
second-order exchange diagrams, based on the potential in 
Eqs.~(\ref{eq:pot})--(\ref{eq:mes-exch}). In Eqs.~(\ref{eq:Pi1ex}) 
and (\ref{eq:Q1alpha}) we have seen that 
\begin{equation}
  \Pi^{(1)\text{ex}}_{\alpha}(q,\omega) =
    -\left(\frac{m_N}{q}\right)^{2}\frac{k_F^4}{(2\pi)^4}
    \left[{\cal Q}_\alpha^{(1)}(0,\psi)
    - {\cal Q}_\alpha^{(1)}(\bar{q},\psi)
    + {\cal Q}_\alpha^{(1)}(0,\psi+\bar{q})
    - {\cal Q}_\alpha^{(1)}(-\bar{q},\psi+\bar{q})\right]
    ,
\end{equation}
where 
\begin{equation}
  {\cal Q}_\alpha^{(1)}(\bar{q},\psi) = 2 \int_{-1}^1 dy
   \frac{1}{\psi-y+i\eta_{\omega}}\int_{-1}^1 dy' \, {W_\alpha}''(y,y';\bar{q})
    \frac{1}{y-y'+\bar{q}} ,
\end{equation}
whereas from Eqs.~(\ref{eq:Pi2ex}) and (\ref{eq:Q2aap}) one has
\begin{eqnarray}
  && \Pi^{(2)\text{ex}}_{\alpha\alpha'}(q,\omega) =
    \left(\frac{m_N}{q}\right)^{3}\frac{k_F^6}{(2\pi)^6}
    \left[{\cal Q}_{\alpha\alpha'}^{(2)}(0,0;\psi)
        - {\cal Q}_{\alpha\alpha'}^{(2)}(0,\bar{q};\psi)
        - {\cal Q}_{\alpha\alpha'}^{(2)}(\bar{q},0;\psi)
        + {\cal Q}_{\alpha\alpha'}^{(2)}(\bar{q},\bar{q};\psi) \right.
  \nonumber \\
  &&\qquad \left.
        - {\cal Q}_{\alpha\alpha'}^{(2)}(0,0;\psi+\bar{q})
        + {\cal Q}_{\alpha\alpha'}^{(2)}(0,-\bar{q};\psi+\bar{q})
        + {\cal Q}_{\alpha\alpha'}^{(2)}(-\bar{q},0;\psi+\bar{q})
        - {\cal Q}_{\alpha\alpha'}^{(2)}(-\bar{q},-\bar{q};\psi+\bar{q})
    \right] , \nonumber \\
\end{eqnarray}
where 
\begin{equation}
  {\cal Q}_{\alpha\alpha'}^{(2)}(\bar{q}_1,\bar{q}_2;\psi) = \int_{-1}^1 dy
    \frac{1}{2}\int_{0}^{1-y^2} dx \, 
    {\cal G}_{\alpha}(x,y+\bar{q}_1;\psi+\bar{q}_1)
    \frac{1}{\psi-y+i\eta_{\omega}}
    {\cal G}_{\alpha'}(x,y+\bar{q}_2;\psi+\bar{q}_2)
\end{equation}
and
\begin{equation}
  {\cal G}_{\alpha}(x,y;\psi) = \int_{-1}^1 dy'
    \frac{1}{\psi-y'+i\eta_{\omega}}W'_\alpha(x,y;y') .
\end{equation}
For a meson-exchange potential the quantities that can be calculated 
analytically are those given by Eqs.~(\ref{eq:Walpha}), 
(\ref{eq:WalphaTN}), (\ref{eq:Wpalpha}) and (\ref{eq:Wppalpha}), namely
\begin{mathletters}
\begin{eqnarray}
  W_{\alpha}(x,y;x',y') &=& \int_0^{2\pi}\frac{d\varphi}{2\pi}
   V_{\alpha}(\bbox{k}-\bbox{k}')\phantom{S_{zz}(\widehat{\bbox{k}-\bbox{k}'})}
    \qquad\text{(non-tensor)} \\
  W_{\alpha}(x,y;x',y') &=& \int_0^{2\pi}\frac{d\varphi}{2\pi}
    V_{\alpha}(\bbox{k}-\bbox{k}') S_{zz}(\widehat{\bbox{k}-\bbox{k}'}) 
    \qquad\text{(tensor)} 
\end{eqnarray}
\end{mathletters}
and
\begin{eqnarray}
  W'_\alpha(x,y;y') &=& \frac{1}{2}\int_0^{1-{y'}^2}dx' \, W_\alpha(x,y;x',y')
    \\
  W''_\alpha(y,y';\bar{q}) &=& \frac{1}{2}\int_0^{1-y^2}dx\,
    \frac{1}{2}\int_0^{1-{y'}^2}dx' \, W_\alpha(x,y+\bar{q};x',y') .
\end{eqnarray}
In any channel $\alpha$ the potential is expressed as a combination 
of the terms displayed in Eq.~(\ref{eq:mes-exch}). Then, for each 
of them one finds
\begin{mathletters}
\begin{eqnarray}
  W_{\delta}(x,y;x',y') &=& \left\{
    \begin{array}{ll}
      g_{\delta} , 
        & \quad \ell=0 \\
      g_{\delta} (\lambda^2-\mu^2) w_a(\lambda|x,y;x',y') ,
        & \quad \ell=1 \\
      g_{\delta} (\lambda^2-\mu^2)^2 w_b(\lambda|x,y;x',y') , 
        & \quad \ell=2
    \end{array}
    \right. \\
  W_{\text{MD}}(x,y;x',y') &=& \left\{
    \begin{array}{ll}
      g_{\text{MD}}\, \mu^2 w_a(\mu|x,y;x',y') , 
        & \quad \ell=0 \\
      g_{\text{MD}}\, \mu^2 [w_a(\mu|x,y;x',y')-w_a(\lambda|x,y;x',y')] , 
        & \quad \ell=1 \\
      g_{\text{MD}}\, \mu^2 [w_a(\mu|x,y;x',y')-w_a(\lambda|x,y;x',y') & \\
        \quad - (\lambda^2-\mu^2) w_b(\lambda|x,y;x',y')] , 
        & \quad \ell=2
    \end{array}
    \right. \\
  W_{\text{TN}}(x,y;x',y') &=& \left\{
    \begin{array}{ll}
      g_{\text{TN}} \{[3(y-y')^2+\mu^2] w_a(\mu|x,y;x',y') - 1\} , 
        & \quad \ell=0 \\
      g_{\text{TN}} \{[3(y-y')^2+\mu^2] w_a(\mu|x,y;x',y') & \\
        \quad - [3(y-y')^2+\lambda^2] w_a(\lambda|x,y;x',y')\} , 
        & \quad \ell=1 \\
      g_{\text{TN}} \{[3(y-y')^2+\mu^2] & \\
        \quad\times [w_a(\mu|x,y;x',y') - w_a(\lambda|x,y;x',y')] & \\
        \quad - (\lambda^2-\mu^2) 
        [3(y-y')^2+\lambda^2] w_b(\lambda|x,y;x',y')\} , 
        & \quad \ell=2 ,
    \end{array}
    \right. 
\end{eqnarray}
\end{mathletters}
where again $\ell$ labels the power of the form factors, we have 
introduced the dimensionless form factor cut-off, 
$\lambda=\Lambda/k_F$, and meson mass, $\mu=m/k_F$, and we have defined 
\begin{mathletters}
\begin{eqnarray}
  w_a(\lambda|x,y;x',y') &=&
    \{ [ \lambda^2+(y-y')^2 +x+x' ]^2 - 4xx' \}^{-1/2}
    \\
  w_b(\lambda|x,y;x',y') &=&
    \frac{\lambda^2+(y-y')^2 +x+x'}
         { \{ [ \lambda^2+(y-y')^2 +x+x' ]^2 - 4xx' \}^{3/2} }
    .
\end{eqnarray}
\end{mathletters}

For $W'_{\alpha}$ one finds
\begin{mathletters}
\begin{eqnarray}
  W'_{\delta}(x,y;y') &=& \left\{
    \begin{array}{ll}
      g_{\delta} (1-{y'}^2)/2, 
        & \quad \ell=0 \\
      g_{\delta} (\lambda^2-\mu^2) w'_a(\lambda|x,y;y') ,
        & \quad \ell=1 \\
      g_{\delta} (\lambda^2-\mu^2)^2 w'_b(\lambda|x,y;y') , 
        & \quad \ell=2
    \end{array}
    \right. \\
  W'_{\text{MD}}(x,y;y') &=& \left\{
    \begin{array}{ll}
      g_{\text{MD}}\, \mu^2 w'_a(\mu|x,y;y') , 
        & \quad \ell=0 \\
      g_{\text{MD}}\, \mu^2 [w'_a(\mu|x,y;y')-w'_a(\lambda|x,y;y')] , 
        & \quad \ell=1 \\
      g_{\text{MD}}\, \mu^2 [w'_a(\mu|x,y;y')-w'_a(\lambda|x,y;y') & \\
        \quad - (\lambda^2-\mu^2) w'_b(\lambda|x,y;y')] , 
        & \quad \ell=2
    \end{array}
    \right. \\
  W'_{\text{TN}}(x,y;y') &=& \left\{
    \begin{array}{ll}
      g_{\text{TN}} \{[3(y-y')^2+\mu^2] w'_a(\mu|x,y;y') - (1-{y'}^2)/2\} , 
        & \quad \ell=0 \\
      g_{\text{TN}} \{[3(y-y')^2+\mu^2] w'_a(\mu|x,y;y') & \\
        \quad - [3(y-y')^2+\lambda^2] w'_a(\lambda|x,y;y')\} , 
        & \quad \ell=1 \\
      g_{\text{TN}} \{[3(y-y')^2+\mu^2] [w'_a(\mu|x,y;y')
        - w'_a(\lambda|x,y;y')] & \\
        \quad - (\lambda^2-\mu^2) 
        [3(y-y')^2+\lambda^2] w'_b(\lambda|x,y;y')\} , 
        & \quad \ell=2 ,
    \end{array}
    \right. 
\end{eqnarray}
\end{mathletters}
where
\begin{mathletters}
\begin{eqnarray}
  && w'_a(\lambda|x,y;y') = \frac{1}{2} \nonumber \\
  && \quad \times \ln\left|
    \frac{ \lambda^2+(y-y')^2 +1-{y'}^2-x+
    \sqrt{ [ \lambda^2+(y-y')^2 +1-{y'}^2+x ]^2 - 4(1-{y'}^2)x } }
    {2 [ \lambda^2+(y-y')^2 ] } \right| 
    \nonumber \\
  \\
  && w'_b(\lambda|x,y;y') = \frac{1}{4}\frac{1}{\lambda^2+(y-y')^2}
    \left[1-\frac{\lambda^2+(y-y')^2-1+{y'}^2+x }
    { \sqrt{ [ \lambda^2+(y-y')^2 +1-{y'}^2+x ]^2 - 4(1-{y'}^2)x } } \right]
    .
    \nonumber \\
\end{eqnarray}
\end{mathletters}

Finally, for $W''_{\alpha}$ one finds
\begin{mathletters}
\begin{eqnarray}
  W''_{\delta}(y,y';\bar{q}) &=& \left\{
    \begin{array}{ll}
      g_{\delta} [(1-y^2)/2] [(1-{y'}^2)/2], 
        & \quad \ell=0 \\
      g_{\delta} (\lambda^2-\mu^2) w''_a(\lambda|y,y';\bar{q}) ,
        & \quad \ell=1 \\
      g_{\delta} (\lambda^2-\mu^2)^2 w''_b(\lambda|y,y';\bar{q}) , 
        & \quad \ell=2
    \end{array}
    \right. \\
  W''_{\text{MD}}(y,y';\bar{q}) &=& \left\{
    \begin{array}{ll}
      g_{\text{MD}}\, \mu^2 w''_a(\mu|y,y';\bar{q}) , 
        & \quad \ell=0 \\
      g_{\text{MD}}\, \mu^2 
          [w''_a(\mu|y,y';\bar{q})-w''_a(\lambda|y,y';\bar{q})] , 
        & \quad \ell=1 \\
     g_{\text{MD}}\, \mu^2 [w''_a(\mu|y,y';\bar{q})-w''_a(\lambda|y,y';\bar{q})
        - (\lambda^2-\mu^2) w''_b(\lambda|y,y';\bar{q})] , 
        & \quad \ell=2
    \end{array}
    \right. \nonumber \\ \\
  W''_{\text{TN}}(y,y';\bar{q}) &=& \left\{
    \begin{array}{ll}
      g_{\text{TN}} \{[3(y-y'+\bar{q})^2+\mu^2] w''_a(\mu|y,y';\bar{q}) & \\
        \quad -[(1-y^2)/2] [(1-{y'}^2)/2] \} , 
        & \quad \ell=0 \\
      g_{\text{TN}} \{[3(y-y'+\bar{q})^2+\mu^2] w''_a(\mu|y,y';\bar{q}) & \\
        \quad - [3(y-y'+\bar{q})^2+\lambda^2] w''_a(\lambda|y,y';\bar{q})\} , 
        & \quad \ell=1 \\
      g_{\text{TN}} \{[3(y-y'+\bar{q})^2+\mu^2] [w''_a(\mu|y,y';\bar{q})
        - w''_a(\lambda|y,y';\bar{q})] & \\
        \quad - (\lambda^2-\mu^2) 
        [3(y-y'+\bar{q})^2+\lambda^2] w''_b(\lambda|y,y';\bar{q})\} , 
        & \quad \ell=2 ,
    \end{array}
    \right. 
\end{eqnarray}
\end{mathletters}
where
\begin{mathletters}
\begin{eqnarray}
  && w''_a(\lambda|y,y';\bar{q}) = \frac{1}{8} \Big\{ \nonumber \\
  &&-[ \lambda^2+(y-y'+\bar{q})^2 +2-y^2-{y'}^2 ]
    -2(2-y^2-{y'}^2)\ln|2[\lambda^2+(y-y'+\bar{q})^2]| \nonumber \\
  &&+\sqrt{[ \lambda^2+(y-y'+\bar{q})^2 +2-y^2-{y'}^2 ]^2-4(1-y^2)(1-{y'}^2)} 
    \nonumber \\
  &&+2(1-y^2) \nonumber \\
  && \times\ln\left|\lambda^2+(y-y'+\bar{q})^2+y^2-{y'}^2
    +\sqrt{[ \lambda^2+(y-y'+\bar{q})^2 +2-y^2-{y'}^2 ]^2-4(1-y^2)(1-{y'}^2)} 
    \right| \nonumber \\
  &&+2(1-{y'}^2) \nonumber \\
  && \times\ln\left|\lambda^2+(y-y'+\bar{q})^2-y^2+{y'}^2
    +\sqrt{[ \lambda^2+(y-y'+\bar{q})^2 +2-y^2-{y'}^2 ]^2-4(1-y^2)(1-{y'}^2)} 
    \right| \Big\} \nonumber \\
    \\
  && w''_b(\lambda|y,y';\bar{q}) = \frac{1}{8} \nonumber \\
  && \times \frac{ \lambda^2+(y-y'+\bar{q})^2 +2-y^2-{y'}^2 
    -\sqrt{[ \lambda^2+(y-y'+\bar{q})^2 +2-y^2-{y'}^2 ]^2-4(1-y^2)(1-{y'}^2)} }
    {\lambda^2+(y-y'+\bar{q})^2}
    .
    \nonumber \\
\end{eqnarray}
\end{mathletters}


\begin{references}

\bibitem{Jou96}    J. Jourdan,
                   Nucl.\ Phys.\ {\bf A603} (1996) 117.

\bibitem{Ang96}    M. Anghinolfi {\em et al\/.},
                   Nucl.\ Phys.\ {\bf A602} (1996) 405.

\bibitem{Wil97}    C. F. Williamson {\em et al\/.},
                   Phys.\ Rev.\ C {\bf 56} (1997) 3152.

\bibitem{Del85}    A. Dellafiore, F. Lenz, and F. A. Brieva, 
                   Phys.\ Rev.\ {\bf C31} (1985) 1088.

\bibitem{Del87}    F. A. Brieva and A. Dellafiore, 
                   Phys.\ Rev.\ {\bf C36} (1987) 899.

\bibitem{Shi89}    T. Shigehara, K. Shimizu, and A. Arima, 
                   Nucl.\ Phys.\ {\bf A492} (1989) 388.

\bibitem{Hor90}    C. J. Horowitz and J. Piekarewicz,
                   Nucl.\ Phys.\ {\bf A511} (1990) 461.

\bibitem{Bub91}    M. Buballa, S. Drozdz, S. Krewald, and J. Speth,
                   Ann.\ Phys.\ (N.Y.) {\bf 208} (1991) 346.

\bibitem{Bof93}    S. Boffi, C. Giusti, and F. D. Pacati,
                   Phys.\ Rep.\ {\bf 226} (1993) 1.

\bibitem{Weh93}    K. Wehrberger,
                   Phys.\ Rep. {\bf 225} (1993) 273.

\bibitem{Ama94}    J. E. Amaro, G. C\`o, and A. M. Lallena,
                   Nucl.\ Phys.\ {\bf A578} (1994) 365.

\bibitem{Cai95}    J. C. Caillon and J. Labarsouque,
                   Nucl.\ Phys.\ {\bf A595} (1995) 189.

\bibitem{ Kim95}   H. Kim, C. J. Horowitz, and M. R. Frank,
                   Phys.\ Rev.\ {\bf C51} (1995) 792.

\bibitem{Bar96a}   M. B. Barbaro, A. De Pace, T. W. Donnelly,
                   and A. Molinari, 
                   Nucl.\ Phys.\ {\bf A596} (1996) 553.

\bibitem{Bes96}    J. Besprosvany,
                   Nucl.\ Phys.\ {\bf A601} (1996) 269.

\bibitem{Fab97}    A. Fabrocini,
                   Phys.\ Rev.\ {\bf C55} (1997) 338.

\bibitem{Cen97a}   R. Cenni, F. Conte, and P. Saracco,
                   Nucl.\ Phys.\ {\bf A623} (1997) 391.

\bibitem{Gil97}    A. Gil, J. Nieves, and E. Oset,
                   Nucl.\ Phys.\ {\bf A627} (1997) 599.

\bibitem{DeP98}    A. De Pace,
                   Nucl.\ Phys.\ {\bf A635} (1998) 163.

\bibitem{Fet71}    A. L. Fetter and J. D. Walecka,
                   {\em Quantum Theory of Many-Particle Systems}
                   (McGraw-Hill, New York, 1971).

\bibitem{Wal95}    J. D. Walecka,
                   {\em Theoretical Nuclear and Subnuclear Physics}
                   (Oxford University Press, Oxford, 1995).

\bibitem{Mac87}    R. Machleidt, K. Holinde, and Ch. Elster, 
                   {\em Phys. Rep.\/} {\bf 149} (1987) 1.

\bibitem{Bar96b}   M. B. Barbaro, A. De Pace, T. W. Donnelly,
                   and A. Molinari, 
                   Nucl.\ Phys.\ {\bf A598} (1996) 503.

\bibitem{Amo96}    P. Amore, M. B. Barbaro, and A. De Pace
                   Phys.\ Rev.\ C {\bf 53} (1996) 2801.

\bibitem{Nak84}    K. Nakayama, S. Krewald, J. Speth, and W. G. Love,
                   Nucl.\ Phys.\ {\bf A431} (1984) 419.

\bibitem{Nak87}    K. Nakayama, S. Drozdz, S. Krewald, and J. Speth,
                   Nucl.\ Phys.\ {\bf A470} (1987) 573.

\bibitem{Alb93}    W. M. Alberico, M. B. Barbaro, A. De Pace, T. W. Donnelly,
                   and A. Molinari, 
                   Nucl.\ Phys.\ {\bf A563} (1993) 605.

\bibitem{Bar94}    M. B. Barbaro, A. De Pace, T.W. Donnelly, and A. Molinari, 
                   Nucl.\ Phys.\ {\bf A569} (1994) 701.
                                                                
\bibitem{Gal58}    V. M. Galitskii and A. B. Migdal, 
                   Sov.\ Phys.\ JEPT {\bf 34} (1958) 7.

\bibitem{Abr63}    A. A. Abrikosov, L. P. Gorkov, and I. E. Dzyaloshinski, 
                   {\em Methods of Quantum Field Theory in Statistical Physics}
                   (Dover, New York, 1963).

\bibitem{Alb91}    W. M. Alberico, A. De Pace, A. Drago, and A. Molinari,
                   Rivista Nuovo Cimento {\bf 14} (1991) 1.

\bibitem{Alb90}    W. M. Alberico, T. W. Donnelly, and A. Molinari,
                   Nucl.\ Phys.\ {\bf A512} (1990) 541.

\bibitem{Ose82}    E. Oset, H. Toki, and W. Weise, 
                   Phys.\ Rep. {\bf 83} (1982) 281.

\bibitem{Len80}    F. Lenz, E. J. Moniz, and K. Yazaki, 
                   Ann.\ Phys.\ (N.Y.) {\bf 129} (1980) 84.

\bibitem{Fes92}    H. Feshbach, 
                   {\em Theoretical Nuclear Physics: Nuclear Reactions} 
                   (Wiley, New York, 1992).

\bibitem{Bal63}    R. Balescu,
                   {\em Statistical Mechanics of Charged Particles}
                   (Interscience, New York, 1963) p. 399;
                   N. I. Muskhelishvili,
                   {\em Singular Integral Equations}
                   (Noordhoff, Groningen, 1953) pp. 56--61;
                   G. D. White, K. T. R. Davies, and P. J. Siemens,
                   Ann.\ Physics {\bf 187} (1988) 198.

\bibitem{DeP97}    A. De Pace, C. Garc\'\i a-Recio, and E. Oset,
                   Phys.\ Rev.\ {\bf C55} (1997) 1394.

\bibitem{Spe77}    J. Speth, E. Werner, and W. Wild, 
                   Phys.\ Reports\ {\bf 33} (1977) 127.

\bibitem{Cen97b}   R. Cenni, T.W. Donnelly, and A. Molinari,
                   Phys.\ Rev.\ {\bf C56} (1997) 276.

\bibitem{MJMrev}   M.J. Musolf, T.W. Donnelly, J. Dubach,
                   S.J. Pollock, S. Kowalski, and E. J. Beise,
                   Phys.\ Rep. {\bf 239} (1994) 1.

\bibitem{Pre78}    C.Y. Prescott {\em et al\/.},
                   Phys.\ Lett.\ {\bf B77} (1978) 1347.

\bibitem{Pre79}    C.Y. Prescott {\em et al\/.},
                   Phys.\ Lett.\ {\bf B84} (1979) 524.

\bibitem{SAMPLE}   B. Mueller {\em et al\/.},
                   Phys.\ Rev.\ Lett.\ {\bf 78} (1997) 3824.

\bibitem{HAPPEX}   K Aniol {\em et al\/.},
                   Phys.\ Rev.\ Lett.\ {\bf 82} (1999) 1096.                   

\bibitem{DDS}      T.W. Donnelly, J. Dubach and I. Sick,
                   Nucl.\ Phys.\ {\bf A503} (1989) 589.

\bibitem{Ama98}    J.E.Amaro, M.B.Barbaro, J.A.Caballero, T.W.Donnelly, and 
                   A.Molinari,
                   Nucl.\ Phys.\ {\bf A643} (1998) 349.

\bibitem{Alb88}    W. M. Alberico, A. Molinari, T. W. Donnelly,
                   E.L. Kronenberg, and J.W. Van Orden, 
                   Phys.\ Rev.\ {\bf C38} (1988) 1801.

\bibitem{Amo97}    P. Amore, R. Cenni, T.W. Donnelly, and A. Molinari,
                   Nucl.\ Phys.\ {\bf A615} (1997) 353.

\bibitem{Day90}    D.B. Day, J.S. McCarthy, T.W. Donnelly, and I. Sick,
                   Annu.\ Rev.\ Nucl.\ Part.\ Sci.\ {\bf 40} (1990) 411.

\bibitem{DonS99}   T.W. Donnelly and I. Sick, 
                   Phys.\ Rev.\ Lett.\ {\bf 82} (1999) 3212.

\bibitem{Don99}    T.W. Donnelly and I. Sick, 
                   Phys.\ Rev.\ {\bf C60} (1999) 065502.

\bibitem{Bar98}    M. B. Barbaro, R. Cenni, A. De Pace, T. W. Donnelly,
                   and A. Molinari, 
                   Nucl.\ Phys.\ {\bf A643} (1998) 137.

\bibitem{Don91}    T. W. Donnelly, M.J. Musolf, W. M. Alberico, M. B. Barbaro, 
                   A. De Pace, and A. Molinari, 
                   Nucl.\ Phys.\ {\bf A541} (1992) 525.

\bibitem{Ser86}    B.D. Serot, and J.D. Walecka,
                   Adv.\ Nucl.\ Phys\ {\bf 16} (1986) 1.

\bibitem{Gal71}    S. Galster {\em et al\/.},
                   Nucl.\ Phys.\ {\bf B32} (1971) 221.

\bibitem{Alb98}    W. M. Alberico, G. Chanfray, J. Delorme, M. Ericson,
                   and A. Molinari,
                   Nucl.\ Phys.\ {\bf A634} (1998) 233.


\end{references}
\end{document}